\documentclass[12pt]{article}   
\pdfoutput=1
\usepackage{geometry,bbm}   
\usepackage{ulem}
\usepackage[toc,page]{appendix}             		
\usepackage{graphicx,cancel}
\usepackage[usenames,dvipsnames]{color}
\usepackage{slashed}					
\usepackage{amssymb,amsmath}
\usepackage{hyperref}
\usepackage{mathtools}
\usepackage{cite}
\usepackage{commath}

\usepackage{tikz-cd}
\usepackage{youngtab}

\usepackage{graphicx}
\usepackage{bm}
\def\bea{\begin{eqnarray}}
\def\eea{\end{eqnarray}}
\usepackage{latexsym}
\usepackage{epsfig,amssymb,euscript}
\usepackage{amsmath}
\usepackage{array,calc,epsfig}
\usepackage{bbold}

\newcommand{\wb}{\overline}

\newcommand{\be}{\begin{equation}}
\newcommand{\ee}{\end{equation}}

\numberwithin{equation}{section}

\include{macros}

\begin{document}

\begin{titlepage}

\begin{flushright}
SISSA 23/2019/FISI
\end{flushright}
\bigskip

\begin{center}
{\LARGE
{\bf
On supersymmetry breaking vacua from \\ \vskip 15pt  D-branes at orientifold singularities
}}
\end{center}

\bigskip
\begin{center}
{\large
Riccardo Argurio$^{1}$, Matteo Bertolini$^{2,3}$, Shani Meynet$^{2}$, Antoine Pasternak$^{1}$}
\end{center}

\renewcommand{\thefootnote}{\arabic{footnote}}

\begin{center}
\vspace{0.2cm}
$^1$ {Physique Th\'eorique et Math\'ematique and International Solvay Institutes, \\ Universit\'e Libre de Bruxelles; C.P. 231, 1050 Brussels, Belgium\\}
$^2$ {SISSA and INFN - 
Via Bonomea 265; I 34136 Trieste, Italy\\}
$^3$ {ICTP - 
Strada Costiera 11; I 34014 Trieste, Italy\\}
\vskip 5pt
{\texttt{rargurio@ulb.ac.be, bertmat@sissa.it, \\ 
smeynet@sissa.it, antoine.pasternak@ulb.ac.be}}

\end{center}

\vskip 5pt
\noindent
\begin{center} {\bf Abstract} \end{center}
\noindent
We present a large class of models of D-branes at (orientifold) Calabi-Yau singularities which enjoy dynamical supersymmetry breaking at low energy, by means of either the $SU(5)$ or 3--2 supersymmetry breaking models. Once embedded in a warped throat or, equivalently, in a large $N$ theory, all models display an instability along a Coulomb branch direction towards supersymmetry preserving vacua. Interestingly, the nature of the runaway mechanism is model-independent and has a precise geometrical interpretation. This naturally suggests the properties a Calabi-Yau singularity should have in order for such instability not to occur.  
\vspace{1.6 cm}
\vfill

\end{titlepage}

\newpage
\setcounter{tocdepth}{2} 
\tableofcontents

\section{Introduction}
\label{intro}

An interesting class of supersymmetric gauge theories is the one that can be engineered with D-branes at Calabi-Yau (CY) singularities. They take the general form of quiver gauge theories, with a number of gauge groups and bi-fundamental chiral matter fields joining them. 

In this paper, we will consider four-dimensional theories with the minimal amount of supersymmetry, that is ${\cal N}=1$. We will restrict to toric CYs, a large and well known class of singularities whose geometry and dual gauge theories can be conveniently described using dimer techniques \cite{Franco:2005rj,Franco:2005sm,Kennaway:2007tq}.

When all gauge groups have the same rank, these theories are superconformal and constitute, in fact, a large portion of known AdS/CFT duals. The dynamics becomes richer when fractional branes are added to the set-up. Fractional branes are allowed because there exist non-trivial cycles at the singularities on which higher dimensional branes can wrap, provided they leave no uncanceled tadpoles. This corresponds to allowing for the ranks of the gauge groups in the quiver to be different, provided there are no gauge anomalies. 

Such theories with different ranks have a non-trivial evolution in energy, {\it i.e.}~a renormalization group (RG) flow, which often takes the form of a cascade of Seiberg dualities, the prototype example being the conifold theory \cite{ Klebanov:1998hh,Klebanov:2000nc,Klebanov:2000hb}. Effectively, the ranks of the gauge groups decrease, in steps, as one goes to lower energies. In the brane picture, this corresponds to a decrease in the effective number of (regular) D3-branes as one moves deep inside the dual bulk geometry.

A natural question one would like to answer is what the endpoint of the RG flow is. Typically, one is left at low energy with a smaller quiver theory, where only a fewer number of gauge groups are non-trivial. In other words, this is as if only fractional branes were left, and no regular branes. This latter set-up  can also be usefully considered when the geometry does not allow to define a cascade. This reduced quiver gauge theory usually does not have a superconformal fixed point, but rather one of the following behaviors typical of ${\cal N}=1$ gauge theories: confinement, an effectively ${\cal N}=2$ Coulomb branch, or dynamical supersymmetry breaking (DSB). In the latter case, the supersymmetry breaking can be of runaway type, metastable, or stable. We will be interested in the possibility of obtaining a fully stable supersymmetry breaking vacuum. 

Generically, the supersymmetry breaking behavior at the end of a cascade is of runaway type  \cite{Berenstein:2005xa,Franco:2005zu,Bertolini:2005di,Intriligator:2005aw}.\footnote{We will not discuss the possible presence of metastable vacua since it is simpler, both in QFT and on the geometric side, to study the state of lowest energy. See \cite{Kachru:2002gs,Argurio:2006ny,Argurio:2007qk} for realizations of metastable vacua at the end of a cascade.}  Promising set-ups where such instabilities might be cured are the ones involving orientifolds. As opposed to purely geometric backgrounds, orientifolds are known to allow for a variety of non-generic dynamical effects (see \cite{Blumenhagen:2006ci} for a review), including the possibility of lifting massless moduli \cite{Argurio:2007vqa,Aharony:2007pr}. Indeed, in \cite{Franco:2007ii} two instances of fractional brane configurations at orientifold singularities reproducing exactly the matter content of the so-called uncalculable $SU(5)$ DSB model \cite{Affleck:1983vc} were provided. 

These models stood out as the only D-brane constructions leading to a reliable stable DSB vacuum, until in \cite{Buratti:2018onj} it was shown, in the set-up involving an orientifold of the $\mathbb{C}^3/\mathbb{Z}_{6'}$ orbifold, that the DSB vacuum is actually not stable. This arises as one tries to embed the supersymmetry breaking configuration in a decoupled, UV complete D-brane system (which is realized as a cascade or, more generally, a large-$N$ gauge theory).  Adding $N$ regular D3-branes to the supersymmetry breaking configuration one can see that a Coulomb branch runaway direction opens up and the vacuum energy is set to zero, {\it i.e.}~the lowest energy state is a supersymmetric vacuum.

The problem of finding stable supersymmetry breaking states in well-defined string theory set-ups is of course of the utmost importance, both in the context of string compactifications, in which such configurations arise as warped throats, as well as in the gauge/gravity duality framework. The evidence provided in \cite{Buratti:2018onj} was taken as a negative result and an indication in favor of a new  swampland conjecture, dubbed locally AdS weak gravity conjecture.\footnote{See \cite{Vafa:2005ui} for the original idea about the swampland and \cite{Brennan:2017rbf,Palti:2019pca} for recent reviews, plus references therein.}

In the present paper, we aim at understanding how generic is the situation analyzed in \cite{Buratti:2018onj}, by
finding more set-ups with a putative DSB vacuum, and then checking whether they also display some instability towards a supersymmetric  vacuum. We will restrict to toric singularities with an orientifold projection which is necessary, as argued above, to allow for DSB models on fractional brane configurations. In short, our results are that the uncalculable $SU(5)$ model can be recovered in a number of set-ups, both based on simple $\mathbb{C}^3$ orbifolds as well as on more general toric singularities. We also find that in some of these orientifold singularities it is possible to find brane configurations which support the so-called 3-2 model \cite{Affleck:1984xz}. To our knowledge, this is the first occurence of this model in a top-down string theory set-up.\footnote{Note that in the absence of an orientifold projection, it proved impossible to find brane configurations reproducing the 3-2 model without adding ad hoc terms \cite{Argurio:2006ew}.} Finally, in several models that we analyze, there is a built-in UV complete duality cascade, whose IR dynamics is of such DSB type.

In all these examples a runaway Coulomb branch direction opens up as soon as one tries to embed these brane configurations in a large-$N$ theory, and the true vacuum turns out to be supersymmetric, eventually, as in \cite{Buratti:2018onj}. This comes both as bad and good news. Bad, because it confirms the difficulty to find fully stable non-supersymmetric vacua in D-brane constructions. Good, because it shows that the dynamical mechanism underlying such instability appears to be model-independent. This is suggestive in view of a first principle  understanding of this phenomenon and for the possibility of providing a general argument proving the existence, or the nonexistence, of configurations leading to truly stable DSB vacua. One thing one could ask, for example, is whether it is possible to avoid this Coulomb branch runaway phenomenon by focusing on (orientifolds of) singularities that do not admit such potentially dangerous moduli to start with (which, in terms of branes, means the absence of so-called ${\cal N}=2$ fractional branes). We have not been able to find an example in this class that allows for either an $SU(5)$ or a 3-2 model, but our scan is by no means complete, so we cannot yet exclude the existence of set-ups allowing for DSB and no runaways.

The paper is structured as follows. In section \ref{sum} we present a summary of our main results, to provide the reader with a simple illustration of the physical picture that emerges from the examples we have explored. We then proceed to discuss in details several cases. In section \ref{C3Z6} we review the orientifold of the $\mathbb{C}^3/\mathbb{Z}_{6'}$ singularity, already analyzed in \cite{Buratti:2018onj}, finding not only the $SU(5)$ model but also a 3-2 model, and their instability. In section \ref{PdP4} we consider the orientifold of a pseudo del Pezzo singularity, $PdP4$, where we also recover both an $SU(5)$ and a 3-2 model. The instability is also present, though the presence of anomalous dimensions makes the analysis slightly more subtle. In section \ref{new} we present a large number of new orientifold set-ups, generalizing previous ones, based on orbifolds or blow ups of del Pezzo CY singularities, where $SU(5)$ and possibly 3-2 models are found.  In all such set-ups we identify, again, a runaway Coulomb branch instability. These results might suggest the existence of some form of no-go theorem against the stability of ${\cal N}=2$ fractional branes in these DSB set-ups. Indeed, in section \ref{other} we provide such a no-go theorem, showing that whenever ${\cal N}=2$ fractional branes are allowed, they inevitably destabilize the otherwise stable DSB vacua. We then look at singularities which do not allow for ${\cal N}=2$ fractional branes, in order to avoid from the start the mechanism of instability. A brief and partial scan yielding negative results hints that in such cases an $SU(5)$ or a 3-2 model cannot actually be easily found. We end-up in section \ref{final} with a critical discussion of our results, and outline possible generalizations of our analysis and new directions worth to be pursued. An appendix contains a brief review on dimer techniques that we use in the main body of the paper.

\section{DSB vacua and their instability}
\label{sum}

In this section we present the basic approach and our main results, so that the reader can have a clear picture before we embark on the detailed analysis of a sizable set of examples.

Our aim is to engineer in string theory, via branes at singularities, supersymmetric gauge theories which admit a stable dynamically supersymmetry breaking (DSB) vacuum. It is then natural to ask first which DSB models we  have a chance of being able to engineer. Known DSB models are rather specific gauge theories, and we have to match their properties to the ones of the gauge theories one can engineer with branes.

D-branes at singularities yield quiver gauge theories. In particular, since all matter fields can be related to open strings, they all have two gauge theory indices, corresponding to the two ends of the open strings. Without orientifold projection we have only two possibilities: matter fields are either in the adjoint representation (if both ends are on the same D-brane) or in the bifundamental representation (if they join two different D-branes). In the presence of an orientifold, there is the additional possibility of having fields in the symmetric or anti-symmetric representations, or their conjugates. 

A generic property of DSB models is that they have a rather contrived matter content, so that there cannot be mass terms and in addition classical flat directions should be lifted by the superpotential. In particular, known DSB models, with only few notable exceptions, are chiral gauge theories. We will focus on the following two well-known models, since they involve matter in at most two-index representations of the gauge groups:

\begin{description}
	\item[$\mathbf{SU(5)}$ one family model] This model \cite{Affleck:1983vc} has an $SU(5)$ gauge group and one GUT-like chiral family ${\tiny \yng(1,1)}\, \oplus \, \tiny\overline{\yng(1)}$ (or in other words ${\bf 10} \oplus {\bf \bar 5}$). No chiral gauge invariant can be written, hence it has no superpotential and no classical flat directions. With arguments based on 't Hooft anomaly matching, its vacuum is believed to break supersymmetry in a purely strongly coupled fashion. The supersymmetry breaking vacuum energy density is given in terms of its dynamical scale, $E_\mathrm{vac} \sim \Lambda_{SU(5)}^4$.
	
	\item[3-2 model] This other model \cite{Affleck:1984xz} involves two gauge groups, $SU(3)$ and $SU(2)$ respectively, and one chiral family, resembling the ones of the Standard Model: under $SU(3)\times SU(2)$ matter fields transform as $({\bf 3},{\bf 2})\oplus ({\bf\bar 3},1)\oplus ({\bf\bar 3},1)\oplus (1,{\bf 2})$. This model has a number of flat directions, but a cubic superpotential lifts them all. After taking into account non-perturbatively generated contributions to the superpotential, it turns out there is a conflict between F-terms and D-terms so that no supersymmetric vacuum can be found. The actual minimum breaks supersymmetry dynamically, where now $E_\mathrm{vac} \sim \Lambda_{SU(3)}^4$ or $\Lambda_{SU(2)}^4$, depending on which group confines first.
\end{description}
Other known DSB models cannot be engineered with brane constructions, such as the $SO(10)$ model \cite{Affleck:1984mf} (because of the spinor representation) and the 4-1 model \cite{Dine:1995ag,Poppitz:1995fh} (because of the $U(1)$ charges). Further models would deserve a closer look, but doing this is beyond the scope of the present paper.

The $SU(5)$ model has a matter field in the anti-symmetric representation, hence to recover it an orientifold projection is necessary. It turns out that also for recovering the 3-2 model an orientifold is needed. This is related to the fact that the two matter fields in the $({\bf\bar 3},1)$ representation are set apart by the superpotential. While in non-orientifolded quivers such pairs of similar fields always come in doublets of a global symmetry, in an orientifolded theory they can simply be taken apart by identifying one of the anti-fundamentals of $SU(3)$ as being in the anti-symmetric representation. Hence, we conclude that in order to recover the basic features of both the $SU(5)$ and the 3-2 models we need to have an orientifold projection.

After the orientifold projection, it turns out that, generically, the anomaly cancellation conditions result in constraints on the various ranks of the form 
\begin{equation}\label{acc4}
\sum_i N_i = \sum_{j'} N_{j'} +4\ ,
\end{equation}
where the two sums run on two different sets of gauge theory nodes, and strictly speaking the $N$s are not the ranks but the dimensions of the fundamental representation of $SU(N)$,\footnote{As usual in quiver gauge theories derived from branes at singularities, we assume that all the $U(1)$ factors have become free at low enough energies, hence acting as global symmetries.} $SO(N)$ or $USp(N)$ groups (the latter two possibilities being possible only after the orientifold projection). The imbalance of 4 units in eq.~\eqref{acc4} is due to the orientifold charge, which contributes to tadpole cancellations.

Two simple ways to satisfy eq.~\eqref{acc4} are the following. We can take one $N_i=5$, one $N_{j'}=1$ and all other ranks to vanish, so that the remaining gauge group is $SU(5)\times SU(1)$ or $SU(5)\times SO(1)$. The trivial factor actually allows for a bifundamental between the two nodes to be interpreted as a (anti)fundamental of $SU(5)$. If the latter also has an anti-symmetric matter field, then the field content is exactly the one of the $SU(5)$ DSB model. 

The other simple solution to \eqref{acc4} is to take one $N_i=3$, another one $N_{j\neq i}=2$, one $N_{j'}=1$, and again all other factors to vanish, leading to the gauge group $SU(3)\times SU(2)/USp(2) \times SU(1)/SO(1)$. The 3-2 model is recovered if there are bifundamentals linking the three gauge groups, together with a cubic superpotential term, and in addition an antisymmetric of $SU(3)$ which provides for the remaining (anti)triplet, necessary for anomaly cancellation. 

In some of the examples that we will review below, some additional decoupled gauge singlets will be present, or even additional decoupled gauge sectors, which themselves do not break supersymmetry. We will even encounter an example with two decoupled $SU(5)$ models. 

The conclusion we can draw is that there is a sizable number of orientifold singularities that allow for configurations with a small number of fractional branes reproducing a gauge theory with a stable DSB vacuum.

We can now ask whether these configurations are stable also after considering their possible UV completions. A natural one is to add regular D3-branes. This is done simply by increasing by a common $N$ the ranks of all gauge groups. As one can check, the anomaly cancellation conditions \eqref{acc4} are still satisfied. It is then easy to show that because of the underlying superconformal fixed point of the parent (non-orientifolded) theory, if one performs scale matching on the node that eventually drives supersymmetry breaking in the IR, one finds that 
\be
\label{match1}
\Lambda_{IR}=\Lambda_{UV}~.
\ee 
In other words, moving the regular D3-branes out of the singularity is still a flat direction, even in the presence of the fractional branes generating the DSB vacuum energy. Note that adding regular D3-branes can also be related, in the models that allow for it, to a duality cascade, though the analysis in the presence of an orientifold can be subtle, see \cite{Argurio:2017upa} for a recent discussion. The upshot is that regular D3-branes do not destabilize the DSB vacuum.

On the other hand, in all examples of DSB we have found, a mechanism similar to the one discussed in \cite{Buratti:2018onj} takes place. A regular D3-brane can be split into two complementary fractional branes of ${\cal N}=2$ kind. The latter have a one-dimensional Coulomb branch that takes them out of the singularity. One can then see what happens when higgsing the quiver in two steps, putting the two sets of ${\cal N}=2$ branes on their Coulomb branches one after the other. In the partially higgsed configuration, the scale matching depends on the VEV $v$ related to the position of the first set of branes. The second VEV $v'$, related to the position of the second set of branes, will then compensate the first so that the final scale matching reads
\begin{equation}\label{scale}
\Lambda_{IR}= \left( \frac{v'}{v}\right)^\alpha \Lambda_{UV}\ ,
\end{equation}
where $\alpha$ is some model-dependent non-vanishing number. If the two sets of branes move together as a regular brane, $v=v'$ and we recover eq.~\eqref{match1}. But if the two sets split, then we see that for $\alpha$ positive there is a runaway towards a supersymmetric vacuum with zero energy at $v'=0$ or equivalently $v=\infty$, viceversa for $\alpha$ negative. We thus conclude that the models are unstable in their UV completion, because the vacuum energy can be brought to zero by moving in (or out) the singularity some ${\cal N}=2$ fractional branes. 

As already mentioned, there is a natural way in which one could try to avoid this destabilizing mechanism, and ensure stability of the DSB vacuum. It would suffice to find a D-brane configuration allowing for a DSB model in a singularity that does not accommodate ${\cal N}=2$ fractional branes. Indeed, other kinds of fractional branes have no flat directions, hence regular branes would not be able to split and separate from the singularity. We have investigated a certain number of such singularities but we were not able to find any DSB model in them. In fact, the number of examples we have analyzed is by no means exhaustive, and a more complete scan of CY singularities with no fractional ${\cal N}=2$ branes clearly deserves to be performed.

In the rest of the paper we analyze a series of singularities, showing how the general pattern described above emerges. 

\section{The $\mathbb{C}^3/\mathbb{Z}_{6'}$ singularity}
\label{C3Z6}

\begin{figure}[ht]
  \centerline{\includegraphics[width=0.5\linewidth]{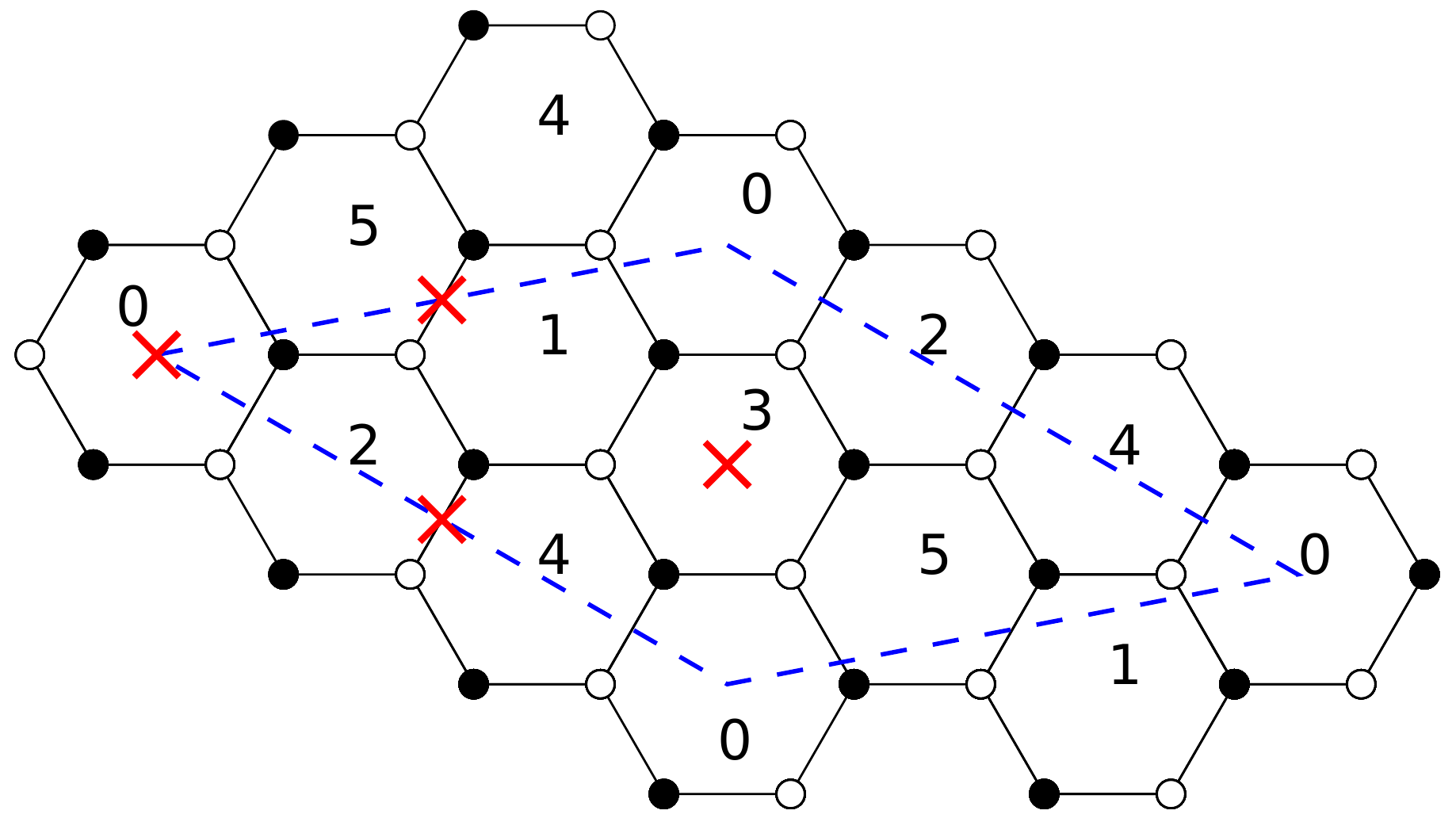}}
  \caption{The $\mathbb{C}^3/\mathbb{Z}_{6'}$ dimer. The theory is chiral, with 6 gauge factors and 18 bifundamental chiral superfields $X_{ij}$. The parallelogram is a possible choice of unit cell. Red crosses represent  fixed points under the orientifold action.}
  \label{fig:boat1}
\end{figure}

As a warm-up, let us start considering the (fixed point) orientifold of the orbifold  $\mathbb{C}^3/\mathbb{Z}_{6'}$, already analyzed in \cite{Buratti:2018onj}. The orbifold action is defined by the $\mathbb{Z}_{6'}$ acting on the complex coordinates of $\mathbb{C}^3$ as
\begin{equation}
\label{defZ6}
\theta~:~~z_i \rightarrow e^{2\pi i v_i} z_i~,
\end{equation}
with  $i=1,2,3$ and $v=(1,2,3)/6)$. The dimer associated to the  $\mathbb{C}^3/\mathbb{Z}_{6'}$ orbifold singularity is reported in figure \ref{fig:boat1}, including the fixed points with respect to which we will eventually take the orientifold projection.  

From the dimer one can read the field content of the theory as well as the superpotential which is
\begin{eqnarray}
W&=&X_{152}+X_{143}+X_{032}+X_{053}+X_{254}+X_{104} \nonumber \\ && 
-X_{052}-X_{031}-X_{142}-X_{043}-X_{253}-X_{154} ~,  
\end{eqnarray}
where, for the ease of notation, we have defined $X_{lmn} \equiv X_{lm}X_{mn}X_{nl}$, and $X_{lm}$ is in the $(\wb{\tiny \yng(1)} _{\, l}, \tiny \yng(1) _{\, m})$ representation, where $l,m,n=0,1,...,5$. Following the general rules summarized in appendix \ref{append1}, one can see that there do not exist deformation fractional branes but ${\cal N}=2$ fractional branes. For instance, the strip 0-2-4 and its complement 1-3-5, are ${\cal N}=2$  fractional branes. 

We now perform an orientifold projection via point reflection. The unit cell has an even number of white nodes hence, following appendix \ref{append1},  we have to choose orientifold charges with an even number of $+$ signs. A convenient choice is $(+,+,-,-)$ starting from the fixed point on face 0 and going clockwise. The orientifold projection gives the following identifications between faces
\begin{eqnarray}
0 \iff 0 \hspace{1 cm} 1 \iff 5 \hspace{1 cm} 2 \iff 4 \hspace{1 cm} 3 \iff 3.
\end{eqnarray}
The daughter theory has hence gauge group 
\be
SO(N_0)\times SU(N_1)\times SU(N_2)\times USp(N_3)~,
\ee
and matter in the following representations
\begin{eqnarray}
&&X_1=( {\tiny \yng(1)}_{\,0} , \wb{\tiny \yng(1)}_{\,1})~~,~~ X_2=( {\tiny \yng(1)}_{\,0} , \wb{\tiny \yng(1)}_{\,2})~~,~~X_3=( {\tiny \yng(1)}_{\,0} ,  {\tiny \yng(1)}_{\,3}), \nonumber
\\
&&Y_1=(  {\tiny \yng(1)}_{\,1} ,   {\tiny \yng(1)}_{\,3})~~\;,~~ Y_2=(  {\tiny \yng(1)}_{\,2} ,  {\tiny \yng(1)}_{\,3})~~\,,~~ Z_1=(  {\tiny \yng(1)}_{\,1} ,   {\tiny \yng(1)}_{\,2})\nonumber
\\
&&Z_2=( \wb  {\tiny \yng(1)}_{\,1} , \wb{\tiny \yng(1)}_{\,2})~~\;,~~  W=( {\tiny \yng(1)}_{\,1} ,\wb {\tiny \yng(1)}_{\,2})~~\,,~~ A={\tiny \yng(1,1)_{\,2}}~~,~~  S=\wb {\tiny \yng(2)}_{\,1}~. 
\label{repz6}
\end{eqnarray}
Imposing anomaly cancellation condition on the two $SU$ factors we get
\begin{eqnarray}
\begin{cases}
N_0+N_1-N_2-N_3+4=0  \hspace{1 cm} SU(N_1)
\\
N_0+N_1-N_2-N_3+4=0 \hspace{1 cm} SU(N_2)
\end{cases}
\nonumber \nonumber
\end{eqnarray}
which is one and the same condition, that is
\begin{eqnarray}
\label{acc1}
N_0+N_1-N_2-N_3+4=0~.
\end{eqnarray}

\subsubsection*{SU(5) model}

\begin{figure}[ht]
  \centerline{\includegraphics[width=0.35\linewidth]{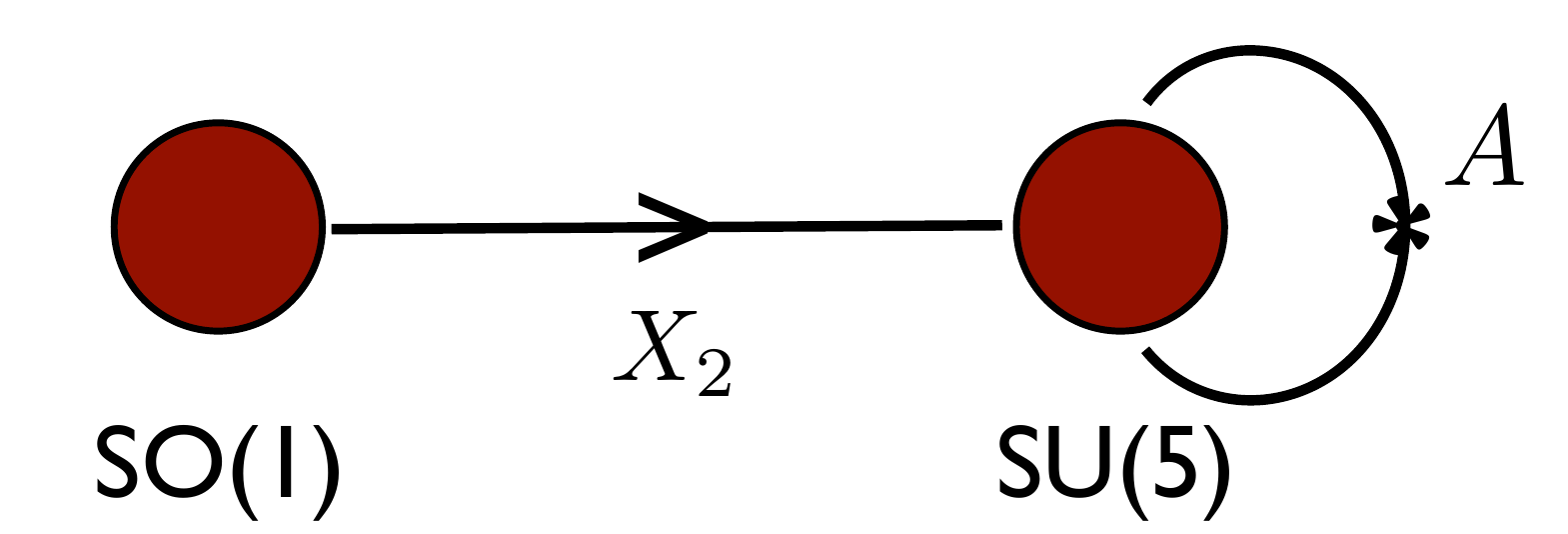}}
  \caption{The quiver of the $SU(5)$ model at the $\mathbb{C}^3/\mathbb{Z}_{6'}$  orbifold singularity. Matter fields follow the definitions in eq.~\eqref{repz6}. The asterisk refers to the anti-symmetric representation.}
  \label{C3U5}
\end{figure}

An interesting choice of ranks compatible with the constraint \eqref{acc1} is $N_0=1, N_1=0, N_2=5, N_3=0$. 
With this choice, the theory becomes exactly the one describing the uncalculable $SU(5)$ DSB model (the $SO(1)$ becomes a flavor index), which breaks supersymmetry dynamically in a stable vacuum. The corresponding quiver is reported in figure \ref{C3U5}. 

As recently argued in \cite{Buratti:2018onj}, in the decoupling limit this DSB vacuum becomes actually unstable. In such limit the effective matter-coupled gauge theory becomes
\be
\label{UVZ6}
SO(N+1) \times SU(N) \times SU(N+5) \times USp(N)~,
\ee
which actually corresponds to adding $N$ regular D3-branes at the singularity. This is a much richer theory than $SO(1) \times SU(5)$ and it might have, possibly, many vacua. One should then ask whether in the larger moduli space of \eqref{UVZ6} there are instabilities which make the supersymmetry breaking vacuum unstable. This can be easily understood by scale matching. 

The vacuum energy of the putative supersymmetry breaking vacuum will be $\sim \Lambda^4$, with $\Lambda$ the intrinsic scale of the $SU(5)$ model. The higgsing of the $N$ regular branes can be obtained giving a VEV of scale $v$ to the gauge invariant operator\footnote{Hereafter, we assume that all fields appearing in the gauge invariants have a rank $N$ piece in the upper left part, to ensure the correct higgsing pattern. If the rank $N$ pieces are all proportional to the identity, then an $SU(N)$ diagonal gauge group is preserved. It can be checked to have ${\cal N}=4$ SUSY to a good approximation, and to decouple from the rest of the quiver. We will not consider it further.} 
\be
\Phi=\text{Tr} (X_1 W Y_2 X_3^t)~.
\ee
This makes the theory \eqref{UVZ6} flow to $SO(1) \times SU(5)$, namely the DSB $SU(5)$ model.

We can match the UV and IR scales evaluating the $\beta$ functions of the relevant $SU$ factor above and below the scale $v$.\footnote{The exact $\beta$ function of $SU(N_c)$ supersymmetric gauge theory coupled to chiral matter fields $\Phi_i$ with anomalous dimensions $\gamma_i$ is $\beta(8 \pi^2/g^2) = 3 N_c - \sum_i N_i T(\rho_i)(1-\gamma_i)$, where $N_i$ is the number of fields in the representation $\rho_i$, and $T(\tiny \yng(1))=\frac12$ and $T(\raisebox{-3pt}{\tiny \yng(1,1)})=\frac12 (N_c-2)$. The absence of the denominator with respect to the usual NSVZ expression  \cite{Novikov:1983uc} is due to a choice of normalization for the vector superfield which differs from the canonical one by a factor of $1/g^2$, as usual in the framework of the gauge/gravity correspondence.} Note that because the orbifold theory is a projection of ${\cal N}=4$ SYM, fundamental fields do not acquire anomalous dimensions (equivalently, all superpotential terms are cubic). With obvious notation, we have (here, and in similar formulas thereafter, we omit a factor of $8\pi^2$ for clarity of exposition)
\begin{eqnarray}
\frac{1}{g^2_{SU(5+N)}}&=&\left[3 (N+5)-\frac 12 \left(6N + 4 \right)\right] \ln\left(\frac{\mu}{\Lambda_{UV}}\right)=13\ln\left(\frac{\mu}{\Lambda_{UV}}\right) \nonumber
\\
\frac{1}{g^2_{SU(5)}}&=&\left[1 5-\left(\frac{1}{2}+\frac{3}{2}\right)\right] \ln\left(\frac{\mu}{\Lambda}\right)=13\ln\left(\frac{\mu}{\Lambda}\right)\nonumber~.
\end{eqnarray}
Matching the gauge coupling at $\mu=v$ we get 
\be
\label{scmatZ6}
 \Lambda =\Lambda_{UV} ~.
\ee
This shows that the effective potential does not depend on the VEV of $\Phi$, meaning that regular brane dynamics does not change the nature of the supersymmetry breaking vacuum and its stability (there is no force acting on the $N$ regular branes). 

\begin{figure}[h!]
\centerline{   \includegraphics[width=0.4\linewidth]{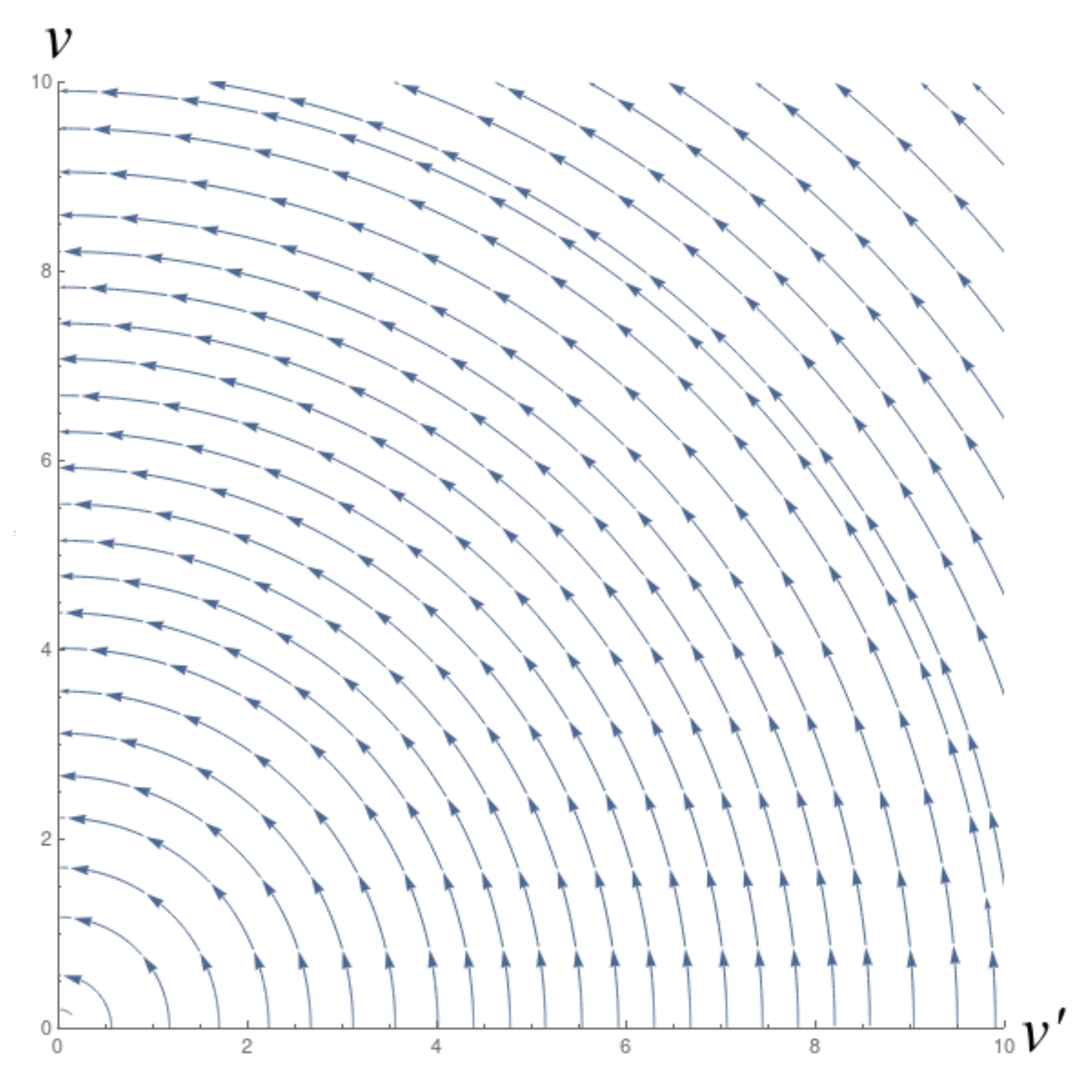}}
   \caption{Plot of $-\mbox{Grad}(V)$ as a function of $v$ and $v'$. The flow goes towards $v'=0$, suggesting that the system, eventually, relaxes to a supersymmetry preserving vacuum at finite distance in field space.}
  \label{fig:potentialflow}
\end{figure}

In fact, it turns out that there exists a different instability channel. This has to do with moduli associated to ${\cal N}=2$ fractional branes, which are massless classically, but become runaway once non-perturbative corrections are taken into account. 
From the dimer in figure~\ref{fig:boat1} we see that a regular brane can be seen as a bound state of two ${\cal N}=2$ fractional branes corresponding to the strips 0-2-4 and 1-3-5, respectively. The classical flat directions correspond to the $z_2$ fixed line that is left invariant by $\theta^3$, see eq.~\eqref{defZ6}. Locally, this is a $\mathbb{C}^2/\mathbb{Z}_2$ singularity. Both these fractional branes survive the orientifold projection, becoming 0-2 and 1-3 strips. Note that (135) and (024) are closed loops which represent operators not present in the superpotential. Hence, their VEVs represent motion along classical flat directions. In the daughter theory these directions are given by the operators\footnote{As before, we assume that the fields in the gauge invariants have a rank $N$ upper left piece, and we do not consider the decoupled effective ${\cal N}=2$ diagonal gauge group. Note that the traces involve squares since $\text{Tr}(X_2 A X_2^t)=0$ because of antisymmetry of $A$, while $\text{Tr}(JY_1 S Y_1^t)=0$ because of antisymmetry of the $USp$-invariant $J$.} 
\begin{eqnarray}
\tilde \Phi=\text{Tr}(X_2 A X_2^t)^2~~~,~~~  \tilde \Phi'= \text{Tr}(JY_1 S Y_1^t)^2~,
\end{eqnarray}
and we will denote the scale of their VEVs as $v$ and $v'$, respectively. The higgsing pattern is then\footnote{This pattern occurs for $v \gg v'$.  One can check that  the end result does not change when inverting the order of the two scales.}
\begin{eqnarray}
&SO(N+1)\times SU(N)\times SU(N+5)\times USp(N) \buildrel v \over \longrightarrow \nonumber
\\
&SO(1)\times SU(N)\times SU(5)\times USp(N) \buildrel v' \over \longrightarrow  SO(1)\times SU(5)~.
\end{eqnarray}
Above and below the scale $v$ the gauge couplings run as 
\begin{eqnarray}
\frac{1}{g^2_{SU(5+N)}}&=&\left[3(N+5)-\frac 12 \left(6N+4\right) \right]\ln\left(\frac{\mu}{\Lambda_{UV}}\right)= 13 \ln\left(\frac{\mu}{\Lambda_{UV}}\right) \nonumber \\
\frac{1}{g^2_{SU(5)_N}}&=&\left[15-\frac 12 \left(4N+4\right)\right] \ln\left(\frac{\mu}{\Lambda_N}\right)=(13-2N) \ln\left(\frac{\mu}{\Lambda_N}\right)\nonumber~.
\end{eqnarray}
Matching them at $\mu=v$ we get 
\be
\Lambda_N^{13-2N}=v^{-2N} \Lambda_{UV}^{13}~. 
\ee
Repeating the same computation above and below the scale $v'$ we have
\begin{eqnarray}
\frac{1}{g^2_{SU(5)_N}}&=&\left[15-\frac 12 \left(4N+4\right)\right] \ln\left(\frac{\mu}{\Lambda_{N}}\right)=(13-2N)\ln\left(\frac{\mu}{\Lambda_{N}}\right) \nonumber
\\
\frac{1}{g^2_{SU(5)}}&=&\left[15-\left(\frac{1}{2}+\frac{3}{2}\right)\right] \ln\left(\frac{\mu}{\Lambda}\right)=13\ln\left(\frac{\mu}{\Lambda}\right)\nonumber~.
\end{eqnarray}
and in turn $\Lambda_N^{13-2N}=v'^{-2N} \Lambda^{13}$. The end result is then 
\be
\label{sm2}
 \Lambda^{13}=\left(\frac{v'}{v}\right)^{2N} \Lambda_{UV}^{13}~.
\ee
This shows that the DSB vacuum is unstable. There exists a one-dimensional supersymmetric moduli space sitting at $v'=0$ and parametrized by $v$. Indeed, one can estimate the minima of the potential in a $(v, v')$ plane and check that any point at $v,v' \not =0$ is driven to the $v'=0$ axis. The gradient flow is reported in figure \ref{fig:potentialflow}.

In principle, one might try to obtain the $SU(5)$ by a different UV completion. For instance, one can add to the DSB configuration $M$ fractional branes populating the second and fourth gauge factors only, corresponding to the strip 1-3-5 in the mother theory, which becomes 1-3 after orientifolding. The theory in this case has gauge group 
\be
SO(1) \times SU(M) \times SU(5) \times USp(M)~.
\ee
This configuration does not change much the fate of the DSB vacuum. Previous analysis shows that, lacking one modulus, $v$ in our conventions, the one-dimensional moduli space of supersymmetry preserving vacua becomes an isolated vacuum. This agrees with known field theory results \cite{Poppitz:1995fh}: the $SU(5)$ factor has extra vector-like matter and the theory does not lead to a supersymmetry breaking vacuum to start with. 

If one instead populates nodes 0 and 2, $N_0=M, N_1=0, N_2=M+ 4, N_3=0$ which in the mother theory corresponds to adding $M$ ${\cal N}=2$ fractional branes associated to the strip 0-2-4, the theory has a runaway direction associated to $v$. Note that this last system has the same gauge and matter content of a known, stable, DSB model \cite{Affleck:1983vc,Poppitz:1995fh}, but it lacks a crucial cubic term in the superpotential whose effect is indeed to stop the runaway associated to $v$. The special case $M=1$ is the only stable DSB model (our original brane construction, in fact). 

\vskip 7pt
To sum up, the $\mathbb{C}^3/\mathbb{Z}_{6'}$ singularity does admit fractional brane configurations whose low energy open string dynamics  enjoys stable DSB vacua. However, once coupled to regular branes, the supersymmetry breaking vacuum becomes unstable towards supersymmetry preserving ones. The $\mathbb{C}^3/\mathbb{Z}_{6'}$ singularity can be embedded into a larger singularity which admits deformation branes \cite{Retolaza:2015nvh} and, as such, a cascade (dual to a warped throat \cite{Franco:2004jz,Franco:2005fd,GarciaEtxebarria:2006aq}). So the above analysis suggests that, at least within this construction, it is not possible to embed the $SU(5)$ DSB model into a warped throat keeping it stable \cite{Buratti:2018onj}.

\subsubsection*{3-2 model}

\begin{figure}[ht]
  \centerline{\includegraphics[width=0.50\linewidth]{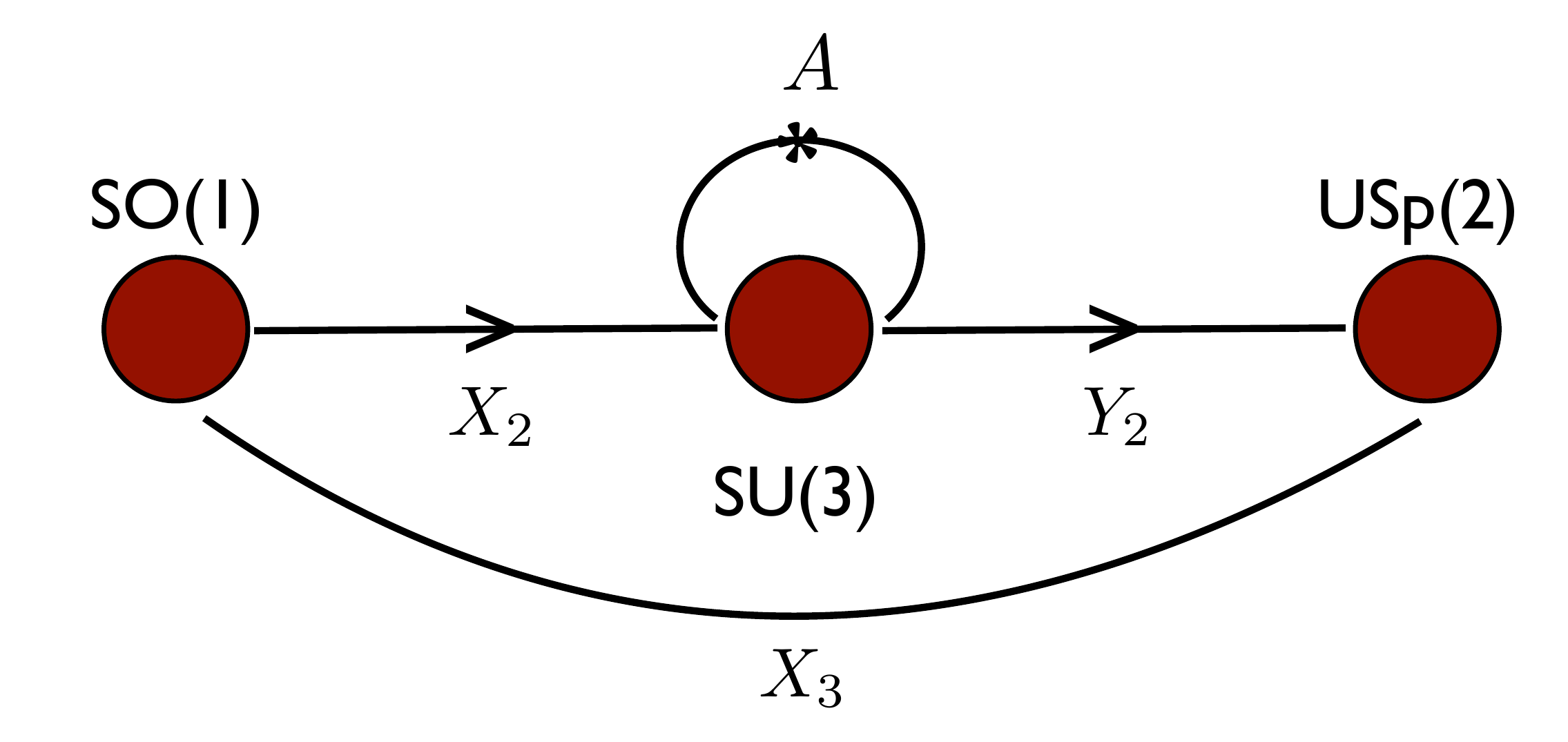}}
  \caption{The quiver of the 3--2 model at the $\mathbb{C}^3/\mathbb{Z}_{6'}$  orbifold singularity. Matter fields follow the definitions in eq.~\eqref{repz6}. Arrows are indicated only when needed (for $USp(2) = SU(2)$ the fundamental and anti-fundamental are equivalent representations). The asterisk refers to the anti-symmetric representation, as in figure \ref{C3U5}.}
  \label{C3U32}
\end{figure}

Looking at eq.~\eqref{acc1}, one can see that another possible anomaly free rank assignment is given by $N_0=1, N_1=0, N_2=3, N_3=2$ 
which corresponds to the following gauge theory  
\begin{eqnarray}
&& SO(1)\times SU(3) \times USp(2) ~.
\end{eqnarray}
Using the fact that $USp(2)=SU(2)$ and that for $SU(3)$ ${\tiny \yng(1,1)}=\wb{\tiny \yng(1)}$, from \eqref{repz6} we see that the matter content is
\begin{eqnarray}
&& X_2=( {\tiny \yng(1)}_{\,0} , \wb{\tiny \yng(1)}_{\,2})=\wb D, \hspace{1 cm} X_3=( {\tiny \yng(1)}_{\,0} ,  {\tiny \yng(1)}_{\,3})=L, \nonumber
\\
&& Y_2=(  {\tiny \yng(1)}_{\,2} ,  {\tiny \yng(1)}_{\,3})=Q, \hspace{1 cm} A={\tiny \yng(1,1)_{\,2}}=\wb U ~,
\end{eqnarray}
with tree level superpotential 
\begin{eqnarray}
W=\wb D QL ~.
\end{eqnarray}
This reproduces exactly the DSB 3-2 model \cite{Affleck:1984xz}! The corresponding quiver is reported in figure \ref{C3U32}.

Again, one could ask what is the fate of this DSB vacuum in the full theory. In the present model, we have to perform the scale matchings on the gauge group that is most strongly coupled, since the DSB vacuum energy will be expressed in terms of its dynamical scale.
We start by considering a regime where supersymmetry breaking is driven by the non-perturbative contributions in the $SU(3)$ gauge group \cite{Affleck:1983mk}. 

As in the $SU(5)$ model, upon adding regular branes and higgsing them, the vacuum shows no instability. 
Indeed, giving a VEV to $\Phi=\text{Tr} (X_1 W Y_2 X_3^t)$, the gauge coupling running above and below the matching scale is 
\begin{eqnarray}
\frac{1}{g^2_{SU(3+N)}}&=&\left[3(N+3)-\frac12\left(6N+4\right)\right] \ln\left(\frac{\mu}{\Lambda_{UV}}\right) =7\ln\left(\frac{\mu}{\Lambda_{UV}}\right) \nonumber
\\
\frac{1}{g^2_{SU(3)}}&=&\left[9-\left(\frac{1}{2}+\frac{1}{2}+1\right)\right] \ln\left(\frac{\mu}{\Lambda_3}\right)=7\ln\left(\frac{\mu}{\Lambda_3}\right)\nonumber~,
\end{eqnarray}
and hence $\Lambda = \Lambda_{UV}$. 

However, the theory has ${\cal N}=2$ fractional branes, and following the same two-steps higgsing pattern as before, namely 
\begin{eqnarray}
&SO(N+1)\times SU(N)\times SU(N+3)\times USp(N+2)  \buildrel v \over \longrightarrow  \nonumber
\\
&SO(1)\times SU(N)\times SU(3)\times USp(N+2)  \buildrel v' \over \longrightarrow  SO(1)\times SU(3)\times USp(2) ~,
\end{eqnarray}
we get, above and below the scale $v$
\begin{eqnarray}
\frac{1}{g^2_{SU(3+N)}}&=&\left[3(N+3)-\frac12\left(6N+4\right) \right]\ln\left(\frac{\mu}{\Lambda_{UV}}\right)=7\ln\left(\frac{\mu}{\Lambda_{UV}}\right) \nonumber \\
\frac{1}{g^2_{SU(3)_N}}&=&\left[9-\frac12\left(4N+4\right)\right] \ln\left(\frac{\mu}{\Lambda_N}\right)=(7-2N) \ln\left(\frac{\mu}{\Lambda_N}\right)\nonumber~,
\end{eqnarray}
and, at scale $v'$
\begin{eqnarray}
\frac{1}{g^2_{SU(3)_N}}&=&\left[9-\frac12\left(4N+4\right)\right] \ln\left(\frac{\mu}{\Lambda_N}\right)=(7-2N) \ln\left(\frac{\mu}{\Lambda_N}\right) \nonumber
\\
\frac{1}{g^2_{SU(3)}}&=&\left[9-\left(\frac{1}{2}+\frac{1}{2}+1\right)\right] \ln\left(\frac{\mu}{\Lambda_3}\right)=7\ln\left(\frac{\mu}{\Lambda_3}\right)\nonumber~,
\end{eqnarray}
which gives in the end
\be 
\Lambda_3^{\,7}=\left(\frac{v'}{v}\right)^{2N} \Lambda_{UV}^{7}~.
\ee

We can now repeat the analysis when the supersymmetry breaking dynamics is driven by the strong coupling scale of $SU(2)$. We have
\begin{eqnarray}
\frac{1}{g^2_{USp(N+2)}}&=&\left[3\frac{(N+4)}{2}-\frac12\left(3N+4\right)\right] \ln\left(\frac{\mu}{\Lambda_{UV}}\right)=4\ln\left(\frac{\mu}{\Lambda_{UV}}\right) \nonumber
\\
\frac{1}{g^2_{USp(2)}}&=&\left[6-\left(\frac{1}{2}+\frac{3}{2}\right)\right] \ln\left(\frac{\mu}{\Lambda_2}\right)=4\ln\left(\frac{\mu}{\Lambda_2}\right)\nonumber~,
\end{eqnarray}
and hence $\Lambda_2=\Lambda_{UV}$ by higgsing the $N$ regular branes. Along the ${\cal N}=2$ directions we get instead the matching 
\begin{eqnarray}
\frac{1}{g^2_{USp(2+N)}}&=&\left[3\frac{(N+4)}{2}-\frac12\left(3N+4\right)\right] \ln\left(\frac{\mu}{\Lambda_{UV}}\right)=4\ln\left(\frac{\mu}{\Lambda_{UV}}\right) \nonumber \\
\frac{1}{g^2_{USp(2+N)_N}}&=&\left[3\frac{(N+4)}{2}-\frac12\left(N+4\right)\right] \ln\left(\frac{\mu}{\Lambda_N}\right)=(4+N) \ln\left(\frac{\mu}{\Lambda_N}\right)\nonumber~
\end{eqnarray}
at scale $v$ and 
\begin{eqnarray}
\frac{1}{g^2_{USp(2+N)_N}}&=&\left[3\frac{(N+4)}{2}-\frac12\left(N+4\right)\right] \ln\left(\frac{\mu}{\Lambda_N}\right)=(4+N) \ln\left(\frac{\mu}{\Lambda_N}\right) \nonumber
\\
\frac{1}{g^2_{USp(2)}}&=&\left[6-\left(\frac{1}{2}+\frac{3}{2}\right)\right] \ln\left(\frac{\mu}{\Lambda_2}\right)=4\ln\left(\frac{\mu}{\Lambda_2}\right)\nonumber~,
\end{eqnarray}
at scale $v'$.  The final relation one gets between the UV and IR scale is now 
\be
\Lambda_2^{\,4}=\left(\frac{v}{v'}\right)^{N} \Lambda_{UV}^{4}~.
\ee 
This result is analogous to the one obtained before (even though the roles of $v$ and $v'$ are exchanged). We conclude that the Coulomb branch is unstable and it is so independently of the regime in which the 3-2 model finds itself.

\section{The $\mathbf{PdP4}$ singularity}
\label{PdP4}

\begin{figure}[ht]
\centerline{  \includegraphics[width=0.25\linewidth]{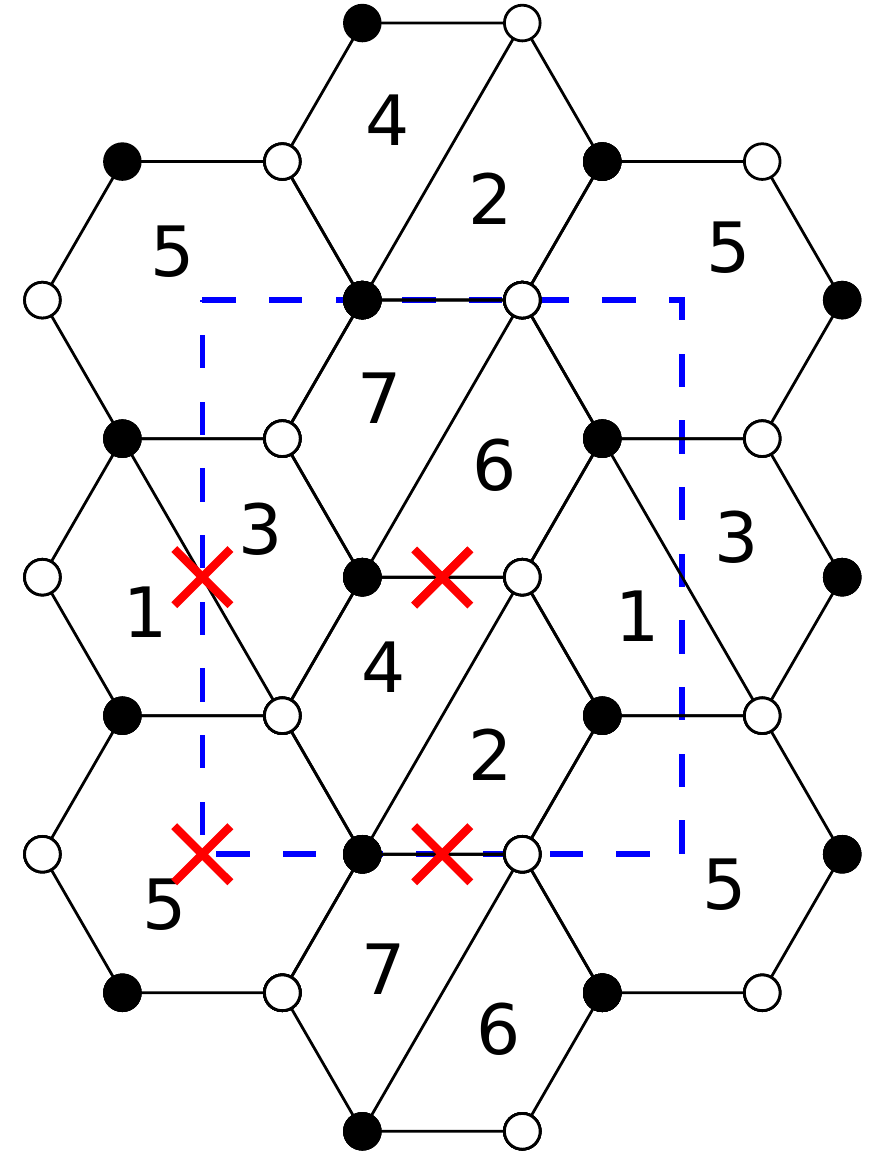}}
  \caption{$PdP4$ dimer. Red crosses represent  fixed points under the orientifold action.}
  \label{fig:boat2}
\end{figure}

We now want to consider a different model, based on the pseudo del Pezzo 4 singularity, $PdP4$ for short. We choose, for definiteness, the phase $I$, following the conventions of \cite{Feng:2002fv}, where pseudo del Pezzo singularities were introduced. The dimer is depicted in figure \ref{fig:boat2}. 

As for the orbifold model, from the dimer one can extract the field theory content and the corresponding superpotential, which now reads
\begin{eqnarray}
\label{WPdP4}
W=X_{1543}+X_{375}+X_{1642}+X_{5276}-X_{152}-X_{7643}-X_{5427}-X_{5316}~,
\end{eqnarray} 
with by now familiar index conventions. Note that the superpotential admits also quartic terms now, implying that (some) fundamental fields have large anomalous dimensions. This is related to an important qualitative difference with respect to the previous example. In this model there exist deformation fractional branes, together with ${\cal N}=2$ ones. For instance, the strip 1-3-5 is a ${\cal N}=2$ brane, while 2-6 and  1-2-5 are two different kinds of deformation branes.

Upon point reflection with charges $(+,-,+,-)$ starting on the fixed point on face 5 and going clockwise, we get the following identifications between faces
\begin{eqnarray}
\label{idePdP4}
1 \iff 3 \hspace{1 cm} 2 \iff 7 \hspace{1 cm} 4 \iff 6 \hspace{1 cm} 5 \iff 5~.
\end{eqnarray}
The daughter theory has hence gauge group 
\be
SO(N_5)\times SU(N_1)\times SU(N_2)\times SU(N_4)~,
\ee 
and matter in the following representations
\begin{eqnarray}
&&X_1=(\wb {\tiny \yng(1)}_{\,1} ,  {\tiny \yng(1)}_{\,5}), \hspace{1 cm} X_2=( {\tiny \yng(1)}_{\,5} ,  {\tiny \yng(1)}_{\,2}), \hspace{1 cm} X_4=( {\tiny \yng(1)}_{\,5} ,  {\tiny \yng(1)}_{\,4}), \nonumber
\\
&&Y_1=(\wb {\tiny  \yng(1)}_{\,4} , \wb {\tiny  \yng(1)}_{\,1}),  \hspace{1 cm} Y_2=(\wb {\tiny  \yng(1)}_{\,4} ,  {\tiny \yng(1)}_{\,2}),   \hspace{1 cm} Z=(\wb {\tiny  \yng(1)}_{\,2} ,  {\tiny \yng(1)}_{\,1}), \nonumber
\\
&&A_1={\tiny \yng(1,1)_{\,1}},  \hspace{2 cm} A_2=\wb {\tiny  \yng(1,1)}_{\,2},  \hspace{2 cm} S_4={\tiny \yng(2)_{\,4}}~.
\label{orientfields} 
\end{eqnarray}
The anomaly cancellation conditions for the three $SU$ factors are 
\begin{eqnarray}
\begin{cases}
N_2+N_1-N_5-N_4-4=0  \hspace{1 cm} SU(N_1)
\\
N_5+N_4-N_2-N_1+4=0 \hspace{1 cm} SU(N_2)
\\
N_5-N_1-N_2+N_4+4=0 \hspace{1 cm} SU(N_4)
\end{cases} \nonumber~,
\end{eqnarray}
which is the unique condition
\begin{eqnarray}
\label{anpdP4}
&  N_1+N_2-N_4-N_5-4=0 ~.
\end{eqnarray}

\subsubsection*{SU(5) model}

The constraint \eqref{anpdP4} allows to obtain an effective $SU(5)$ DSB model, as for the theory studied in the previous section, by choosing $N_5=1, N_1=5, N_2=N_4=0$.\footnote{We note that there is also the configuration $N_4=1, N_1=5, N_2=N_5=0$ that leads to a $SU(5)$ DSB model, with the extra decoupled singlet $S_4$. The analysis goes through very similarly. Further interchanging the roles of nodes 1 and 2 provides two more trivially equal examples.} 
 
We can now proceed as for the  orbifold $\mathbb{C}^3/\mathbb{Z}_{6'}$ and add to the aforementioned DSB brane configuration $N$ regular D3-branes, which change the theory to $SO(N+1)\times SU(N+5)\times SU(N)\times SU(N)$ (with the  corresponding matter and superpotential terms). 

As already noticed, differently from the orbifold case, where the tree level potential of the parent theory contains only cubic terms, and hence the anomalous dimensions of all fields are zero, in this model we have both cubic and quartic terms, and so we have to take into account non-trivial anomalous dimensions. We can compute such anomalous dimensions in the parent theory. They are fixed by populating the dimer with regular branes and imposing vanishing $\beta$ functions and $R$-charge equal 2 to all superpotential terms (in the present case this corresponds to 7+8=15 equations). 

The symmetries of the dimer help in simplifying the system one has to solve. In particular there exist three $\mathbb{Z}_2$ symmetries acting on faces as $2 \leftrightarrow 3$, $1 \leftrightarrow 7$ and $(4,1,2) \leftrightarrow (6,3,7)$, respectively. Using these symmetries the number of independent anomalous dimensions is just 5 which gives back only 4 independent equations one has to solve, implying in the end a solution with one unfixed modulus. This can be fixed by a-maximization \cite{Intriligator:2003jj} giving finally, for the fields \eqref{orientfields}, the following result
\begin{eqnarray}
&&\gamma_{X_1}=\frac{2}{5}, \hspace{1 cm} \gamma_{X_2}=\frac{2}{5}, \hspace{1 cm} \gamma_{X_4}=-\frac{4}{5}, \nonumber
\\
&&\gamma_{Y_1}=-\frac{4}{5},  \hspace{1 cm} \gamma_{Y_2}=-\frac{4}{5},   \hspace{1 cm} \gamma_{Z}=-\frac{4}{5}, \nonumber
\\
&&\gamma_{A_1}=-\frac{4}{5},  \hspace{1 cm} \gamma_{A_2}=-\frac{4}{5},  \hspace{1 cm} \gamma_{S_4}=\frac{2}{5}~,
\end{eqnarray}
where $\Delta_i = 1 +  \frac 12 \gamma_i$. The orientifold projection may provide $1/N$ corrections to these anomalous dimensions, as fractional branes similarly do. Here and in the following, we will consistently neglect both of them.

We can now proceed as in the previous example by adding $N$ regular branes to the DSB system, higgsing and doing scale matching. The gauge coupling associated to face 1 above and below the scale $v$ of the VEV of the gauge invariant operator  $\Phi=\text{Tr}(X_4 Y_2 Z X_1)$ runs as
\begin{eqnarray}
\frac{1}{g^2_{SU(5+N)}}&=&\bigg[3(N+5)-\bigg(\frac{N+1}{2}\left(1-\frac{2}{5}\right)+\frac{N}{2}\left(1+\frac{4}{5}\right)+\frac{N}{2}\left(1+\frac{4}{5}\right)+ \nonumber
\\
&& \frac{N+3}{2}\left(1+\frac{4}{5} \right)\bigg)\bigg] \ln\left(\frac{\mu}{\Lambda_{UV}}\right)=12\ln\left(\frac{\mu}{\Lambda_{UV}}\right) \nonumber
\\
\frac{1}{g^2_{SU(5)}}&=&\left[15-\left(\frac{3}{10}+\frac{27}{10}\right)\right]  \ln\left(\frac{\mu}{\Lambda}\right)=12 \ln\left(\frac{\mu}{\Lambda}\right)\nonumber~.
\end{eqnarray}
Matching the scale at $\mu=v$ we get
\be
\Lambda= \Lambda_{UV}~.
\ee
Again, the matching of the two scales is exact, {\it i.e.} $\Lambda$ does not depend on $v$. 

Let us now investigate other possible decay channels. In the parent theory, a regular brane can be seen as a bound state  of a ${\cal N}=2$ brane associated to the strip  1-3-5 and its complement, 2-4-6-7. Upon orientifolding these two fractional branes become a 1-5 and a 2-4 strips, respectively. We can see that also in this theory, after the orientifold projection, the two types of ${\cal N}=2$ fractional branes behave differently, one leading to a supersymmetric vacuum and the other triggering supersymmetry breaking into runaway. The details are similar to the orbifold case, and we refrain to repeat the analysis here. The end result, after scale matching, is 
\be 
\Lambda^{12}=\left(\frac{v'}{v}\right)^{\frac{9}{5}N}\Lambda_{UV}^{12}~.
\ee
The dynamics is qualitatively the same as for the $\mathbb{C}^3/\mathbb{Z}_{6'}$ case. The theory enjoys a one-dimensional moduli space of supersymmetry preserving vacua at $v'=0$, parametrized by $v$.

\subsubsection*{3-2 model}

Also this orientifold admits a 3-2 DSB model. Indeed, a different rank assignment which satisfies the anomaly cancellation condition is $N_5=1, N_1=3, N_2=2, N_4=0$ which gives a $SU(3)\times SU(2) \times SO(1)$ gauge theory with matter content
 \begin{eqnarray}
 X_1=(\wb {\tiny \yng(1)}_{\,1} ,  {\tiny \yng(1)}_{\,5}) = \wb D, \hspace{1 cm} &&X_2=( {\tiny \yng(1)}_{\,5} ,  {\tiny \yng(1)}_{\,2}) = L, \hspace{1 cm},  \hspace{1 cm} Z=(\wb {\tiny  \yng(1)}_{\,2} ,  {\tiny \yng(1)}_{\,1}) = Q, \nonumber \\
A_1={\tiny \yng(1,1)_{\,1}} = \wb U,  && \hspace{1 cm} A_2=\wb {\tiny  \yng(1,1)}_{\,2} = S~,
\end{eqnarray}
where we used again that the two index antisymmetric representation of $SU(3)$ is equal to the antifundamental, and that the two index antisymmetric of $SU(2)$ is actually the singlet representation. The matter content is precisely the one of the 3-2 model (up to a decoupled singlet $S$) and, from \eqref{WPdP4}, one can also see that the only term surviving in the superpotential is precisely
\be
\label{sup32}
W= \wb D Q L~.
\ee 
As for the $SU(5)$ model, the addition of regular branes does not destabilize the DSB vacuum. As before, however, the theory has ${\cal N}=2$ fractional branes, which eventually do destabilze the vacuum, as in the orbifold case. In particular, it is possible to show, by scale matching, that the strong coupling scale which controls the DSB vacuum energy ($\Lambda_3$ of the $SU(3)$ factor or $\Lambda_2$ of the $SU(2)$ factor) is affected by the higgsing procedure and the vacuum energy relaxes to zero. The scale matchings give, in the two cases
\be
 \Lambda_3^{6}=\left(\frac{v'}{v}\right)^{\frac{9}{5}N} \Lambda_{UV}^{6}~,\qquad~ \Lambda_2^{3}=\left(\frac{v}{v'}\right)^{\frac{6}{5}N} \Lambda_{UV}^{3}
\ee 
which, again, show that the vacuum is unstable, eventually.

The $PdP4$ case is different from the previous orbifold case in that it has a natural warped throat UV completion. Indeed, it contains deformation fractional branes, so that
the parent theory admits a cascade of Seiberg dualities. For instance, in the conformal case, $N_i=N$ for any $i$, it is straightforward to show that starting from node 1 and following the sequence $1\to2\to4\to5\to6\to7\to3$ we get back to the starting point. Then, if we add $M$ (deformation) fractional branes on nodes 1-2-5, we trigger a cascade. Performing the previous sequence six times, we find that the number of regular branes is diminished by seven times the number of fractional ones
\begin{eqnarray}
SU_1(N'+M)\times SU_2(N'+M) \times SU_3(N') \times SU_4(N') \times SU_5(N'+M) \times SU_6(N') \times SU_7(N') \nonumber
\end{eqnarray}
where $N'=N-7M$. Upon orientifolding,\footnote{See again \cite{Argurio:2017upa} for the subtleties of performing a duality cascade in an orientifolded theory.} the fractional brane configuration that could give rise to the one containing the 3-2 model discussed above should in fact be 
\be
N_5=1+2M \hspace{0,5 cm} N_1=3+M \hspace{0,5 cm} N_2=2+M \hspace{0,5 cm} N_4=0~,
\ee
which is indeed compatible with the anomaly condition \eqref{anpdP4}. This can be seen in the parent theory as a bound state of $M$ fractional branes 1-2-5 and $M$ fractional branes 3-5-7, both of deformation type,  one mirror of the other through the orientifold projection. Now, however, this superposition of deformation branes can be alternatively seen as being composed of two sets of ${\cal N}=2$ fractional branes, 1-3-5 and 2-5-7, whose Coulomb branch survives the orientifold projection. 
A straightforward repetition of the scale matchings previously discussed shows that even this different UV completion has a Coulomb branch instability, eventually.

\section{Other DSB set-ups}
\label{new}

In this section we want to generalize the previous analysis and show that the DSB $SU(5)$ and 3-2 models arise in a large class of CY orientifold singularities, either $\mathbb{C}^3$ orbifolds or Pseudo del Pezzo's, of which previous examples are protoypes. As we will see, in all these models the same instability channel displayed above emerges. 

\subsection{del Pezzo singularities}
\label{Pdpgen}

We start focusing on non-orbifold singularities, like the one discussed in section \ref{PdP4}, and limit ourselves to toric CY's whose dual gauge theories admit at most eight gauge factors. The complete list of corresponding toric diagrams and dimers can be found in \cite{Franco:2017jeo}, to which we refer for details. 

Most of these singularities are obtained as blow ups of del Pezzo singularities. Toric CYs at del Pezzo singularities are complex cones over del Pezzo surfaces $dPn$ with $n=0,\dots,3$  \cite{Franco:2005rj}. By blowing up at smooth points of the del Pezzo one obtains larger CY singularities, dubbed Pseudo del Pezzo's, following the terminolgy of  \cite{Feng:2002fv}. The blow up corresponds to unhiggsing in the dual field theory.

Within this class, we list below those singularities which, after suitable orientifold projection, admit an anomaly free rank assignment giving a $SU(5)$ or 3--2 dynamical supersymmetry breaking model. The general procedure to obtain a consistent orientifold projection from a dimer is summarized in appendix \ref{append1}, whose conventions we follow. 

For each singularity we present the dimer, including the unit cell and the orientifold action. Orientifold charges are reported as a string of plus and minus signs with the following conventions. For point reflection, by starting from the bottom left corner of the unit cell and going clockwise. For lines, the first sign is for the central line and the second for the one on the edge of the fundamental cell. We also present the gauge group and the matter content of the orientifolded theory, the anomaly cancellation conditions (ACC) for the $SU$ gauge factors and the rank assignment leading to interesting DSB configurations (as far as the 3--2 model, it is understood that also the correct cubic superpotential term is reproduced). Finally, for each singularity, we indicate the Coulomb branch directions whose quantum dynamics we have analyzed, following the two-steps Higgsing pattern discussed in previous sections. For the sake of clarity, these are indicated in terms of faces of the dimer after the orientifold action has been taken into account. Our end results are summarized in the table below.
\begin{center}
	\begin{tabular}{c|c|c|}
	&  $SU(5)$ model & 3-2 model \\ \hline
		$PdP3_c$ &  $\circ$ & $\circ$ \\ \hline
		$PdP3_b$(line) &  $\circ$ & $\times$ \\ \hline
		$PdP3_b$(point)  &  $\circ$ & $\times$ \\ \hline
		$PdP4_b$ &  $\circ$ & $\circ$ \\ \hline
		$PdP5$ &  $\circ$ & $\times$ \\ \hline
		$PdP5_b$ &  $\circ$ & $\times$ \\ \hline
		$PdP5'_a$ &  $\circ$ & $\times$ \\ \hline
		$PdP5'_b$ &  $\circ$ & $\times$ \\ \hline
	\end{tabular}
\end{center}

\begin{itemize}

\item $PdP3_c$ $(+,-,+,-)$
\begin{figure}[ht]
 \centerline{   \includegraphics[width=0.25\linewidth]{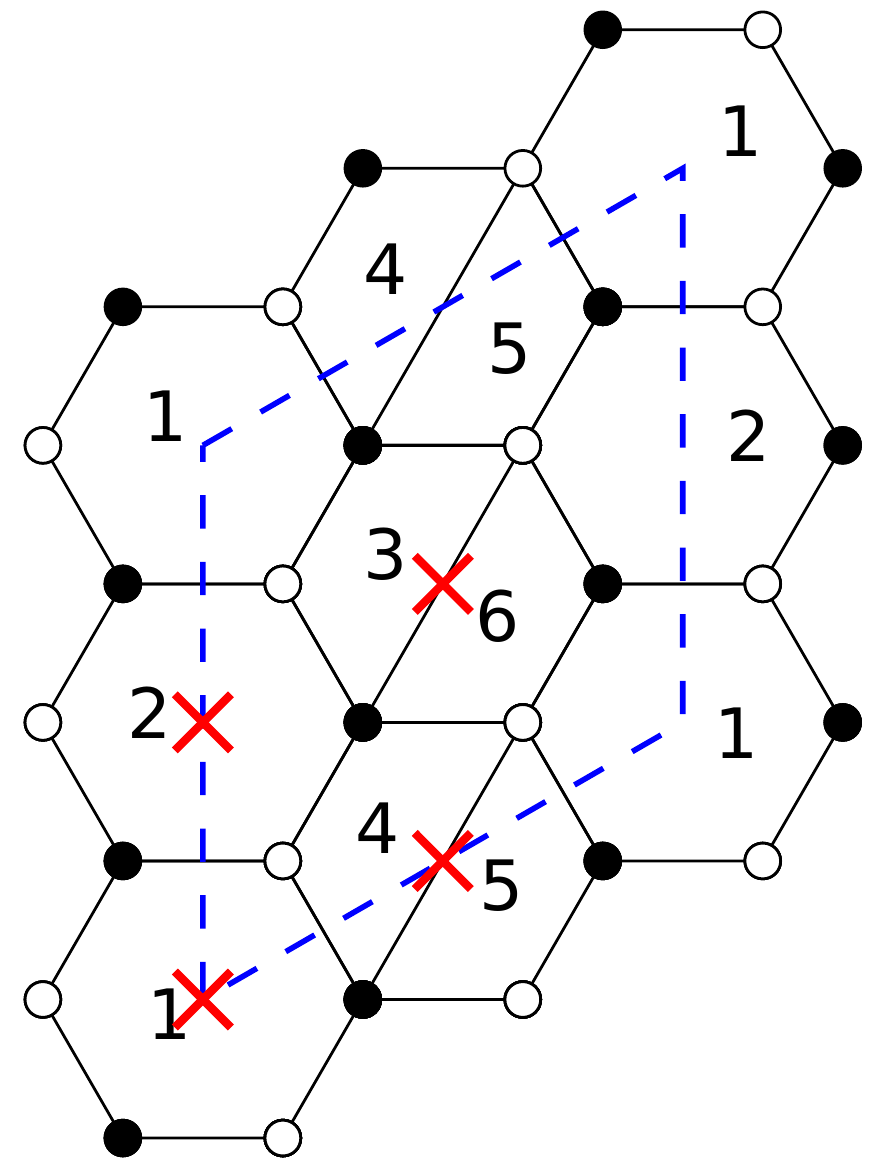}}
   \caption{$PdP3_c$ dimer with orientifold points.}
  \label{fig:PdP3_c p dimer}
\end{figure}

- Gauge group:
\be SO(N_1) \times USp(N_2) \times SU(N_3) \times SU(N_4) \ee

- Matter content:
\begin{eqnarray}
&&X_1=({\tiny  \yng(1)}_{\, 1}, \wb {\tiny  \yng(1)}_{\, 3}), \hspace{1 cm} X_2=( {\tiny  \yng(1)}_{\, 3}, {\tiny  \yng(1)}_{\, 2}), \hspace{1 cm} Z=( {\tiny  \yng(1)}_{\, 3} , {\tiny  \yng(1)}_{\, 4}) \nonumber
\\
&&Y_1=( {\tiny  \yng(1)}_{\, 4} , {\tiny  \yng(1)}_{\, 1}),  \hspace{1 cm} Y_2=({\tiny  \yng(1)}_{\, 2} ,\wb {\tiny  \yng(1)}_{\, 4}),   \hspace{1 cm} V=({\tiny  \yng(1)}_{\, 2} , {\tiny  \yng(1)}_{\, 1}), \nonumber
\\
&&\wb A_4={\tiny \wb{\yng(1,1)}_{\, 4}},  \hspace{2 cm} \wb S_3={\tiny  \wb {\yng(2)}_{\, 3}} .
\end{eqnarray}

- ACC:
\begin{eqnarray}
\begin{cases}
N_4-N_1+N_2-N_3-4=0  & \hspace{0.5 cm} SU(N_3) ~\mbox{and}~ SU(N_4)
\end{cases}
\end{eqnarray}

- DSB configurations:

$SU(5)$ model:  $N_1=1$, $N_4=5$; $N_3=1$, $N_4=5$ gives an additional singlet, $ \wb S_3$. 

3-2 model: $N_1=1$, $N_2=2$ and $N_4=3$.

- Coulomb branch directions: 
\be
1-4~~,~~2-3
\ee

\item $PdP3_b$ $(+,-)$
\begin{figure}[ht]
 \centerline{   \includegraphics[width=0.27\linewidth]{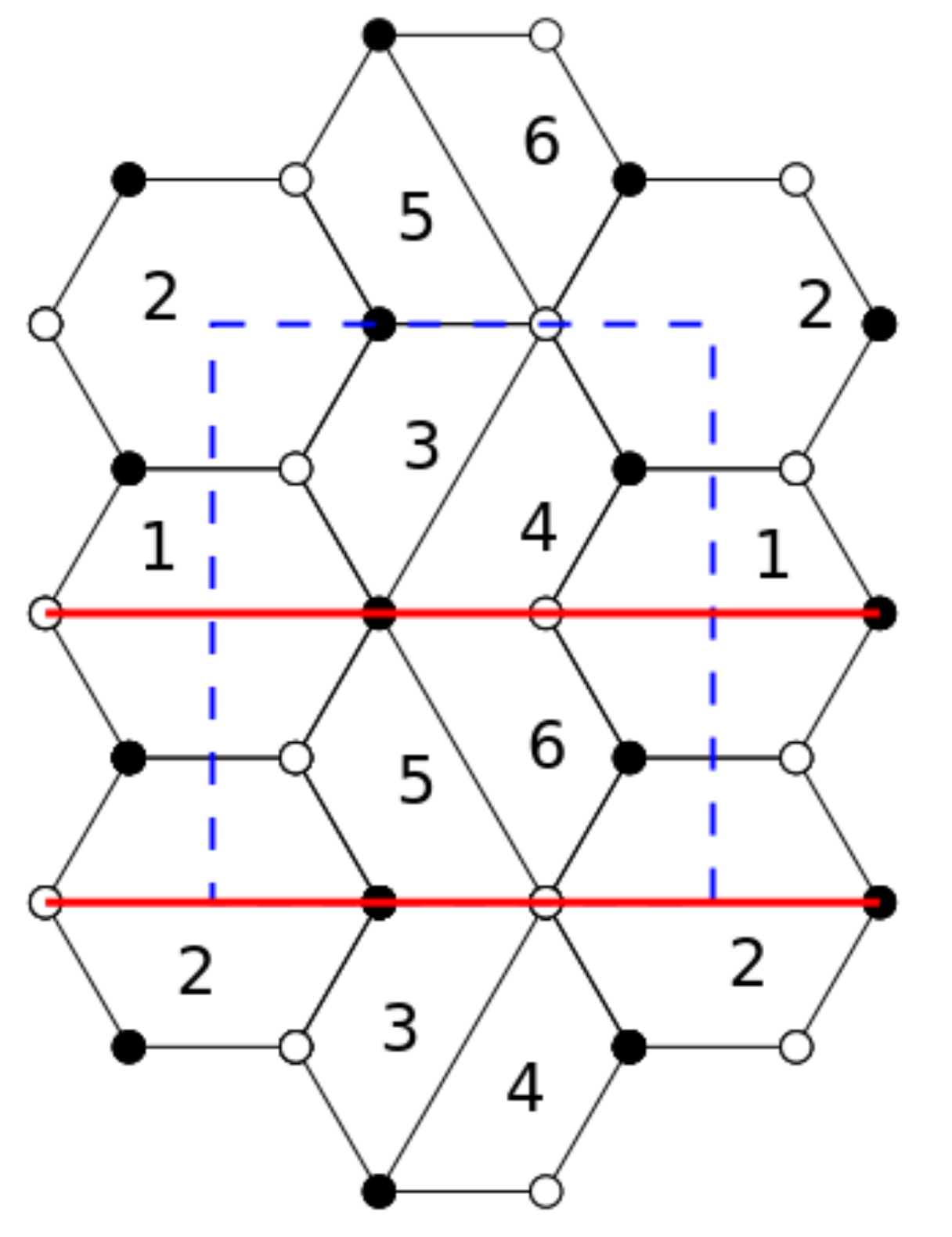}}
   \caption{$PdP3_b$ dimer with orientifold lines.}
  \label{fig:PdP3_c l dimer}
\end{figure}

- Gauge group:
\be SO(N_1) \times USp(N_2) \times SU(N_3) \times SU(N_4) \ee

- Matter content:
\begin{eqnarray}
&&Z=({\tiny  \yng(1)}_{\, 1}, {\tiny  \yng(1)}_{\, 2}), \hspace{1 cm} X_1=({\tiny  \yng(1)}_{\, 1}, {\tiny  \yng(1)}_{\, 3}), \hspace{1 cm} X_2=(\wb {\tiny  \yng(1)}_{\, 3} ,{\tiny  \yng(1)}_{\, 2}) \nonumber
\\
&&V=(\wb {\tiny  \yng(1)}_{\, 3} , {\tiny  \yng(1)}_{\, 4}),  \hspace{1 cm} Y_2=(\wb {\tiny  \yng(1)}_{\, 4} , {\tiny  \yng(1)}_{\, 2}),   \hspace{1 cm} Y_1=({\tiny  \yng(1)}_{\, 1} , {\tiny  \yng(1)}_{\, 4}), \nonumber
\\
&&\wb S_4={\tiny \wb {\yng(2)}_{\, 4}},  \hspace{2 cm} A_3={\tiny  \yng(1,1)_{\, 3}} .
\end{eqnarray}
- ACC:
\begin{eqnarray}
\begin{cases}
N_1-N_2+ N_3 -N_4-4=0 &  \hspace{0.5 cm}  SU(N_3) ~\mbox{and}~ SU(N_4)
\end{cases}
\end{eqnarray}
- DSB configurations:

$SU(5)$ model: $N_3=5$ and $N_4=1$. This model has an extra singlet due to the symmetric tensor of $SU(N_4)$, $\wb S_4$.\footnote{$N_3=5$ and $N_2=1$ is not a valid configuration since $N_2$, being related to a $USp$ group, has to be even.}

- Coulomb branch directions: 
\be
3-4~~,~~1-2
\ee

Note that this is the only example we found of a {\it line} orientifold admitting an anomaly free rank assignment leading to a DSB configuration.

\item $PdP3_b$ $(+,-,+,-)$

\begin{figure}[ht]
\centerline{   \includegraphics[width=0.26\linewidth]{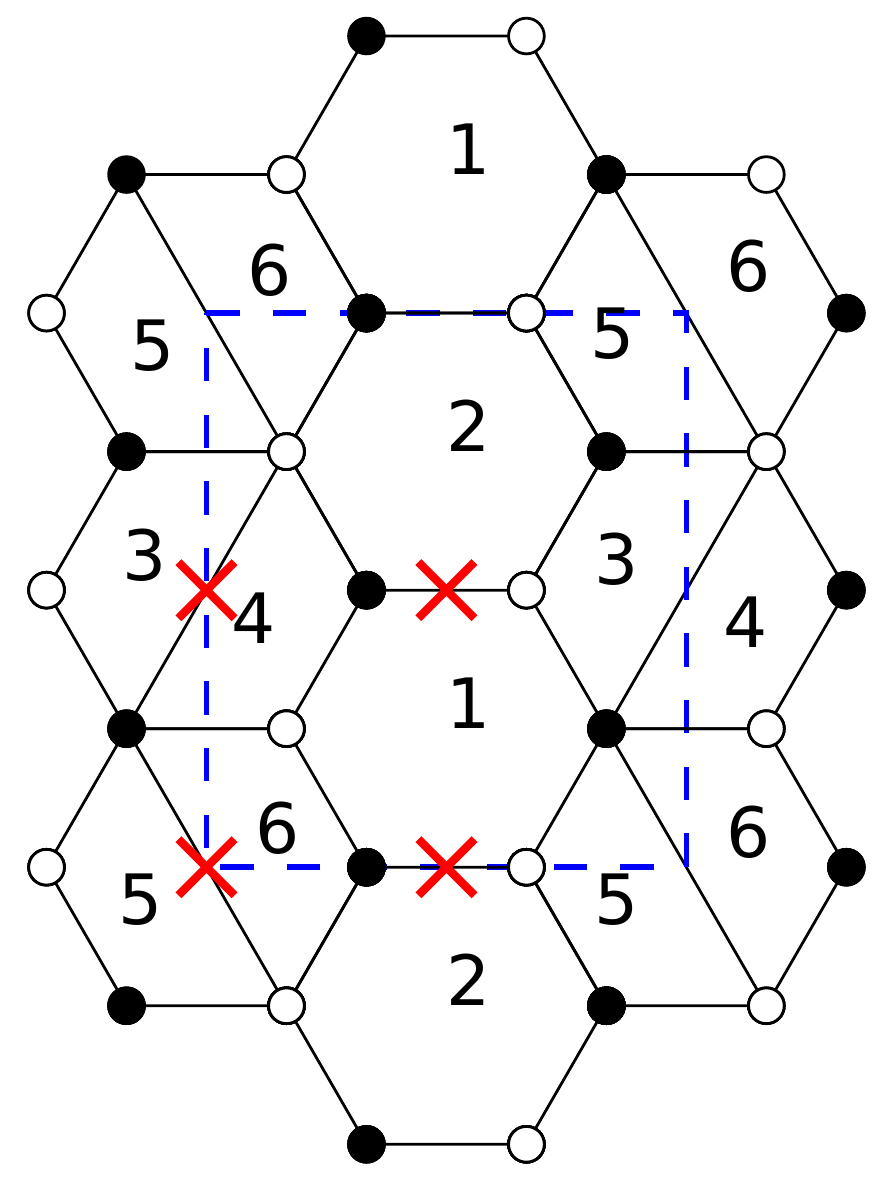}}
   \caption{$PdP3_b$ dimer with orientifold points.   }
  \label{fig:PdP3_b dimer}
\end{figure}

- Gauge group:
\be 
SU(N_1) \times SU(N_3) \times SU(N_5) 
\ee
- Matter content:
\begin{eqnarray}
&&Z=(\wb {\tiny  \yng(1)}_{\, 5}, {\tiny  \yng(1)}_{\, 3}), \hspace{1 cm} X_1=(\wb {\tiny  \yng(1)}_{\, 1}, {\tiny  \yng(1)}_{\, 3}), \hspace{1 cm} X_2=( {\tiny  \yng(1)}_{\, 1} , {\tiny  \yng(1)}_{\, 5}) \nonumber
\\
&&Y_1=(\wb {\tiny  \yng(1)}_{\, 3} , \wb {\tiny  \yng(1)}_{\, 1}),   \hspace{1 cm} Y_2=(\wb {\tiny  \yng(1)}_{\, 5} , {\tiny  \yng(1)}_{\, 1}), \nonumber
\\
&&\wb S_1={\tiny \wb {\yng(2)}_{\, 1}},\hspace{1.6 cm}  A_1={\tiny {\yng(1,1)}_{\, 1}}, \nonumber
\\
&& S_5={\tiny  {\yng(2)}_{\, 5}}, \hspace{1.6 cm} \wb A_3={\tiny \wb {\yng(1,1)}_{\, 3}} .
\end{eqnarray}
- ACC:
\begin{eqnarray}
\begin{cases}
N_3-N_5-4=0  &  \hspace{0.5 cm}  SU(N_1)\,,\,SU(N_3)~\mbox{and}~ SU(N_5)
\end{cases}
\end{eqnarray}
- DSB configurations:

$SU(5)$ model: $N_3=5$ and $N_5=1$. This model has an extra singlet due to the symmetric tensor of $SU(N_5)$, $ S_5$.

- Coulomb branch directions: 
\be
3-5~~,~~1
\ee

\item $PdP4_b$ $(-,-,+,+)$ 

\begin{figure}[ht]
 \centerline{   \includegraphics[width=0.33\linewidth]{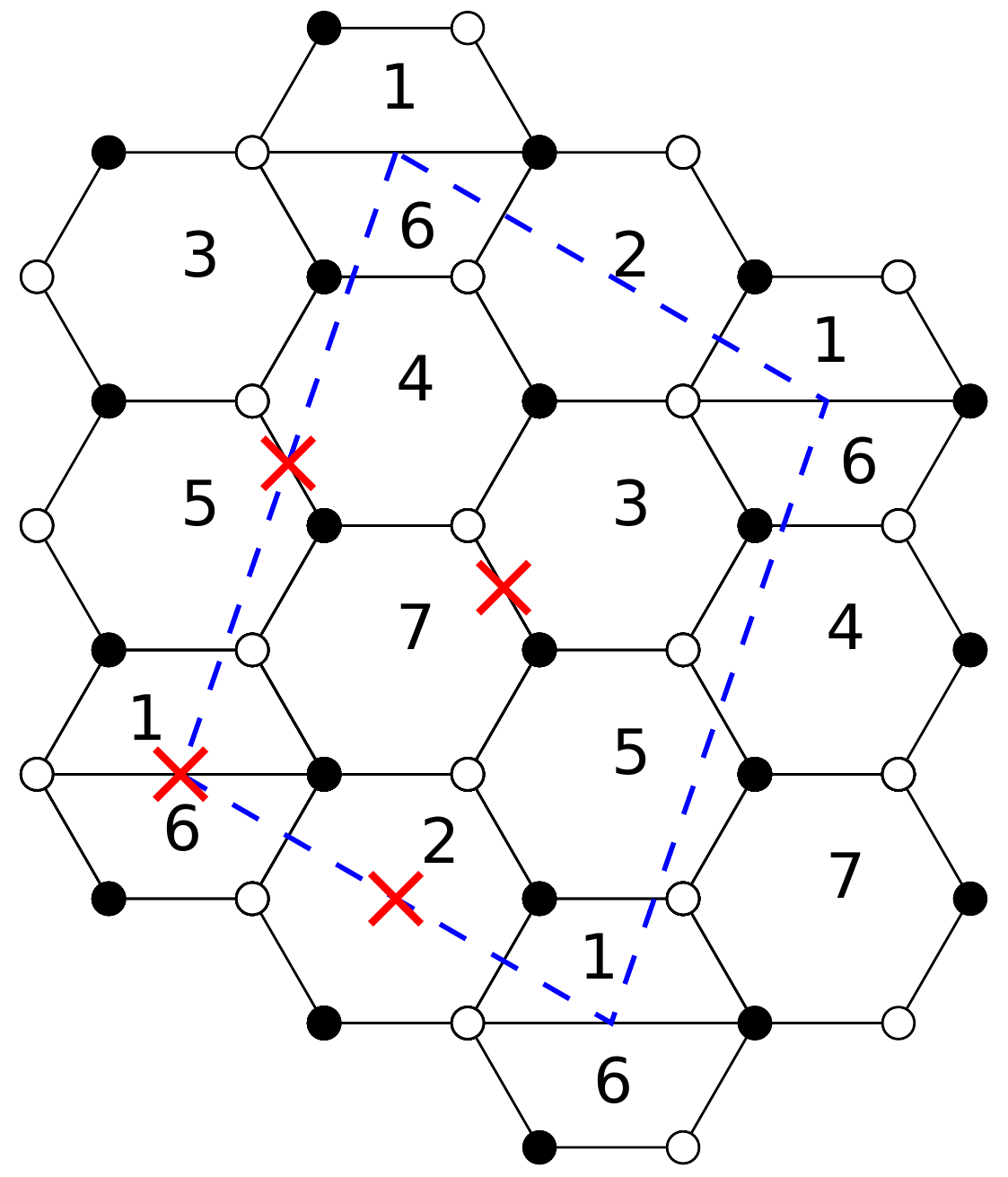}}
    \caption{$PdP4_b$ dimer with orientifold points.}
  \label{fig:PdP4_b dimer}
\end{figure}

- Gauge group:
\be SU(N_1)\times SO(N_2) \times SU(N_3) \times SU(N_4) \ee

Matter content:
\begin{eqnarray}
&&X_1=({\tiny  \yng(1)}_{\, 2}, {\tiny  \yng(1)}_{\, 1}), \hspace{1 cm} X_3=({\tiny  \yng(1)}_{\, 2}, {\tiny  \yng(1)}_{\, 3}), \hspace{1 cm} X_4=(\wb {\tiny  \yng(1)}_{\, 4} ,{\tiny  \yng(1)}_{\, 2}) \nonumber
\\
&&Y_1=(\wb {\tiny  \yng(1)}_{\, 3} , {\tiny  \yng(1)}_{\, 1}),  \hspace{1 cm} Y_2=(\wb {\tiny  \yng(1)}_{\, 1} , {\tiny  \yng(1)}_{\, 4}),   \hspace{1 cm} Z_1=(\wb {\tiny  \yng(1)}_{\, 3} , {\tiny  \yng(1)}_{\, 4}), \nonumber
\\
&&Z_2=(\wb {\tiny  \yng(1)}_{\, 4} , {\tiny  \yng(1)}_{\, 3}),  \hspace{1 cm} Z_3=(\wb {\tiny  \yng(1)}_{\, 4} ,\wb {\tiny  \yng(1)}_{\, 3}), \nonumber
\\
&&\wb A_1={\tiny \wb {\yng(1,1)}_{\, 1}},  \hspace{2 cm} S_3={\tiny  \yng(2)_{\, 3}},  \hspace{2 cm} A_4={\tiny  \yng(1,1)_{\, 4}} .
\end{eqnarray}
- ACC:
\begin{eqnarray}
\begin{cases}
N_1-N_2-N_3+N_4-4=0 &  \hspace{0.5 cm}  SU(N_1)\,,\,SU(N_3)~\mbox{and}~ SU(N_4)
\end{cases}
\end{eqnarray}
- DSB configurations:

$SU(5)$ model I: $N_1=5$ and $N_2=1$; equivalently $N_4=5$ and $N_2=1$.

$SU(5)$ model II: $N_1=5$ and $N_3=1$. This model has an extra singlet due to the symmetric tensor of $SU(N_3)$, $S_3$.

3-2 model: $N_1=3$, $N_2=1$ and $N_4=2$. There is again an extra singlet due to the antisymmetric tensor of $SU(N_4)$, $A_4$. The roles of nodes 1 and 4 can be interchanged, providing another equivalent model.

- Coulomb branch directions: 
\be
1-2~~,~~3-4
\ee

\item $PdP5$ $(-,+,+,-)$
\begin{figure}[ht]
 \centerline{   \includegraphics[width=0.5\linewidth]{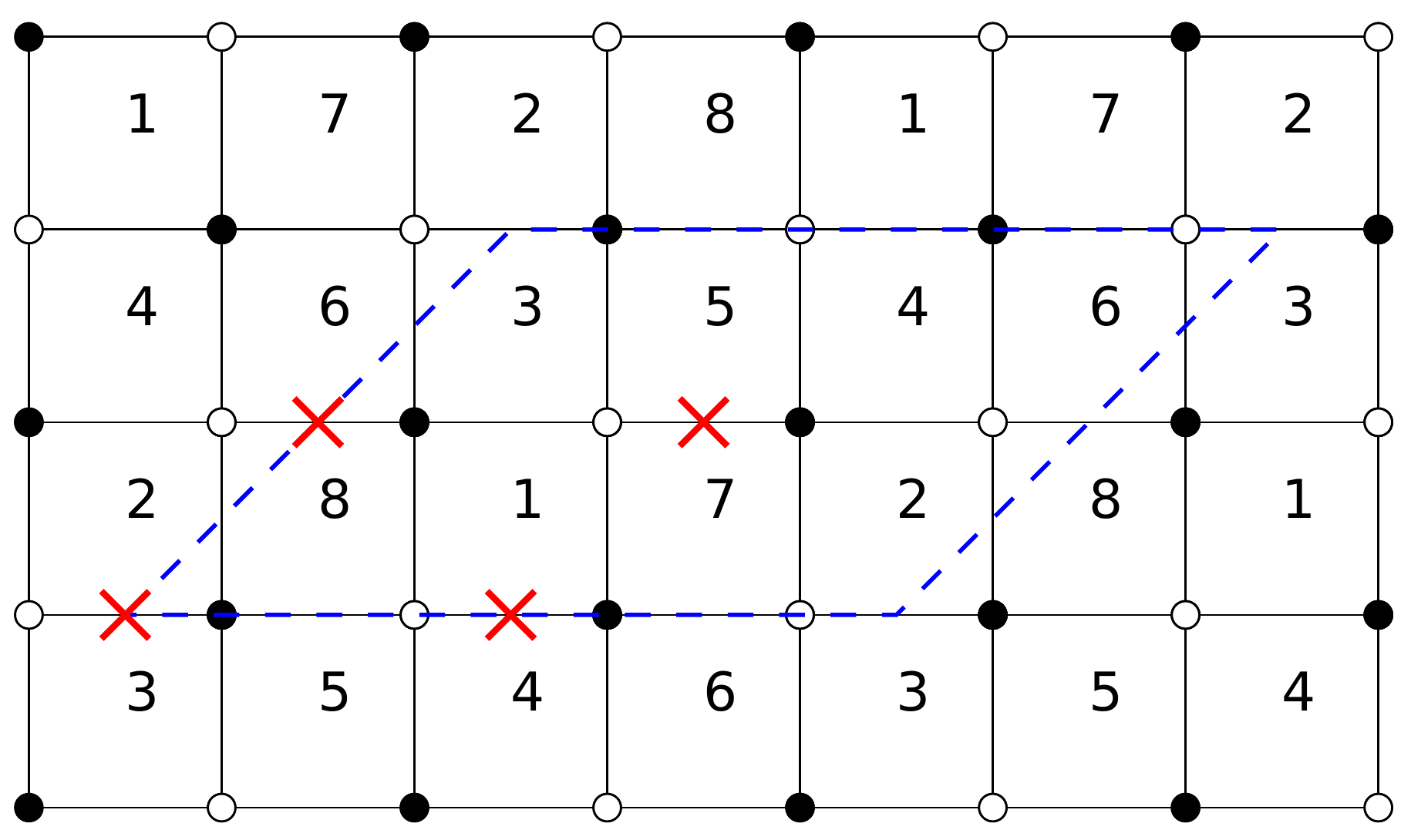}}
     \caption{$PdP5$ dimer with orientifold points.}
  \label{fig:PdP5 dimer}
\end{figure}

- Gauge group:
\be SU(N_1) \times SU(N_2) \times SU(N_5) \times SU(N_6) \ee

- Matter content:
\begin{eqnarray}
&&X_1=(\wb {\tiny  \yng(1)}_{\, 1},\wb {\tiny  \yng(1)}_{\, 5}), \hspace{1 cm} X_2=({\tiny  \yng(1)}_{\, 5}, {\tiny  \yng(1)}_{\, 6}), \hspace{1 cm} X_3=(\wb {\tiny  \yng(1)}_{\, 1} ,\wb {\tiny  \yng(1)}_{\, 6}) \nonumber
\\
&&X_4=( {\tiny  \yng(1)}_{\, 2} , {\tiny  \yng(1)}_{\, 1}),  \hspace{1 cm} X_5=(\wb {\tiny  \yng(1)}_{\, 2} , \wb {\tiny  \yng(1)}_{\, 5}),   \hspace{1 cm} X_6=(\wb {\tiny  \yng(1)}_{\, 2} , \wb {\tiny  \yng(1)}_{\, 6}), \nonumber
\\
&&A_2={\tiny \yng(1,1)_{\, 2}},  \hspace{1 cm} S_6={\tiny  {\yng(2)}_{\, 6}}, \hspace{1 cm} S_5={\tiny \yng(2)_{\, 5}},  \hspace{1 cm} A_1={\tiny   {\yng(1,1)}_{\, 1}} ~,
\end{eqnarray}
- ACC:
\begin{eqnarray}
\begin{cases}
N_1+ N_2-N_5-N_6-4=0  & \hspace{0.5 cm}  SU(N_1)\,,\,SU(N_2)\,,\,SU(N_5)~\mbox{and}~ SU(N_6)
\end{cases}
\end{eqnarray}
- DSB configurations:

$SU(5)$ models: $N_1=5$ or $N_2=5$ and $N_5=1$ or $N_6=1$. In all configurations there is an additional singlet arising from the symmetric representation at nodes 5 or 6.

Note also that in this model it is straightforward to exclude the existence of a 3-2 model, since the superpotential is purely quartic.

- Coulomb branch directions: 
\be
1-5~~,~~2-6
\ee

As a side remark, note that $PdP5$ is actually an orbifold of the conifold, hence it inherits some of its features, such as all anomalous dimensions being equal to $\gamma=-1/2$. This makes the scale matching simpler to check.

\item $PdP5_b$ $(+,-,-,+)$
\begin{figure}[ht]
 \centerline{   \includegraphics[width=0.32\linewidth]{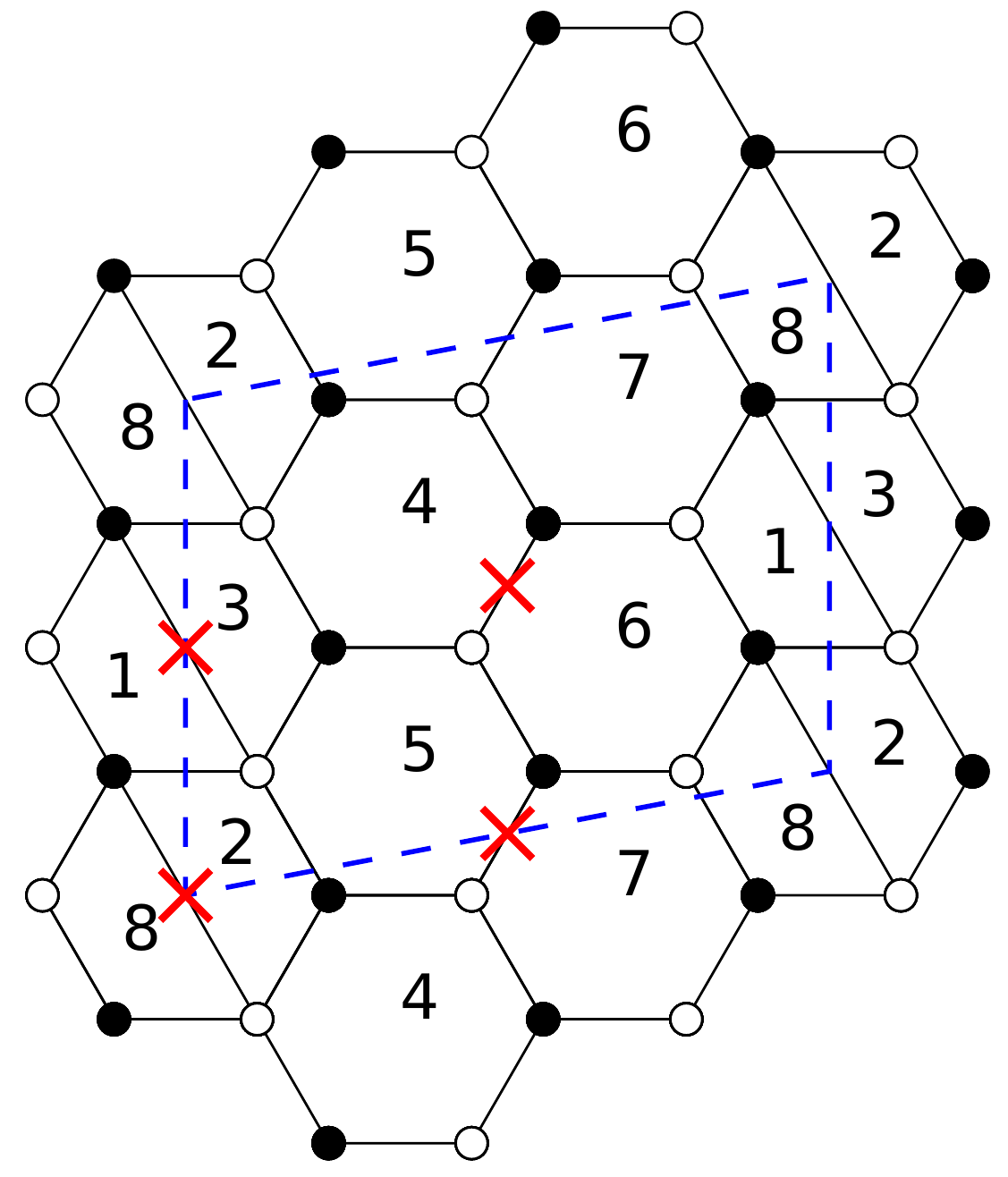}}
     \caption{$PdP5_b$ dimer with orientifold points.}
  \label{fig:PdP5_b dimer}
\end{figure}

- Gauge group:
\be 
SU(N_1) \times SU(N_2) \times SU(N_4) \times SU(N_5) 
\ee
- Matter content:
\begin{eqnarray}
&&X_1=(\wb {\tiny  \yng(1)}_{\, 1}, {\tiny  \yng(1)}_{\, 2}), \hspace{1 cm} X_2=({\tiny  \yng(1)}_{\, 1}, {\tiny  \yng(1)}_{\, 4}), \hspace{1 cm} X_3 = (\wb {\tiny  \yng(1)}_{\, 1} , \wb {\tiny  \yng(1)}_{\, 5}), \nonumber
\\
&&Y_1=(\wb {\tiny  \yng(1)}_{\, 4} , {\tiny  \yng(1)}_{\, 2}), \hspace{1 cm} Y_2=(\wb {\tiny  \yng(1)}_{\, 2} , {\tiny  \yng(1)}_{\, 5}),  \hspace{1 cm} Z_1=(\wb {\tiny  \yng(1)}_{\, 4} , {\tiny  \yng(1)}_{\, 5}),   \nonumber
\\
&&Z_2=(\wb {\tiny  \yng(1)}_{\, 5} , {\tiny  \yng(1)}_{\, 4}), \hspace{1 cm} Z_3=(\wb {\tiny  \yng(1)}_{\, 4} , \wb {\tiny  \yng(1)}_{\, 5}), \nonumber
\\
&&A_1={\tiny \yng(1,1)_{\, 1}},  \hspace{1 cm}  S_5={\tiny   {\yng(2)}_{\, 5}}, \hspace{1 cm} \wb S_2={\tiny \wb{\yng(2)}_{\, 2}},  \hspace{1 cm} A_4={\tiny  \yng(1,1)_{\, 4}} ~.
\end{eqnarray}
- ACC:
\begin{eqnarray}
\begin{cases}
N_1-N_2-N_5+N_4-4=0  &   \hspace{0.5 cm}  SU(N_1)\,,\,SU(N_2)\,,\,SU(N_4)~\mbox{and}~ SU(N_5)
\end{cases}
\end{eqnarray}
- DSB configurations:

$SU(5)$ models: $N_1=5$ and $N_2=1$, or $N_1=5$ and $N_5=1$, or $N_4=5$ and $N_2=1$. They all include a singlet related to the symmetric of the $SU(1)$ node. 

- Coulomb branch directions: 
\be
1-2~~,~~4-5 
\ee
\end{itemize}

This exhausts Pseudo del Pezzo's properly defined. Below we consider two more models in the list reported in \cite{Franco:2017jeo} {(corresponding to the toric diagrams 15 and 16 of their table 6, respectively).

\begin{itemize}
\item $PdP5'_a$ $(-,+,+,-)$

This singularity can be obtained by unhiggsing $PdP3_c$.
\begin{figure}[ht]
 \centerline{   \includegraphics[width=0.32\linewidth]{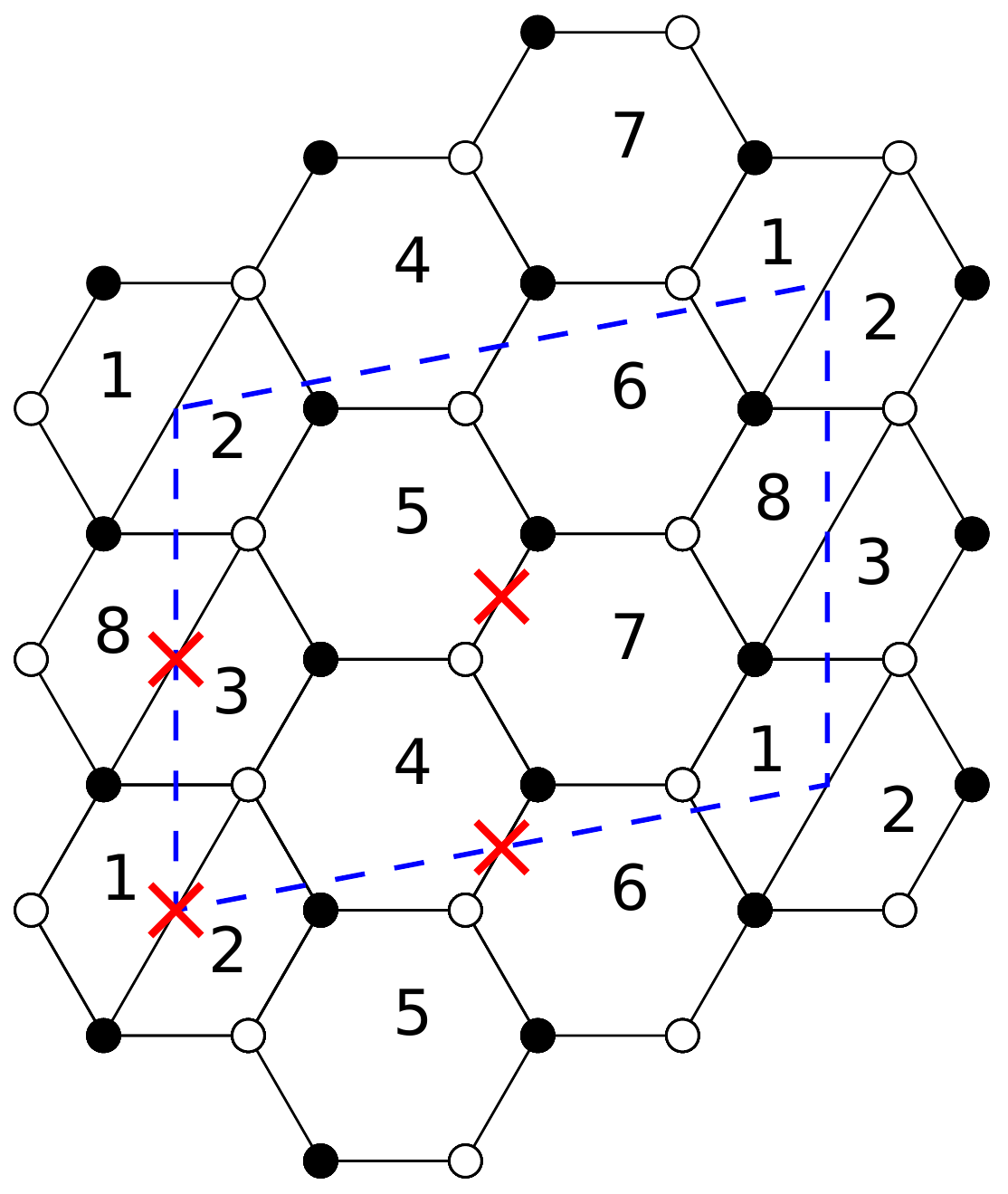}}
     \caption{$PdP5'_a$ dimer with orientifold points.}
  \label{fig:Unknown_a dimer}
\end{figure}

- Gauge group:
\be 
SU(N_1) \times SU(N_3) \times SU(N_4) \times SU(N_5) 
\ee
- Matter content:
\begin{eqnarray}
&&X_1=(\wb {\tiny  \yng(1)}_{\, 3}, {\tiny  \yng(1)}_{\, 1}), \hspace{1 cm} X_2=({\tiny  \yng(1)}_{\, 4}, {\tiny  \yng(1)}_{\, 1}), \hspace{1 cm} X_3=(\wb {\tiny  \yng(1)}_{\, 5} ,\wb {\tiny  \yng(1)}_{\, 1}) \nonumber
\\
&&Y_1=(\wb {\tiny  \yng(1)}_{\, 3} , {\tiny  \yng(1)}_{\, 5}),  \hspace{1 cm} Y_2=(\wb {\tiny  \yng(1)}_{\, 4} , {\tiny  \yng(1)}_{\, 3}),   \hspace{1 cm} Z_1=(\wb {\tiny  \yng(1)}_{\, 5} , {\tiny  \yng(1)}_{\, 4}), \nonumber
\\
&&Z_2=(\wb {\tiny  \yng(1)}_{\, 4} , \wb {\tiny  \yng(1)}_{\, 5}), \hspace{1 cm} Z_3=(\wb {\tiny  \yng(1)}_{\, 4} , {\tiny  \yng(1)}_{\, 5})\nonumber
\\
&&\wb A_1={\tiny \wb{\yng(1,1)}_{\, 1}},  \hspace{1 cm} S_5={\tiny   {\yng(2)}_{\, 5}}, \hspace{1 cm}  S_3={\tiny {\yng(2)}_{\, 3}},  \hspace{1 cm} A_4={\tiny  \yng(1,1)_{\, 4}} ~,
\end{eqnarray}

- ACC:
\begin{eqnarray}
\begin{cases}
N_1-N_3-N_4 +N_5-4=0  & \hspace{0.5 cm}  SU(N_1)~\mbox{and}~ SU(N_3)
\\
N_1-N_3+N_4-N_5-4=0 & \hspace{0.5 cm}  SU(N_4)~\mbox{and}~ SU(N_5)
\end{cases}  
\end{eqnarray}
leading to $N_1=N_3+4$, $N_4=N_5$. 

- DSB configurations:

$SU(5)$ model: $N_1=5$ and $N_3=1$. There is an additional singlet given by $S_3$. 

- Coulomb branch directions: 
\be
1-3~~,~~4-5
\ee

\item $PdP5'_b$ $(-,+,-,+)$

This singularity is again an unhiggsing of $PdP3_c$.
\begin{figure}[ht]
 \centerline{   \includegraphics[width=0.31\linewidth]{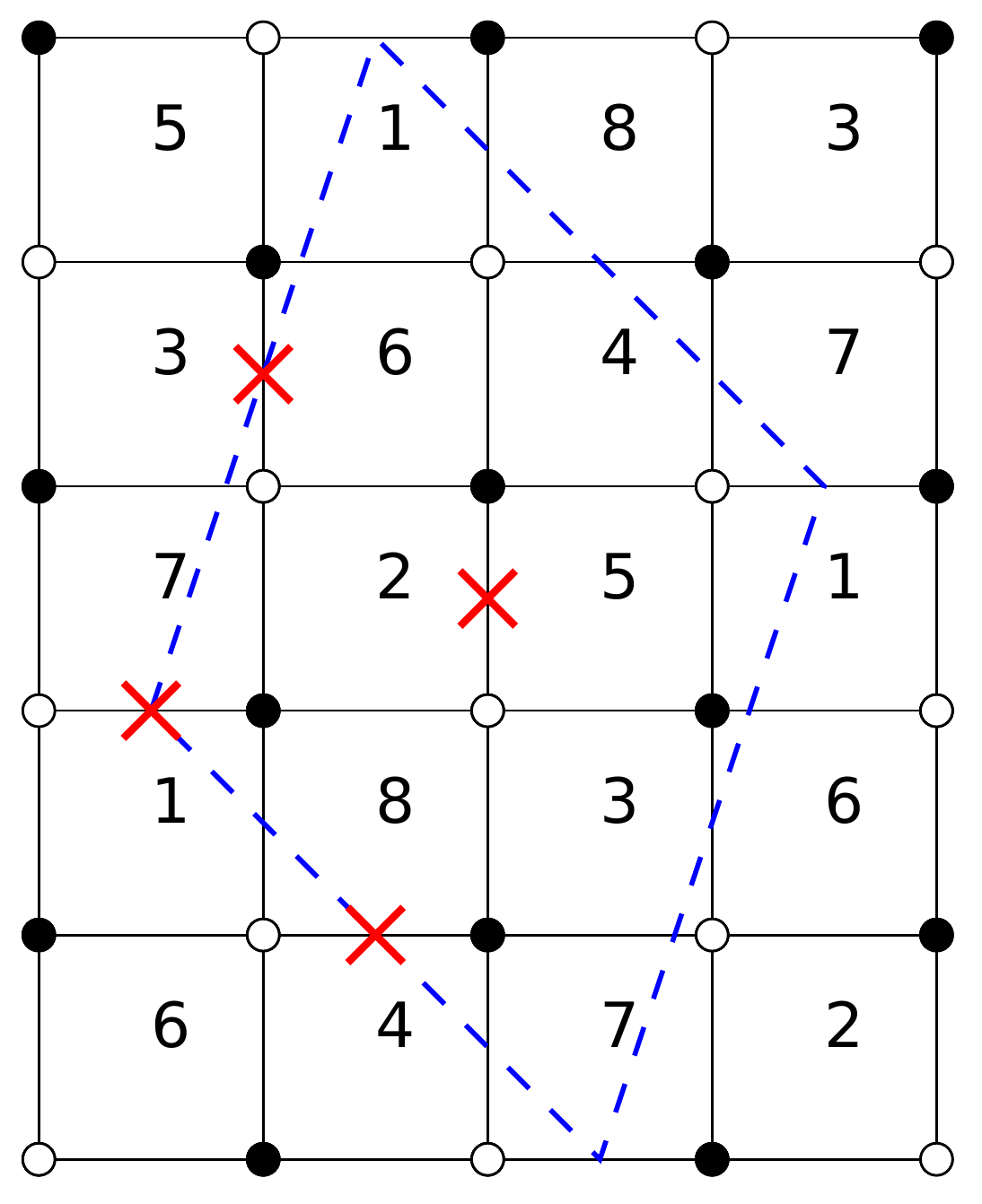}}
      \caption{$PdP5'_b$ dimer with orientifold points.}
  \label{fig:Unknown_b dimer}
\end{figure}

- Gauge group:
\be 
SU(N_1) \times SU(N_2) \times SU(N_3) \times SU(N_4) 
\ee
- Matter content:
\begin{eqnarray}
&&X_1=({\tiny  \yng(1)}_{\, 1}, {\tiny  \yng(1)}_{\, 2}), \hspace{1 cm} X_2=({\tiny  \yng(1)}_{\, 4}, {\tiny  \yng(1)}_{\, 1}), \hspace{1 cm} X_3=(\wb {\tiny  \yng(1)}_{\, 1} , \wb {\tiny  \yng(1)}_{\, 3}) \nonumber
\\
&&Y_1=(\wb {\tiny  \yng(1)}_{\, 2} , \wb {\tiny  \yng(1)}_{\, 3}),  \hspace{1 cm} Y_2=(\wb {\tiny  \yng(1)}_{\, 4} ,\wb {\tiny  \yng(1)}_{\, 2}),   \hspace{1 cm} Z=( {\tiny  \yng(1)}_{\, 4} , {\tiny  \yng(1)}_{\, 3}), \nonumber
\\
&&\wb A_1={\tiny \wb {\yng(1,1)}_{\, 1}},  \hspace{1 cm} A_2={\tiny  \yng(1,1)_{\, 2}}, \hspace{1 cm} S_3={\tiny \yng(2)_{\, 3}},  \hspace{1 cm} \wb S_4={\tiny  \wb{\yng(2)}_{\, 4}} ~,
\end{eqnarray}
- ACC:
\begin{eqnarray}
\begin{cases}
N_1-N_2+N_3 -N_4-4=0  & \hspace{0.5 cm}  SU(N_1)~\mbox{and}~ SU(N_4)
\\
N_1+ N_2-N_3-N_4-4=0 & \hspace{0.5 cm}  SU(N_2)~\mbox{and}~ SU(N_3)
\end{cases} 
\end{eqnarray}
leading to $N_1=N_4+4$, $N_2=N_3$.

- DSB configurations:

$SU(5)$ model: $N_1=5$ and $N_4=1$. We have again the additional singlet $\wb S_4$. 

- Coulomb branch directions: 
\be
1-4~~,~~2-3
\ee
\end{itemize}

One can continue the unhiggsing process and look for more and more singularities admitting fractional brane configurations described by  $SU(5)$ or 3--2 DSB models at low energy. The procedure is easy to understand from the point of view of toric diagrams and we refer the interested reader to \cite{Franco:2017jeo}. 

The above analysis shows that the DSB $SU(5)$ and the 3--2 models are not specific to the $PdP4$ example discussed in section \ref{PdP4}, but actually arise in a large (in principle infinite, see above comment) class of (blown-up) del Pezzo singularities, sensibly enlarging the landscape of D-brane configurations enjoying a stable DSB vacuum at low energy. 

Similarly to the $PdP4$ case, one can then ask what is the fate of these vacua in a large $N$ completion. As anticipated, one can show that the Coulomb branch directions we have indicated and that all these singularities possess, become runaway at the quantum level, and the true vacua are in fact supersymmetric.

\subsection{Orbifolds}
\label{orbgen}

In this subsection we want to generalize the analysis of section \ref{C3Z6} and present other instances of (orientifolds of) $\mathbb{C}^3$ orbifold singularities displaying DSB models. The corresponding dimers can be obtained from the hexagonal tiling of $\mathbb{C}^3$ with algorithms that can be found in \cite{Davey:2010px}. We report below a scan of both $\mathbb{C}^3 / \mathbb{Z}_n$ and $\mathbb{C}^3 / \mathbb{Z}_p\times\mathbb{Z}_q$ orbifolds. 

\subsubsection{Orbifolds $\mathbb{C}^3 / \mathbb{Z}_n$}

Following the same logic of the $\mathbb{C}^3 / \mathbb{Z}_{6\prime}$ case, we extended our analysis to higher orders of the cyclic group $\mathbb{Z}_n$. DSB models can again be found for some orientifold projections. Interestingly, no DSB models were found for $n$ odd.  We summarize our scan for $n$ as large as 30 in the table below.

\begin{center}
	\begin{tabular}{r|c|c|c|}
	& Action on $\mathbb{C}_3$	& $SU(5)$ model & 3-2 model \\ \hline
		$\mathbb{Z}_8$ & (1,2,5) & $\circ$ & $\times$ \\ \hline
		$\mathbb{Z}_{12}$ & (1,4,7) & $\circ$ & $\times$ \\ \hline
		$\mathbb{Z}_{30}$ & (1,10,19) & $\circ$ & $\times$ \\ \hline
			\end{tabular}
\end{center}
In the above table, a triplet $(i,j,k)$ refers to an orbifold action defined as
\begin{equation}
\label{orbac1}
\theta~:~~z_i \rightarrow e^{2\pi i  w_i}  z_i
\end{equation}
with $w = (a,b,c)/n$, where $a+b+c=n$.  It turns out that a necessary condition for allowing interesting DSB models is to focus on orientifold point reflection with two points on top a face. This has the effect of giving an orientifoled theory with two $SO/USp$ gauge factors and $(n-2)/2$ $SU$ factors as
\be
SO/USp \times SU \times \dots \times SU \times SO/USp~.
\ee
Let us summarize the specific features of each case.
\begin{itemize}
\item $\mathbb{Z}_{8}$	$(+,-,+,-)$
\begin{figure}[ht]
	\centerline{   \includegraphics[width=0.45\linewidth]{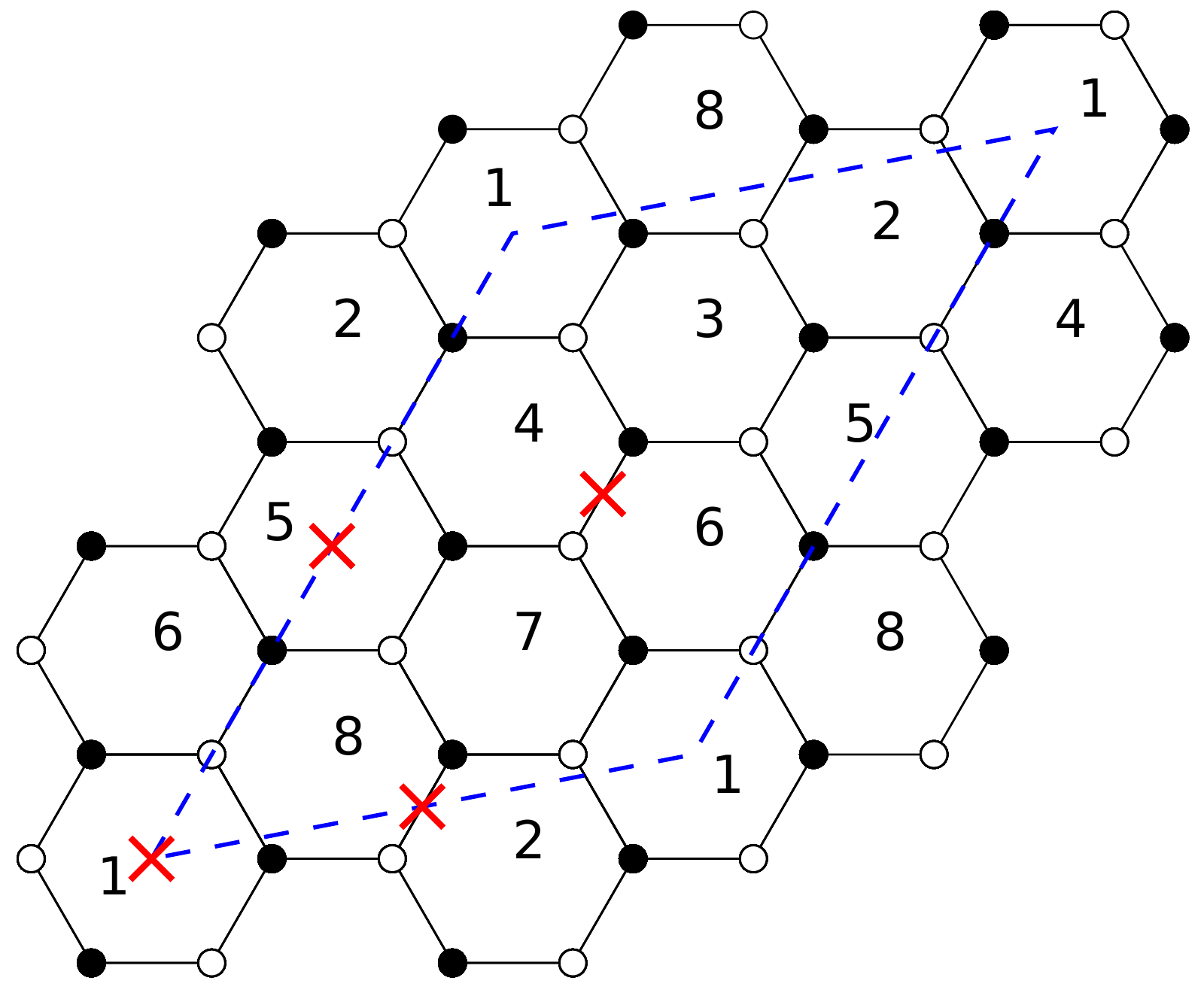}}
	\caption{$\mathbb{C}^3 / \mathbb{Z}_8$ dimer with orientifold points. 	}
	\label{fig:Z8 dimer}
\end{figure}

- Gauge group:
\be 
SO(N_1) \times SU(N_2) \times SU(N_3) \times SU(N_4) \times USp(N_5) 
\ee
- Matter content:
\begin{eqnarray}
&&({\tiny  \yng(1)}_{\, 1}, \wb {\tiny  \yng(1)}_{\, 2}), \hspace{1 cm} ({\tiny  \yng(1)}_{\, 1},\wb {\tiny  \yng(1)}_{\, 3}), \hspace{1 cm} ( {\tiny  \yng(1)}_{\, 1} ,  {\tiny  \yng(1)}_{\, 4}), \hspace{1 cm} ( {\tiny  \yng(1)}_{\, 2} , \wb {\tiny  \yng(1)}_{\, 3}) \nonumber
\\
&&({\tiny  \yng(1)}_{\, 2}, \wb {\tiny  \yng(1)}_{\, 4}), \hspace{1 cm} (\wb {\tiny  \yng(1)}_{\, 2}, {\tiny  \yng(1)}_{\, 5}), \hspace{1 cm} ( {\tiny  \yng(1)}_{\, 2} ,  {\tiny  \yng(1)}_{\, 3}), \hspace{1 cm} ( {\tiny  \yng(1)}_{\, 3} , \wb {\tiny  \yng(1)}_{\, 4}) \nonumber
\\
&&({\tiny  \yng(1)}_{\, 3},  {\tiny  \yng(1)}_{\, 5}), \hspace{1 cm} (\wb{\tiny  \yng(1)}_{\, 3},\wb {\tiny  \yng(1)}_{\, 4}), \hspace{1 cm} ( {\tiny  \yng(1)}_{\, 4} ,  {\tiny  \yng(1)}_{\, 5}), \hspace{1 cm} \wb{\tiny  \yng(1,1)}_{\, 2}, \hspace{.5 cm} {\tiny \yng(2)_{\, 4}}, 
\end{eqnarray}
- ACC:
\begin{eqnarray}
\begin{cases}
N_1+N_2-2N_3 -N_4+N_5-4=0  & \hspace{0.5 cm}  SU(N_2)
\\
N_1-N_5=0 & \hspace{0.5 cm}  SU(N_3)
\\
N_1-N_2-2N_3 +N_4+N_5+4=0  & \hspace{0.5 cm}  SU(N_4)
\end{cases} 
\end{eqnarray}
leading to $N_2=N_4+4$, $N_1=N_3=N_5$.

- DSB configurations:

$SU(5)$ model: $N_2=5$ and $N_4=1$, with an additional singlet at node 4.

- Coulomb branch directions: 
\be
2-4~~,~~1-3-5
\ee

\item $\mathbb{Z}_{12}$	$(-,+,-,+)$

\begin{figure}[ht]
	\centerline{   \includegraphics[width=0.46\linewidth]{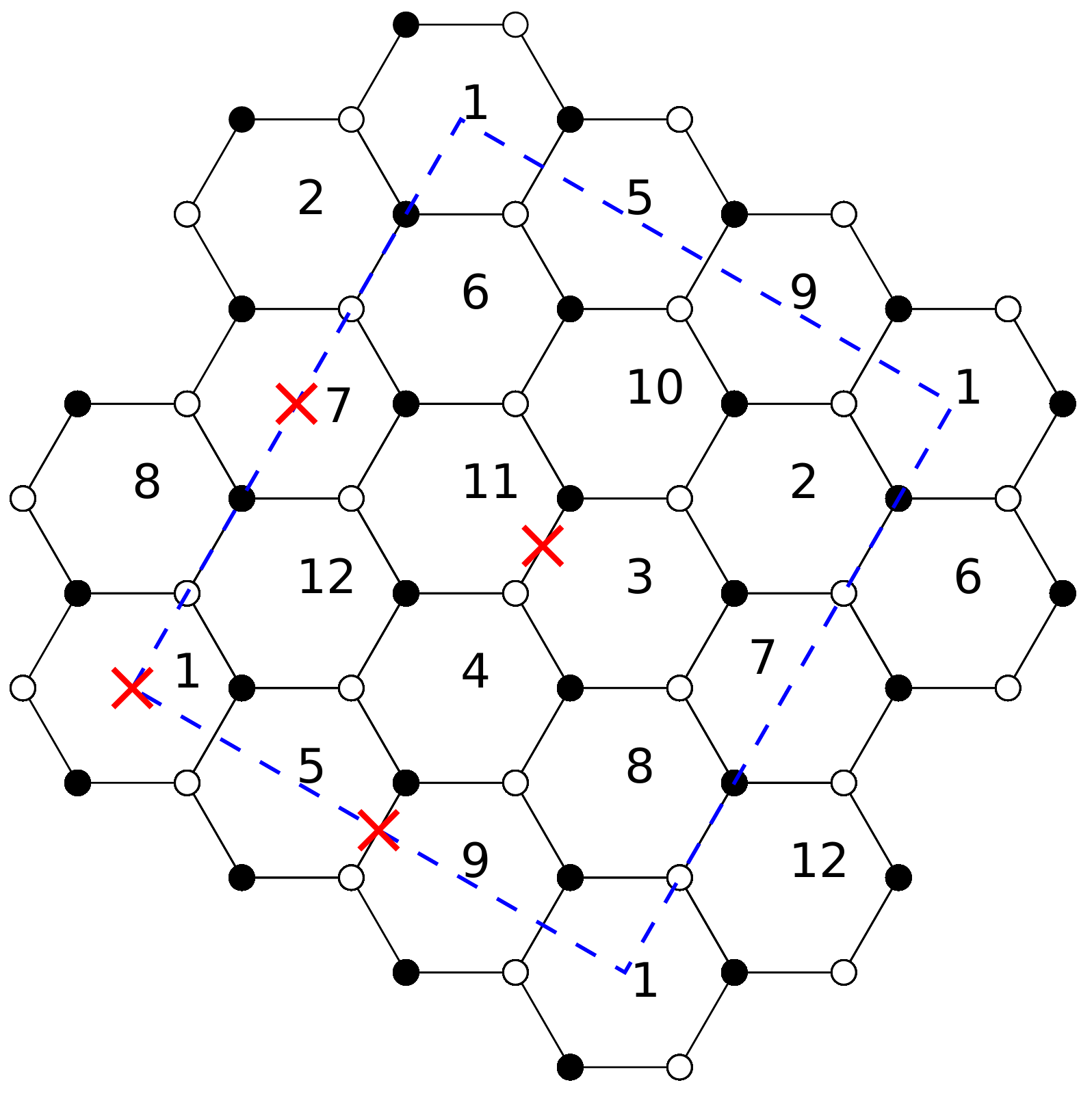}}
	\caption{$\mathbb{C}^3 / \mathbb{Z}_{12}$ dimer with orientifold points. 	}
	\label{fig:Z12 dimer}
\end{figure}

- Gauge group:
\be 
USp(N_1) \times SU(N_2) \times SU(N_3) \times SU(N_4) \times SU(N_5) \times SU(N_6) \times SO(N_7)
\ee
- Matter content:
\begin{eqnarray}
&&({\tiny  \yng(1)}_{\, 1},  {\tiny  \yng(1)}_{\, 2}), \hspace{1 cm} ({\tiny  \yng(1)}_{\, 1}, {\tiny  \yng(1)}_{\, 5}), \hspace{1 cm} ( {\tiny  \yng(1)}_{\, 1} ,  \wb{\tiny  \yng(1)}_{\, 6}), \hspace{1 cm} ( {\tiny  \yng(1)}_{\, 2} ,  {\tiny  \yng(1)}_{\, 4}) \nonumber
\\
&&(\wb{\tiny  \yng(1)}_{\, 2}, \wb {\tiny  \yng(1)}_{\, 5}), \hspace{1 cm} ( {\tiny  \yng(1)}_{\, 2}, {\tiny  \yng(1)}_{\, 7}), \hspace{1 cm} (\wb {\tiny  \yng(1)}_{\, 2} ,  {\tiny  \yng(1)}_{\, 6}), \hspace{1 cm} (\wb {\tiny  \yng(1)}_{\, 2} ,  {\tiny  \yng(1)}_{\, 3}) \nonumber
\\
&&(\wb{\tiny  \yng(1)}_{\, 3},\wb  {\tiny  \yng(1)}_{\, 4}), \hspace{1 cm} ({\tiny  \yng(1)}_{\, 3}, {\tiny  \yng(1)}_{\, 6}), \hspace{1 cm} (\wb {\tiny  \yng(1)}_{\, 3} ,  {\tiny  \yng(1)}_{\, 7}),  \hspace{1 cm} (\wb {\tiny  \yng(1)}_{\, 3} ,  {\tiny  \yng(1)}_{\, 4}) \nonumber
\\
&&({\tiny  \yng(1)}_{\, 4},  {\tiny  \yng(1)}_{\, 5}), \hspace{1 cm} (\wb{\tiny  \yng(1)}_{\, 4}, \wb{\tiny  \yng(1)}_{\, 6}), \hspace{1 cm} (\wb {\tiny  \yng(1)}_{\, 4} ,  {\tiny  \yng(1)}_{\, 5}),  \hspace{1 cm} (\wb {\tiny  \yng(1)}_{\, 5} ,  {\tiny  \yng(1)}_{\, 6}) \nonumber
\\
&&  (\wb {\tiny  \yng(1)}_{\, 6} ,  {\tiny  \yng(1)}_{\, 7}),
\hspace{1.5 cm} {\tiny  \yng(1,1)}_{\, 3}, \hspace{1.9 cm} \wb{\tiny \yng(2)}_{\, 5}.
\end{eqnarray}
- ACC:
\begin{eqnarray}
\begin{cases}
N_1+N_4-N_5 +N_7-N_6-N_3=0  & \hspace{0.5 cm}  SU(N_2)
\\
N_2-2N_4+N_6-N_7+N_3-4=0 & \hspace{0.5 cm}  SU(N_3)
\\
N_2-N_6 =0  & \hspace{0.5 cm}  SU(N_4)
\\
N_1-N_2+2N_4-N_6-N_5-4=0 & \hspace{0.5 cm}  SU(N_5)
\\
N_1-N_2-N_3+N_4-N_5+N_7=0 & \hspace{0.5 cm}  SU(N_6)
\end{cases} 
\end{eqnarray}
leading to $N_2=N_4=N_6$, $N_1=N_5+4$, $N_3=N_7+4$.

- DSB configurations:

$SU(5)$ model: $N_3=5$, $N_7=1$ and $N_1=4$. This is actually an anomaly-free $SU(5) \times USp(4)$ gauge theory, with matter charged under the $SU(5)$ factor only. The $USp(4)$ pure SYM condenses leaving exactly the  uncalculable $SU(5)$ DSB model at low energy.

- Coulomb branch directions: 
\be
3-7~~,~~1-5~~,~~2-4-6
\ee

\item $\mathbb{Z}_{30}$

Due to the large order of this orbifold, we will refrain from listing all its characteristics (and displaying the dimer), but just comment on the outcome.

Upon orientifolding, the gauge group reduces to sixteen gauge groups and the ACC allows for the following choice of non-vanishing ranks
\begin{eqnarray}
SO(1)_{1}\times SU(5)_2\times SU(4)_3 \times SU(4)_4 \times SU(4)_5 \times USp(4)_6~,
\end{eqnarray}
with matter content 
\begin{eqnarray}
&& Q=({\tiny  \yng(1)}_{\, 1} ,\wb {\tiny  \yng(1)}_{\, 2}), \hspace{1 cm} X=(\wb {\tiny  \yng(1)}_{\, 4} , {\tiny  \yng(1)}_{\, 3}), \hspace{1 cm} Y=({\tiny  \yng(1)}_{\, 5} , {\tiny  \yng(1)}_{\, 4}), \nonumber
\\
&&Z=(\wb {\tiny  \yng(1)}_{\, 3} ,\wb {\tiny  \yng(1)}_{\, 5}), \hspace{1 cm} A={\tiny \yng(1,1)_{\, 2}} ~,
\end{eqnarray}
and tree level superpotential
\begin{eqnarray}
W=YXZ ~.
\end{eqnarray}
Each $SU(4)$ factor has four flavors and they all condense on the baryonic branch. Supposing that, say, $SU(4)_3$ condenses first, the superpotential becomes a mass term for the meson $M=XZ$ and the field $Y$, which can then be integrated out. The remaining two $SU(4)$s become pure SYM at low energy and condense, too, leaving again a DSB $SU(5)$ model at low energy.

\end{itemize}

The analysis for orbifolds of order higher than 30 is more complicated, hence we stop our scan at this level. We just mention that, for instance, a $\mathbb{Z}_{40}$ orbifold seems to possess an $SU(5)$ DSB model configuration, though it comes together with a decoupled sector involving 6 more gauge groups. A preliminary analysis suggests that the extra sector eventually confines in a supersymmetric vacuum, but a detailed analysis is clearly beyond the scope of the present scan.

All above examples have also the usual Coulomb branch instability that destabilizes the DSB vacua. Being orientifolds of orbifolds, all anomalous dimensions vanish and it is a simple exercise to check that the scale matchings lead to a dependency on the VEVs in the DSB vacuum energy.

\subsubsection{Orbifolds $\mathbb{C}^3 / \mathbb{Z}_p\times\mathbb{Z}_q$}

One may also consider the product of cyclic groups, {\it i.e.}~the $\mathbb{Z}_p\times\mathbb{Z}_q$ orbifold action. 

Also within this class, at least within our scan, one can find DSB $SU(5)$ models as well as 3-2 models. The end results, for some of the cases we have analyzed, are summarized in the table below. 
\begin{center}
	\begin{tabular}{r|c|c|c|}
	& Action on $\mathbb{C}_3$	& $SU(5)$ model & 3-2 model \\ \hline
		$\mathbb{Z}_2 \times\mathbb{Z}_4 $ & [(0,1,1),(1,0,3)] & $\circ$ & $\times$ \\ \hline
		$\mathbb{Z}_3 \times\mathbb{Z}_3 $ & [(0,1,2),(1,0,2)] & $\circ$ & $\circ$ \\ \hline
		$\mathbb{Z}_2 \times\mathbb{Z}_6 $  & [(0,1,1),(1,0,5)]  & $\circ\circ$ & $\times$ \\ \hline
	\end{tabular}
\end{center}

Starting from $\mathbb{C}^3$, the orbifold action is now defined by two triplets, corresponding to  $\mathbb{Z}_p$  and $\mathbb{Z}_q$ actions, respectively, both defined as \eqref{orbac1}. Similarly, following the conventions of \cite{Davey:2010px}, to which we refer for details,  faces in the dimer have a double-index notation associated to the two independent orbifold actions.

From the dimer one can look for suitable orientifold projections and DSB anomaly free rank assignments. Again, in all cases a Coulomb branch runaway direction is present as soon as one tries to embed the D-brane configurations in a large-$N$ theory.

In the following we list the properties of each case.

\begin{itemize}
	
\item $\mathbb{Z}_2 \times\mathbb{Z}_4 $ $(-,+,-,+)$

\begin{figure}[ht]
	\centerline{   \includegraphics[width=0.24\linewidth]{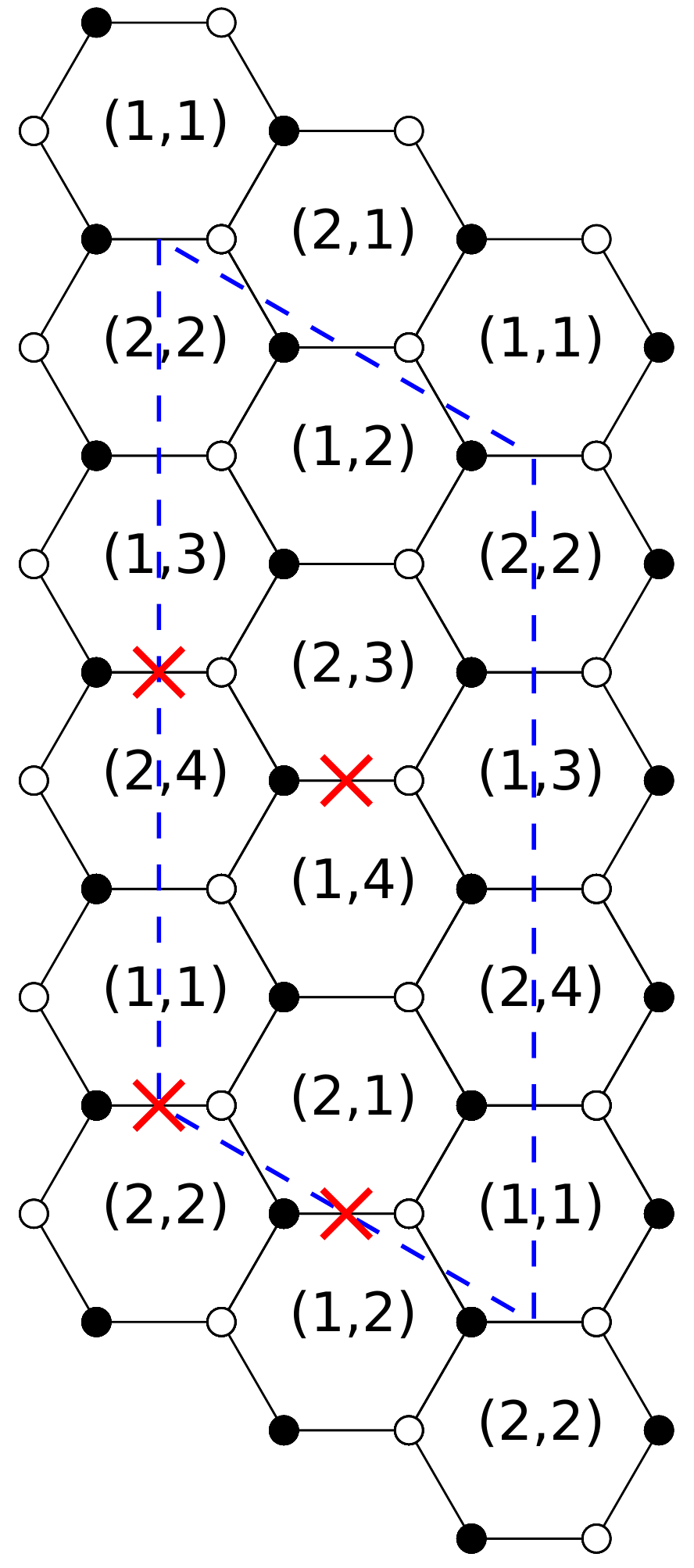}}
	\caption{$\mathbb{C}^3 / \mathbb{Z}_2 \times\mathbb{Z}_4$ dimer with orientifold points.	}
	\label{fig:Z24 dimer}
\end{figure}
We denote the surviving faces by $i\equiv (1,i)$ with $i=1\dots 4$.

- Gauge group:
\be 
SU(N_1) \times SU(N_2) \times SU(N_3) \times SU(N_4)
\ee
- Matter content:
\begin{eqnarray}
&&({\tiny  \yng(1)}_{\, 1}, \wb {\tiny  \yng(1)}_{\, 2}), \hspace{1 cm} ({\tiny  \yng(1)}_{\, 1}, {\tiny  \yng(1)}_{\, 3}), \hspace{1 cm} (\wb {\tiny  \yng(1)}_{\, 1} ,  {\tiny  \yng(1)}_{\, 4}), \hspace{1 cm} ( {\tiny  \yng(1)}_{\, 1} ,  {\tiny  \yng(1)}_{\, 2}) \nonumber
\\
&&(\wb{\tiny  \yng(1)}_{\, 1}, \wb {\tiny  \yng(1)}_{\, 2}), \hspace{1 cm} ( {\tiny  \yng(1)}_{\, 2},\wb {\tiny  \yng(1)}_{\, 3}), \hspace{1 cm} (\wb {\tiny  \yng(1)}_{\, 2} , \wb {\tiny  \yng(1)}_{\, 4}), \hspace{1 cm} ( {\tiny  \yng(1)}_{\, 3} , \wb {\tiny  \yng(1)}_{\, 4}) \nonumber
\\
&&({\tiny  \yng(1)}_{\, 3},  {\tiny  \yng(1)}_{\, 4}), \hspace{1 cm} (\wb{\tiny  \yng(1)}_{\, 3},\wb {\tiny  \yng(1)}_{\, 4}),  \hspace{.8 cm} \wb{\tiny  \yng(1,1)}_{\, 1}, \hspace{.5 cm} {\tiny \yng(2)_{\, 2}}, \hspace{.5 cm} \wb{\tiny \yng(2)}_{\, 3}, \hspace{.5 cm} {\tiny \yng(1,1)_{\, 4}}\ . 
\end{eqnarray}
- ACC:
\begin{eqnarray}
\begin{cases}
N_1+N_4-N_2-N_3 -4=0  & \hspace{0.5 cm}  SU(N_1)\,,\,SU(N_2)\,,\,SU(N_3)~\mbox{and}~ SU(N_4)
\end{cases} 
\end{eqnarray}

- DSB configurations:

$SU(5)$ model: $N_1=5$ and $N_3=1$, or $N_4=5$ and $N_2=1$. Both models have an additional singlet at nodes 3 or 2, respectively.

- Coulomb branch directions: 
\be
1-3~~,~~2-4 \ee

\item $\mathbb{Z}_3 \times\mathbb{Z}_3 $ $(-,-,+,-)$

\begin{figure}[ht]
	\centerline{   \includegraphics[width=0.50\linewidth]{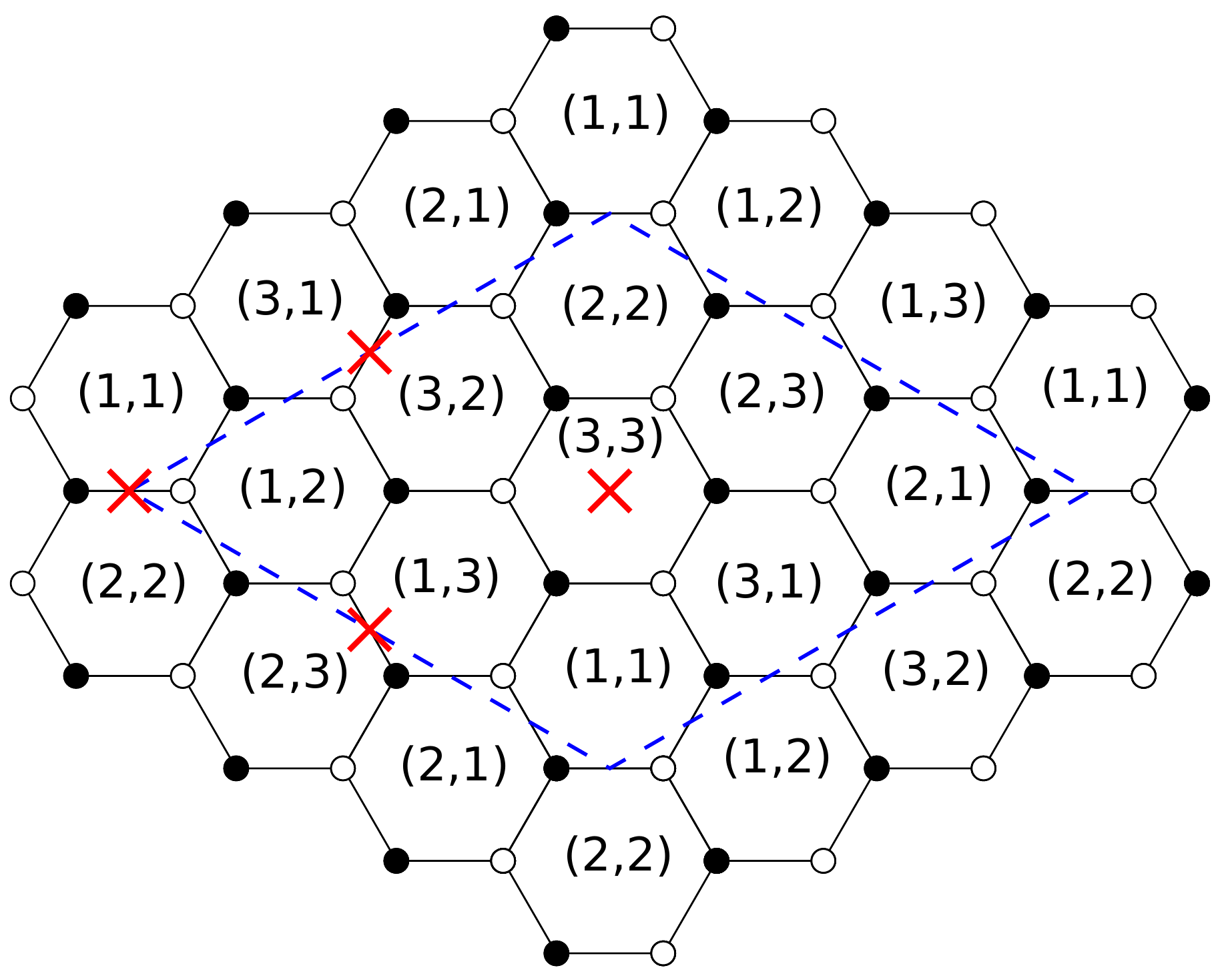}}
	\caption{$\mathbb{C}^3 / \mathbb{Z}_3 \times\mathbb{Z}_3$ dimer with orientifold points. 	}
	\label{fig:Z33 dimer}
\end{figure}
We denote the surviving faces as follows
\begin{eqnarray}
1 \equiv (1,1) \iff (2,2) &\hspace{1 cm}& 2 \equiv (1,2) \iff (2,1)  \nonumber\\
3 \equiv (1,3) \iff (2,3) &\hspace{1 cm}& 4 \equiv (3,1) \iff (3,2) \\
&5 \equiv (3,3)&  \nonumber
\end{eqnarray}

- Gauge group:
\be 
SU(N_1) \times SU(N_2) \times SU(N_3) \times SU(N_4) \times SO(N_5) 
\ee
- Matter content:
\begin{eqnarray}
&&({\tiny  \yng(1)}_{\, 1},  {\tiny  \yng(1)}_{\, 2}), \hspace{1 cm} ({\tiny  \yng(1)}_{\, 1}, {\tiny  \yng(1)}_{\, 5}), \hspace{1 cm} ( \wb{\tiny  \yng(1)}_{\, 1} ,  {\tiny  \yng(1)}_{\, 4}), \hspace{1 cm} (\wb {\tiny  \yng(1)}_{\, 1} ,  {\tiny  \yng(1)}_{\, 3}) \nonumber
\\
&&({\tiny  \yng(1)}_{\, 1}, \wb {\tiny  \yng(1)}_{\, 2}), \hspace{1 cm} (\wb {\tiny  \yng(1)}_{\, 2},\wb {\tiny  \yng(1)}_{\, 3}), \hspace{1 cm} (\wb {\tiny  \yng(1)}_{\, 2} ,  \wb{\tiny  \yng(1)}_{\, 4}), \hspace{1 cm} ( {\tiny  \yng(1)}_{\, 2} , \wb {\tiny  \yng(1)}_{\, 4}) \nonumber
\\
&&({\tiny  \yng(1)}_{\, 2}, \wb {\tiny  \yng(1)}_{\, 3}), \hspace{1 cm} ({\tiny  \yng(1)}_{\, 3}, {\tiny  \yng(1)}_{\, 4}), \hspace{1 cm} ( \wb{\tiny  \yng(1)}_{\, 3} ,  {\tiny  \yng(1)}_{\, 5}), \hspace{1 cm} ( \wb{\tiny  \yng(1)}_{\, 4} ,  {\tiny  \yng(1)}_{\, 5}), \nonumber 
\\
&&\hspace{1.9 cm} \wb{\tiny  \yng(1,1)}_{\, 1}, \hspace{2 cm} {\tiny \yng(1,1)_{\, 3}}, \hspace{2 cm} {\tiny \yng(1,1)_{\, 4}}\ . 
\end{eqnarray}
- ACC:
\begin{eqnarray}
\begin{cases}
N_1-2N_2+N_3 +N_4-N_5-4=0  & \hspace{0.5 cm} SU(N_1)\,,\,SU(N_3)~\mbox{and}~ SU(N_4)
\end{cases} 
\end{eqnarray}
while the ACC on $SU(N_2)$ is trivially satisfied.

- DSB configurations:

$SU(5)$ models: $N_5=1$ and either $N_1=5$, $N_3=5$ or $N_4=5$. 

3-2 models: $N_1=3$, $N_3=2$ and $N_5=1$, and any other permutation of nodes 1, 3 and 4. There is an additional decoupled singlet related to the antisymmetric at the $SU(2)$ node.  

- Coulomb branch directions: 
\be
1-5~~,~~2-3-4\ee

\item $\mathbb{Z}_2 \times\mathbb{Z}_6 $ $(-,-,+,+)$

\begin{figure}[ht]
	\centerline{ \includegraphics[width=0.24\linewidth]{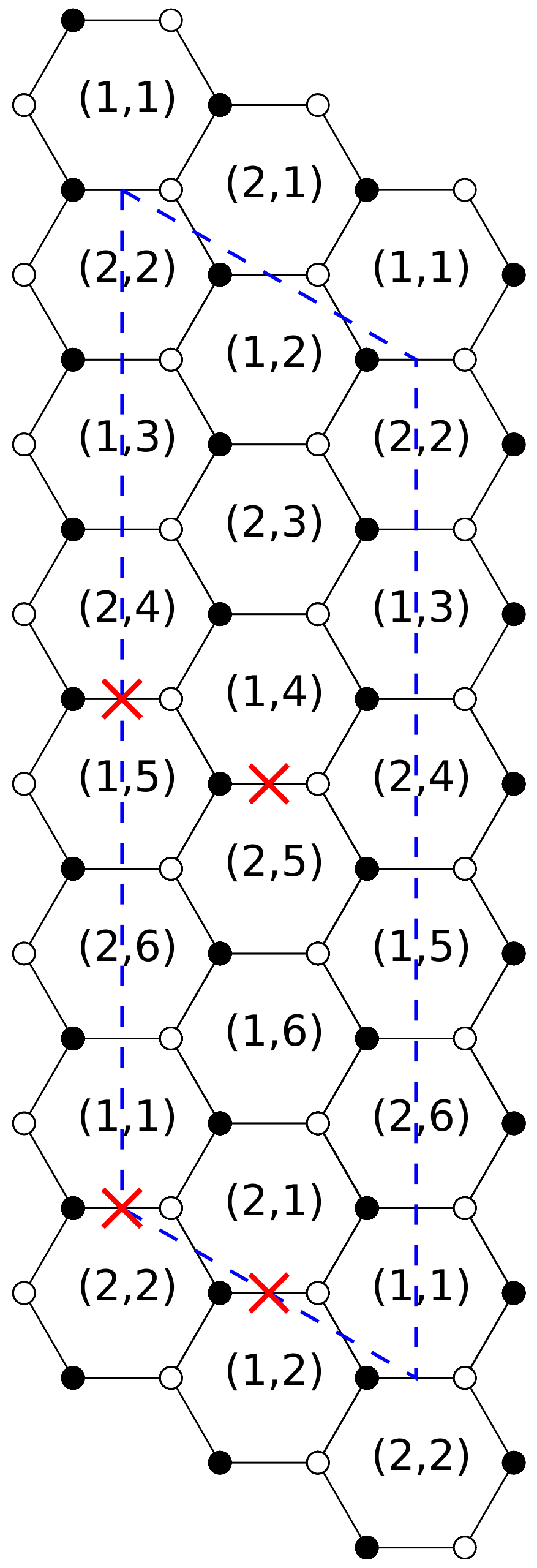}}
	\caption{The dimer of the orbifold $\mathbb{C}^3 / \mathbb{Z}_2\times\mathbb{Z}_6$ with orientifold points. 	}
	\label{fig:dimerZ26}
\end{figure}

Upon the following face identifications
\begin{eqnarray}
1 \equiv (1,1) \iff (2,2) &\hspace{1 cm}& 2 \equiv (1,3) \iff (2,6)  \nonumber\\
3 \equiv (1,5) \iff (2,4) &\hspace{1 cm}& 4 \equiv (1,2) \iff (2,1) \\
5 \equiv (1,6) \iff (2,3) &\hspace{1 cm}& 6 \equiv (1,4) \iff (2,5)  \nonumber
\end{eqnarray} 
the gauge group is $\prod _i ^6 SU(N_i)$ with matter in the following representations
\begin{eqnarray}
&& ({\tiny  \yng(1)}_{\, 1},{\tiny  \yng(1)}_{\, 2}), \hspace{1 cm} (\wb {\tiny  \yng(1)}_{\, 2} , \wb{\tiny  \yng(1)}_{\, 3}),\hspace{1 cm} (\wb {\tiny  \yng(1)}_{\, 4} , \wb {\tiny  \yng(1)}_{\, 5}), \hspace{1 cm}( {\tiny  \yng(1)}_{\, 6} , {\tiny  \yng(1)}_{\, 5}) , \nonumber\\
&& (\wb {\tiny  \yng(1)}_{\, 1} , {\tiny  \yng(1)}_{\, 5}), \hspace{1 cm} (\wb {\tiny  \yng(1)}_{\, 5} ,  {\tiny  \yng(1)}_{\, 3}),\hspace{1 cm} ( \wb {\tiny  \yng(1)}_{\, 3} , {\tiny  \yng(1)}_{\, 6}), \hspace{1 cm}(\wb {\tiny  \yng(1)}_{\, 6} ,  {\tiny  \yng(1)}_{\, 2}) , \nonumber\\
&&  ( \wb {\tiny  \yng(1)}_{\, 2} , {\tiny  \yng(1)}_{\, 4}),\hspace{1 cm} (\wb {\tiny  \yng(1)}_{\, 4} , {\tiny  \yng(1)}_{\, 1}), \hspace{1cm} ( {\tiny  \yng(1)}_{\, 1} , {\tiny  \yng(1)}_{\, 4}),\hspace{1 cm} ( \wb {\tiny  \yng(1)}_{\, 1} , \wb {\tiny  \yng(1)}_{\, 4}),  
\\
&& ( {\tiny  \yng(1)}_{\, 2}, {\tiny  \yng(1)}_{\, 5}),\hspace{1 cm} ( \wb {\tiny  \yng(1)}_{\, 2} , \wb {\tiny  \yng(1)}_{\, 5}),\hspace{1cm} ( {\tiny  \yng(1)}_{\, 3} , {\tiny  \yng(1)}_{\, 6}),\hspace{1 cm} ( \wb {\tiny  \yng(1)}_{\, 3} , \wb {\tiny  \yng(1)}_{\, 6}),  \nonumber
\\
&&\hspace{.5 cm} \wb {\tiny  \yng(1,1)}_{\, 1},  \hspace{2 cm}  {\tiny  \yng(1,1)}_{\, 3} ,  \hspace{2 cm} {\tiny  \yng(2)}_{\, 4} ,  \hspace{1.7 cm} \wb {\tiny  \yng(2)}_{\, 6}  ~. \nonumber
\end{eqnarray}
The ACC reads
\begin{eqnarray}
\begin{cases}
N_2-N_5+N_4-N_1+4=0  & \hspace{0.5 cm}  SU(N_1)~\mbox{and}~SU(N_4)
\\
N_2-N_5+N_6-N_3+4=0  & \hspace{0.5 cm} SU(N_3)~\mbox{and}~SU(N_6) 
\\
N_1-N_3-N_4+N_6=0  & \hspace{0.5 cm}  SU(N_2)~\mbox{and}~SU(N_5)
\end{cases} ~.
\end{eqnarray}
The solution to the ACC allows for a choice of ranks leaving a non-anomalous theory with gauge group
\be SU(5)_1\times SU(1)_2\times SU(5)_3 \ee
and matter content given by
\begin{eqnarray}
&& X=({\tiny  \yng(1)}_{\, 1},{\tiny  \yng(1)}_{\, 3}) ,\hspace{1 cm} A_1 = {\tiny \wb{\yng(1,1)}_{\,1}},  \hspace{1 cm} Y=(\wb {\tiny  \yng(1)}_{\, 2}, \wb {\tiny  \yng(1)}_{\, 3} ),\hspace{1 cm}  A_2 = {\tiny {\yng(1,1)}_{\,2}}~.
\end{eqnarray}

We then end up with two decoupled $SU(5)$ DSB models. Since we now have two independent contributions to the vacuum energy, one could think that the different higgsing scales can conspire in a non-trivial way, possibly leading to a non-zero minimum. 

Again, higgsing by regular branes does not destabilize the supersymmetry breaking vacua. Performing the following (three steps, now) ${\cal N}=2$ brane higgsing pattern 
\begin{eqnarray}
&& SU(5+N)_1\times SU(1+N)_2\times SU(5+N)_3 \times SU(N)_{4}\times SU(N)_{5}\times SU(N)_{6}  \nonumber 
\\
&& \overset{v}{\longrightarrow}  SU(5)_1\times SU(1+N)_2\times SU(5+N)_3 \times SU(N)_{5}\times SU(N)_{6} \nonumber 
\\
&& \overset{v^\prime}{\longrightarrow}  SU(5)_1\times SU(1+N)_2\times SU(5)_3 \times SU(N)_{5} \nonumber 
\\
&& \overset{v^{\prime\prime}}{\longrightarrow}  SU(5)_1\times SU(1)_2\times SU(5)_3 ~, \nonumber 
\end{eqnarray}
we get instead the following scale matching
\begin{eqnarray}
\Lambda_{1,\text{IR}}^{13}  = \left( \frac{v''}{v} \right)^N \Lambda_{1, \text{UV}}^{13}  & \text{and} & \Lambda_{3,\text{IR}}^{13}  = \left( \frac{v''}{v^{\prime}} \right)^N \Lambda_{3, \text{UV}}^{13} ~,
\end{eqnarray}
for the two $SU(5)$ factors, respectively.  The potential hence scales as 
\begin{eqnarray}
V &\sim & \left|\left( \frac{v''}{v} \right)^{N/13} \Lambda_{1, \text{UV}} \right|^4 + \left|\left( \frac{v''}{v^{\prime}} \right)^{N/13} \Lambda_{3, \text{UV}}\right|^4~.
\end{eqnarray}

When trying to minimize the potential with respect to $v$, $v^\prime$ and $v^{\prime \prime}$, the minimum is reached at $v''=0$, and it is a supersymmetry preserving one. In other words, there is no compensation between the two factors in the potential, as one could have in principle hoped for.

\end{itemize}

This ends the list of examples we wanted to present. As anticipated, in all orbifold models we have discussed, similarly to  the models of subsection \ref{Pdpgen}, the supersymmetry breaking vacua are destabilized once one tries to embed the DSB configurations in a large $N$ theory. As one can easily check, the mechanism is  again the same: while regular branes correspond to exact flat directions, ${\cal N}=2$ fractional brane directions become runaway once the dependence of the vacuum energy on the Coulomb branch modulus is taken into account.

\section{A no-go theorem and how to avoid it}
\label{other}

In previous sections we presented several models which allow for brane configurations giving DSB vacua, both at orbifold and del Pezzo-like singularities. However, when properly UV completed, all models include runaway directions, associated to ${\cal N}=2$ fractional branes, which destabilize the non-supersymmetric minima. One might wonder whether it is possible to get rid of such an ubiquitous instability channel. 

The first question one could ask is under which conditions the dangerous Coulomb branch direction can remain flat at the quantum level. In order for this to hold it suffices that the coefficient $\alpha$ in eq.~\eqref{scale} vanishes
\be
\alpha=0~.
\ee
Let us then see if this can happen. Let us start considering the gauge theory prior to the orientifold projection. Generically, if considering $N$ regular D3-brane at the singularity the theory is a SCFT and all $\beta$ functions vanish, that is for each gauge factor the following holds
\begin{eqnarray}
\label{sumgamma0}
\beta_{SU(N)} = 3N-\frac{N}{2}\sum_{i=1}^n (1-\gamma_i)=0~,
\end{eqnarray}
where $\gamma_i$ are the anomalous dimensions of the bi-fundamental fields charged under the given gauge group (recall that in the unorientifolded theory all matter fields are in bifundamental representations).

Let us now add $M$ fractional branes to the $N$ regular ones and focus on those gauge groups to which the fractional branes couple  to. The corresponding $\beta$ function changes as
\begin{eqnarray}
\label{sumgamma1}
\beta_{SU(N+M)} = 3(N+M)-\frac{N}{2}\sum_{i=1}^j (1-\gamma^{(0)}_i)-\frac{N+M}{2}\sum_{i=1}^k (1-\gamma^{(1)}_i)~,
\end{eqnarray}
where $\gamma^{(0)}_i$ are the anomalous dimensions of bifundamental fields charged under groups not coupling to the fractional branes  while $\gamma^{(1)}_i$ are those of fields charged under groups coupling to the fractional branes. Using eq.~\eqref{sumgamma0} and the identity $\sum_i^k (1-\gamma^{(1)}_i)+\sum_i^j (1-\gamma^{(0)}_i)=\sum_i^n (1-\gamma_i)$ the above expression can be re-written as
\begin{eqnarray}
\beta_{SU(N+M)} =\frac{M}{2}\sum_{i=1}^j (1-\gamma^{(0)}_i)~,
\end{eqnarray}
which does not vanish since fractional branes do not support a SCFT. Hence we conclude that
\be
\label{finalunor0}
\sum_{i=1}^j (1-\gamma^{(0)}_i) \neq0~.
\ee
Let us now consider the orientifold action and start with a configuration with regular D3-branes, only. One important point is that $\beta$ functions are now affected by the fact that some ranks are finitely shifted  to balance the O-plane charge. For example, in the $PdP4$ model discussed in section \ref{PdP4}, the orientifolded theory with $N$ regular branes has gauge group $SO(N+1)\times SU(N+5) \times SU(N) \times SU(N)$.

Compared to eq.~\eqref{sumgamma0}, the expression for the $\beta$ function becomes
\begin{eqnarray}
\label{betaor1}
3(N+c)-\sum_{i=1}^n (1-\gamma_i)\frac{N+b_i}{2} = 3c-\sum_{i=1}^n (1-\gamma_i)\frac{b_i}{2}~,
\end{eqnarray}
where $c$ is the extra coefficient of the gauge group we are considering and $b_i$ those of the gauge groups under which bifundamental matter is charged (in our $PdP4$ example $c=5$ for the $SU(N+5)$ group, and bifundamental matter charged also under the $SO(N+1)$ group has $b=1$). Note that the $\beta$ function is no longer vanishing, due to the O-plane charge, and its coefficient does not depend on $N$. 

Let us now perform the two-steps Higgsing which ${\cal N}=2$ fractional branes make possible, as in all models previously considered. Using the same conventions as in previous sections, the gauge coupling running at different scales is
\begin{itemize}
\item UV (above scale $v$)
\begin{eqnarray}
\frac{1}{g^2_{SU(N+c)}}=\left(3(N+c)-\sum_{i=1}^n (1-\gamma_i)\frac{N+b_i}{2}\right)\ln\left(\frac{\mu}{\Lambda_{UV}} \right)\nonumber \\
=\left( 3c-\sum_{i=1}^n (1-\gamma_i)\frac{b_i}{2} \right) \ln\left(\frac{\mu}{\Lambda_{UV}} \right) ~.
\end{eqnarray}

\item Intermediate scale (below scale $v$ and above scale $v'$)
\begin{eqnarray}
\frac{1}{g^2_{SU(c)_N}}=\left(3 c-\sum_{i=1}^k (1-\gamma^{(1)}_i)\frac{b_i}{2}-\sum_{i=1}^j (1-\gamma^{(0)}_i)\frac{N+b_i}{2} \right)\ln\left(\frac{\mu}{\Lambda_{N}} \right) \nonumber
\\
=\left(3c-\sum_{i=1}^n (1-\gamma_i)\frac{b_i}{2}-\sum_{i=1}^j (1-\gamma^{(0)}_i)\frac{N}{2}\right)\ln\left(\frac{\mu}{\Lambda_{N}} \right)~.
\end{eqnarray}

\item IR (below scale $v'$)
\begin{eqnarray}
\frac{1}{g^2_{SU(c)}}=\left( 3c-\sum_{i=1}^n (1-\gamma_i)\frac{b_i}{2} \right)\ln\left(\frac{\mu}{\Lambda} \right)~.
\end{eqnarray}

\end{itemize}
Note that this pattern holds for all groups and all kinds of matter. Indeed, the presence of (anti)symmetric representations gives factors of the form $\frac{N+b}{2}$ in the $\beta$-function, the same as having fundamental matter charged under a $SU(N+b)$ flavor group. 

Matching the scale at $\mu=v$ and $\mu=v'$ gives 
\begin{eqnarray}\label{geninst}
\Lambda^{3c-\sum_{i=1}^n (1-\gamma_i)\frac{b_i}{2}}=\left( \frac{v'}{v} \right)^{\sum_{i=1}^j (1-\gamma^{(0)}_i)\frac{N}{2}}\Lambda_{UV}^{3c-\sum_{i=1}^n (1-\gamma_i)\frac{b_i}{2}}~,
\end{eqnarray}
which implies that
\be
\alpha \propto \sum_{i=1}^j (1-\gamma^{(0)}_i)\frac{N}{2}~.
\ee
In order for $\alpha$ to vanish we need that $\sum_i^j(1-\gamma^{(0)}_i)=0$, which is in contradiction with eq.~\eqref{finalunor0}. This shows that whenever ${\cal N}=2$ fractional branes couple to the DSB nodes, they inevitably become runaway and destabilize the otherwise stable DSB vacuum.

This result suggests that in order to avoid this instability channel one could try to look at singularities which, unlike those we have analyzed, admit deformations or DSB branes and no ${\cal N}=2$ ones, and see whether there could be room for DSB models there.  

A comprehensive survey of toric singularities up to eight gauge groups is provided in \cite{Franco:2017jeo} and we have analyzed, in this finite class,  all singularities having deformation and/or DSB fractional branes only (note that $\mathbb{C}^3$ orbifolds do not belong to this class, since at these singularities a basis of fractional branes, if there are any, always includes ${\cal N}=2$ ones). 

More specifically, following the list provided in \cite{Franco:2017jeo}, the singularities not admitting ${\cal N}=2$ fractional branes are the following ones: for toric diagrams of area 2 (Table 1) singularity number 2; for toric diagrams of area 4 (Table 2) singularities number 6 and 7; for area 5 (Table 3) number 5, for area 6 (Table 4), number 8, 9, 10, 12 and 13; for area 7 (Table 5), number 7, 8 and 9; for area 8 (Table 6), number 1, 3, 4, 5, 7, 8, 13 and 17. In order to obtain the dimer, we used the techniques of \cite{Hanany:2005ss}.

Starting from these singularities, one has to see which do admit orientifold point or line projections. This can be done using the criteria spelled out in \cite{Retolaza:2016alb}. If an orientifold projection is admitted, one performs it and then checks the anomaly cancellation conditions. The latter often do not have any solutions (barring the addition of flavors). If they do have solutions, instead, one has then to see if the corresponding orientifold admits a configuration reproducing a DSB model.

The upshot of our scan is that there exist several possible point and line reflections and in some cases one can also satisfy anomaly cancellation conditions without the addition of extra flavors. When this is the case, however, it turns out that there do not exist configurations leading to any known DSB model and in fact all solutions lead to supersymmetric vacua. This result seems to suggest that the presence of line singularities ({\it i.e.} ${\cal N}=2$ fractional branes) is a key property a CY singularity should have to allow for DSB low energy dynamics but, at the same time, the one that eventually makes the vacua unstable. In the next section we will elaborate further on this point.

\section{Outlook}
\label{final}

The difficulty in finding stable supersymmetry breaking vacua might be seen as a support to the swampland conjecture advocated in \cite{Buratti:2018onj}. However, a more general, systematic approach would be required before reaching any definite conclusion. First, our analysis has focused on a large but still finite class of CY singularities which does not exhaust all possible toric varieties and their orientifolds, not to mention non-toric ones (for which there do not exist, at present, general enough and powerful techniques). Moreover, we have focused on D3-brane configurations only, and refrained from considering set-ups including also flavor D7-branes. These would considerably enlarge the number of potentially interesting models.
Similarly, the inclusion of stringy instantons effects which at orientifold singularities often play a key role, see {\it e.g.}~\cite{Argurio:2007vqa,Aharony:2007pr}, could sensibly change the gauge theory dynamics. 
Finally, we have been focusing on two DSB models and there could be, in principle, other DSB models which could be engineered with D-branes at singularities.

Following a case by case 
approach one can at best provide evidence against the existence of such DSB vacua or, on the contrary, come up with counter-examples. Note, however, that even in the latter case things might not be completely settled. Indeed, we have neglected $1/N$ corrections, in particular  to anomalous dimensions, on the grounds that such corrections would not qualitatively affect the fact that ${\cal N}=2$ branes turn out to be  unstable, see eq.~\eqref{geninst}. On the contrary, the fact that regular D3-branes are always stable depends on the exact matching of the UV and IR $\beta$ functions, which both depend on the anomalous dimensions. One should make sure that $1/N$ corrections  do not spoil this equality. After all, that regular D3-branes still possess a flat moduli space despite broken supersymmetry should not be taken for granted. 

More generally, we believe that a more geometric approach is needed in order to reach any solid conclusion. In this respect, it is very suggestive that the plethora of examples we have found enjoy one and the same instability channel, which is model-independent and has a precise geometrical meaning. String theory might be telling us something here on if and how it is possible to cure it.

Concretely, one could either try to prove that line singularities ({\it i.e.}~${\cal N}=2$ fractional brane directions) are a necessary geometric property of CYs supporting DSB models and conclude, by the no-go theorem presented in section \ref{other}, that such Coulomb branch directions inevitably become runaway when strong coupling dynamics is taken into account. Or, try to understand the geometric properties a CY singularity should have in order to support DSB vacua and be free of such ${\cal N}=2$ fractional brane directions, and see whether such CYs actually exist. Finally, it can very well be that if string theory can cure such runaway it does so by genuinely stringy effects, and a full understanding of the IR dynamics could not avoid including such contributions in the analysis. 

\section*{Acknowledgments} 
We are grateful to Sergio Benvenuti for useful discussions and to Sebastian Franco, Eduardo Garcia-Valdecasas and Angel Uranga for feedback on a preliminary draft version. 
R.A. and A.P. acknowledge support by IISN-Belgium (convention 4.4503.15) and by the F.R.S.-FNRS under the ``Excellence of Science" EOS be.h project n.~30820817, M.B. and S.M. by the MIUR PRIN Contract 2015 MP2CX4 ``Non-perturbative Aspects Of Gauge Theories And Strings" and by INFN Iniziativa Specifica ST\&FI. R.A. is a Research Director and A.P. is a FRIA grantee of the F.R.S.-FNRS (Belgium). We warmly thank each other's institutions for hospitality during the preparation of this work.


\appendix

\section{Dimers and orientifolds}
\label{append1}

In this appendix we review some basic rules which allow to extract the gauge theory living on a stack of D3-branes at Calabi-Yau singularities (with or without orientifold projection) using dimer formalism. 

\subsection{Toric Calabi-Yau singularities and dimers}

A convenient way to describe gauge theories obtained by D3-branes at {\it toric} Calabi-Yau singularities is in terms of dimers. Below we summarize how this goes, following \cite{Franco:2005rj,Franco:2005sm}, to which we refer for details. 

A dimer (or bipartite graph) is a tiling of the plane by means of polygons. Every vertex (or node) is either a white or a black dot, and edges may connect nodes of different colors, only. Physically, this describes NS5-branes tiling a torus $T^2$ or, in a T-dual picture, the gauge theory living on a stack of D3-branes  at a non-compact toric Calabi-Yau singularity. 

From the dimer it is possible to read off the quiver, which encodes the matter content of the theory, as well as the superpotential. The rules are as follows:

\begin{enumerate}
	\item Every face corresponds to a $SU(N)$ gauge group, where $N$ is the number of D3-branes.
	\item Every edge separating two faces corresponds to a bifundamental field. To avoid over counting, one should consider, say, white nodes only and, going counterclockwise, associate every edge (or line) to a field $X_{ij}$ in the 
	$(\wb {\tiny  \yng(1)}_{\, i} , {\tiny  \yng(1)}_{\, j})$ representation. 
	\item Every vertex corresponds to a superpotential term, with a + sign for white nodes and a $-$ sign for black nodes. The term is obtained contracting all fields, {\it i.e.} edges, ending on the node, going counterclockwise for white nodes and clockwise for black ones.
\end{enumerate}

Let us consider, as a concrete example, a pseudo del Pezzo singularity, known as $PdP4$ \cite{Feng:2002fv}, whose dimer is reported in figure \ref{fig:PdP4_1}.
\begin{figure}[ht]
	\centerline{ \includegraphics[width=0.27\linewidth]{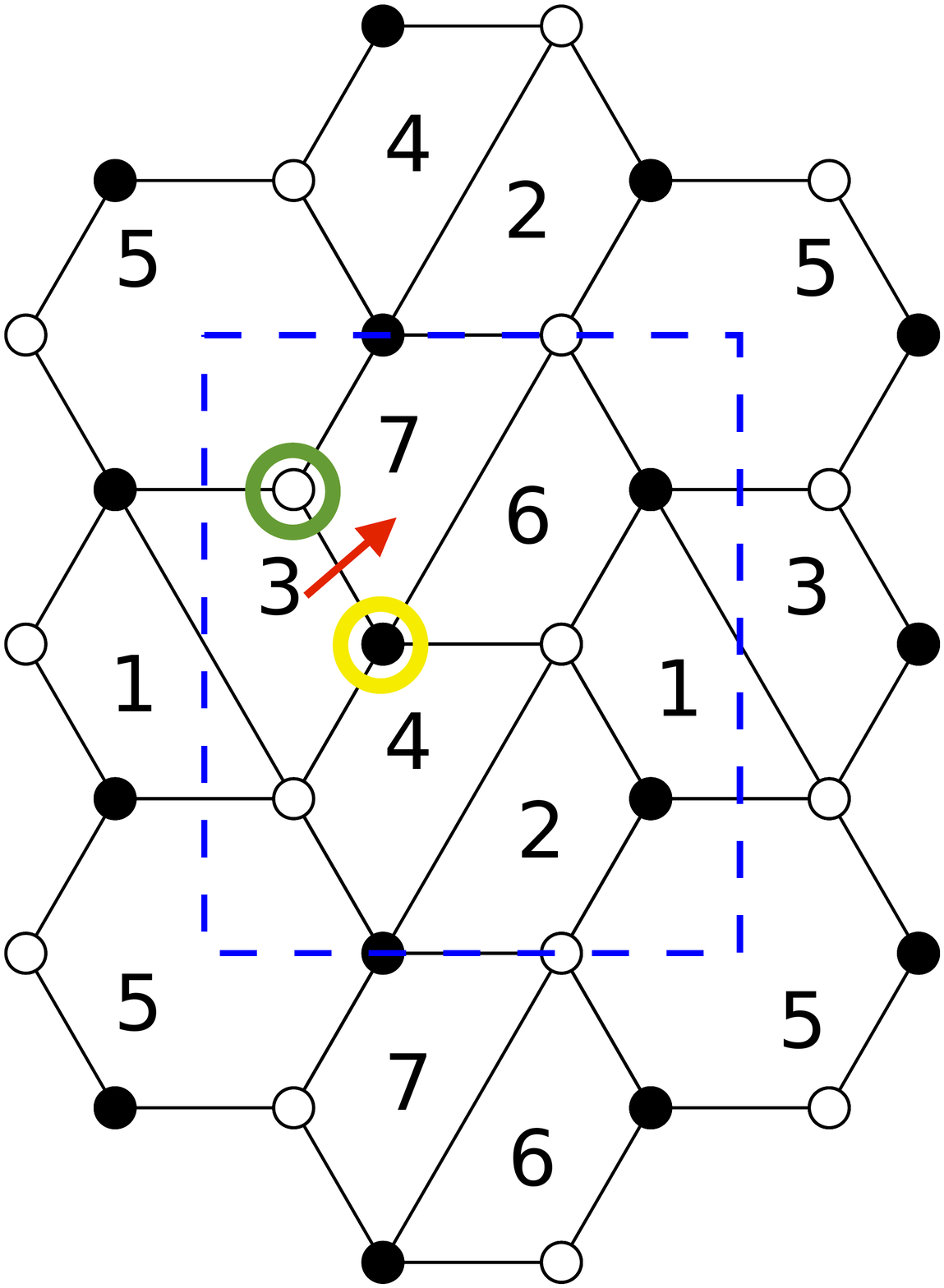}}
	\caption{The dimer of the $PdP4$ singularity. The square represents (a choice of) the fundamental cell.}
	\label{fig:PdP4_1}
\end{figure}

There are 7 distinct faces, so the gauge group is $SU(N)^7$. Matter is described by edges. To get the correct  representations we use the same convention used for quivers. One writes arrows such that white nodes stay on the left of the arrow and black nodes on its right. Arrows' tails represent antifundamentals while heads fundamentals. So the red arrow in the figure represents a chiral superfield, $X_{37}$, in the $(\wb {\tiny  \yng(1)}_{\, 3} , {\tiny  \yng(1)}_{\, 7})$  of the $SU(N)$ gauge groups associated to faces 3 and 7, respectively. Following a similar logic, one can work out all fields in the theory. 

What is left, is to compute the superpotential. Let us first focus on, {\it e.g.}, the white node surrounded by the green circle in the figure. There are three lines ending on the white node and so a cubic superpotential term associated to it. Following the general rule one gets a term proportional to $+\text{Tr}(X_{37}X_{75}X_{53})$, where the trace is both on gauge and flavor indices. For the black node surrounded by the yellow circle, instead, four lines meet, so one expects a quartic term. One has to multiply fields going clockwise and take a $-$ sign, now, so the contribution is proportional to $-\text{Tr} (X_{37}X_{76}X_{64}X_{43})$.

The full quiver and superpotential are written in section \ref{PdP4}.

\vskip 7pt

In general, one can have unbalanced ranks in the $SU$ factors, provided the cancellation of gauge anomalies. From the D-brane perspective, this corresponds to RR tadpole cancellation. Equal rank assignment corresponds to regular branes, which populate all gauge factors democratically (and provide a superconformal field theory). Other rank assignments compatible with RR tadpole cancellation correspond to fractional branes, which can be thought of as D5-branes wrapped on 2-cycles (collapsed at the singularity) whose dual 4-cycles are non-compact.

\subsubsection{Fractional branes and dimers}

Following the classification proposed in \cite{Franco:2005zu}, there exist three classes of fractional branes, which differ by the IR dynamics they trigger: confinement, effective ${\cal N}=2$ SYM or supersymmetry breaking.  The existence of fractional branes, and their nature, can be argued directly from the dimer, as summarized below.

\begin{itemize}
	\item Deformation branes:
	
	These branes correspond to isolated faces in the dimer touching each other at nodes (so, only gauge groups and no bifundamental fields are involved) or to isolated clusters of faces surrounding a given node. The gauge theory is then either a set of decoupled SYM theories, or SYM theories coupled via a superpotential term, respectively. In both cases,  the low energy effective theory leads to confinement and the geometry undergoes a complex structure deformation. 
		
	Two examples of deformation branes are reported in figure \ref{fig:boat10}.
	\begin{figure}[h!]
		\centerline{  \includegraphics[width=0.43\linewidth]{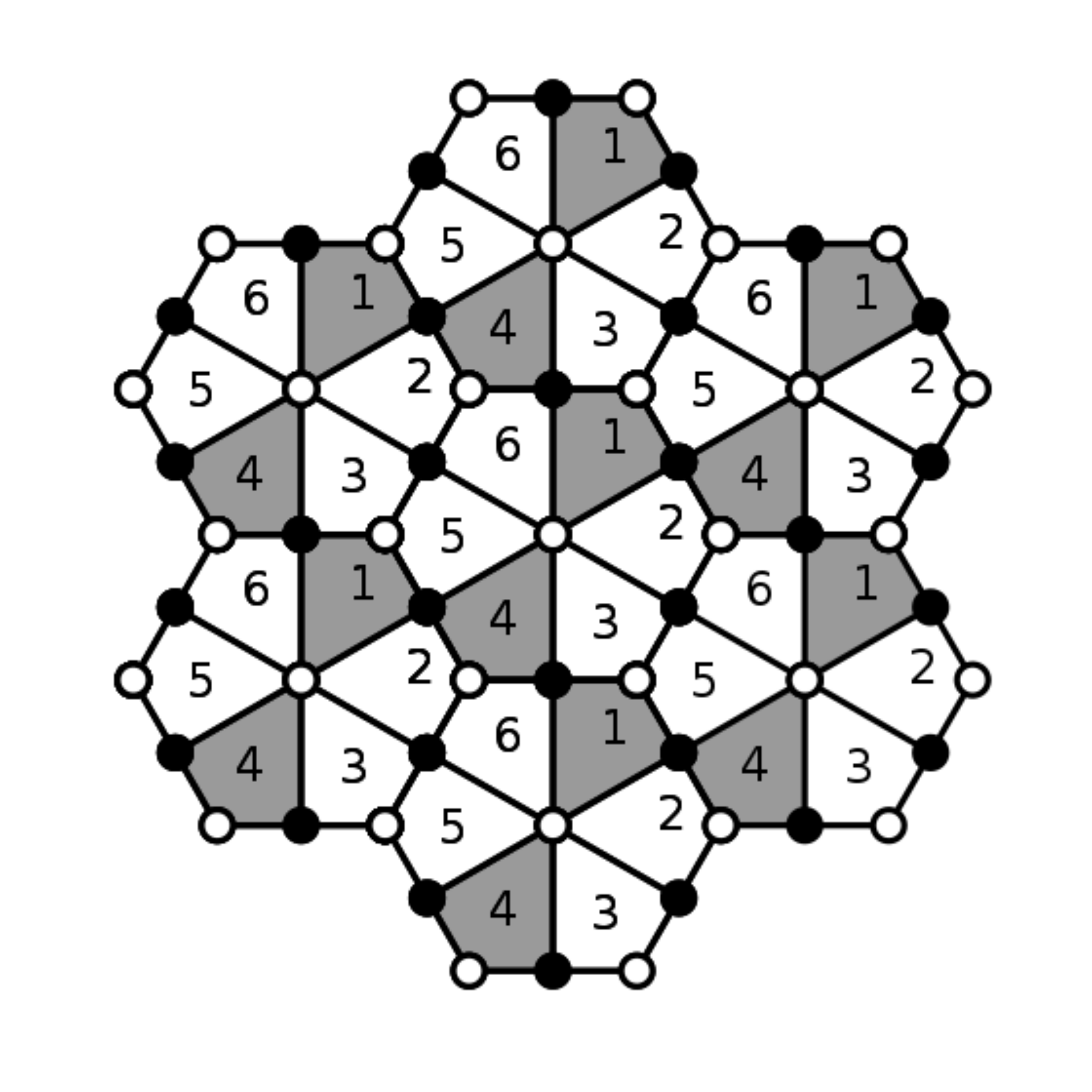}
			\includegraphics[width=0.43\linewidth]{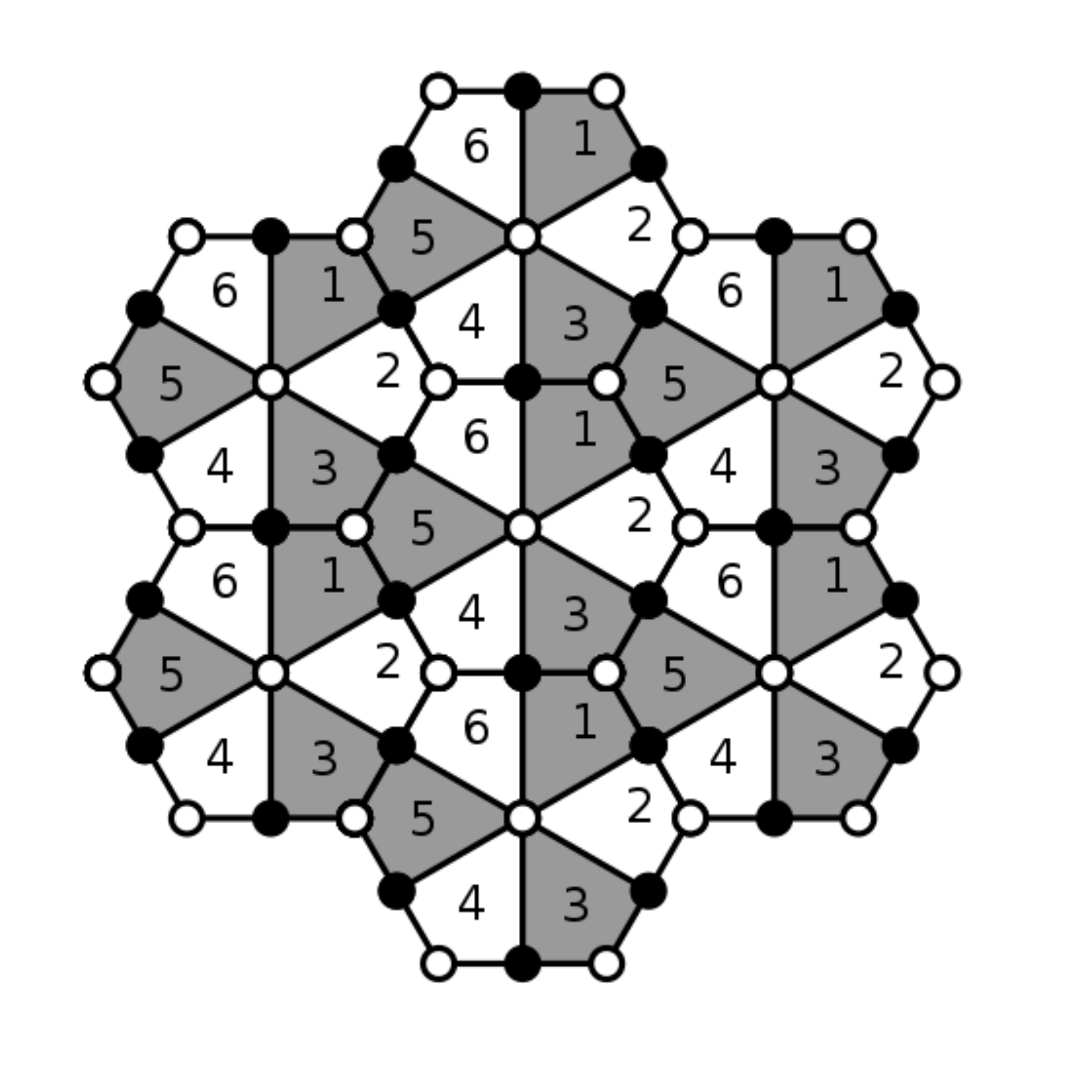}}
		\caption{The dimer of the $dP3$ singularity, which admits both classes of deformation branes. Left figure: deformation fractional branes corresponding to isolated nodes. Right figure: deformation fractional branes corresponding to loops in the quiver.}
		\label{fig:boat10}
	\end{figure}
	Note that single faces are allowed deformation branes only for non-chiral theories. Chiral theories may admit deformation branes but these correspond to clusters or to two or more isolated faces. The $dP3$ dimer in the figure is one such chiral example.
	
	\item ${\cal N}=2$ branes
	
	These other fractional branes correspond to paths along faces keeping, in our conventions, white nodes on the left and going across the unit cell without making any closed loop. This implies that the gauge invariant operator constructed along the closed path does not appear in the superpotential. The VEV of such operator is unconstrained and parametrizes a one-dimensional moduli space, along which the dynamics has ${\cal N}=2$ supersymmetry. 
	
	Geometrically, these branes correspond to D5-branes wrapping a non-trivial (collapsed) two-cycle of a CY 3-fold which locally looks like $K3 \times \mathbb{C}$, where $\mathbb{C}$ corresponds to the flat direction the brane can freely move into.
	
	A simple such example is shown in figure \ref{fig:boat12}.
	\begin{figure}[h!]
		\centerline{ \includegraphics[width=0.43\linewidth]{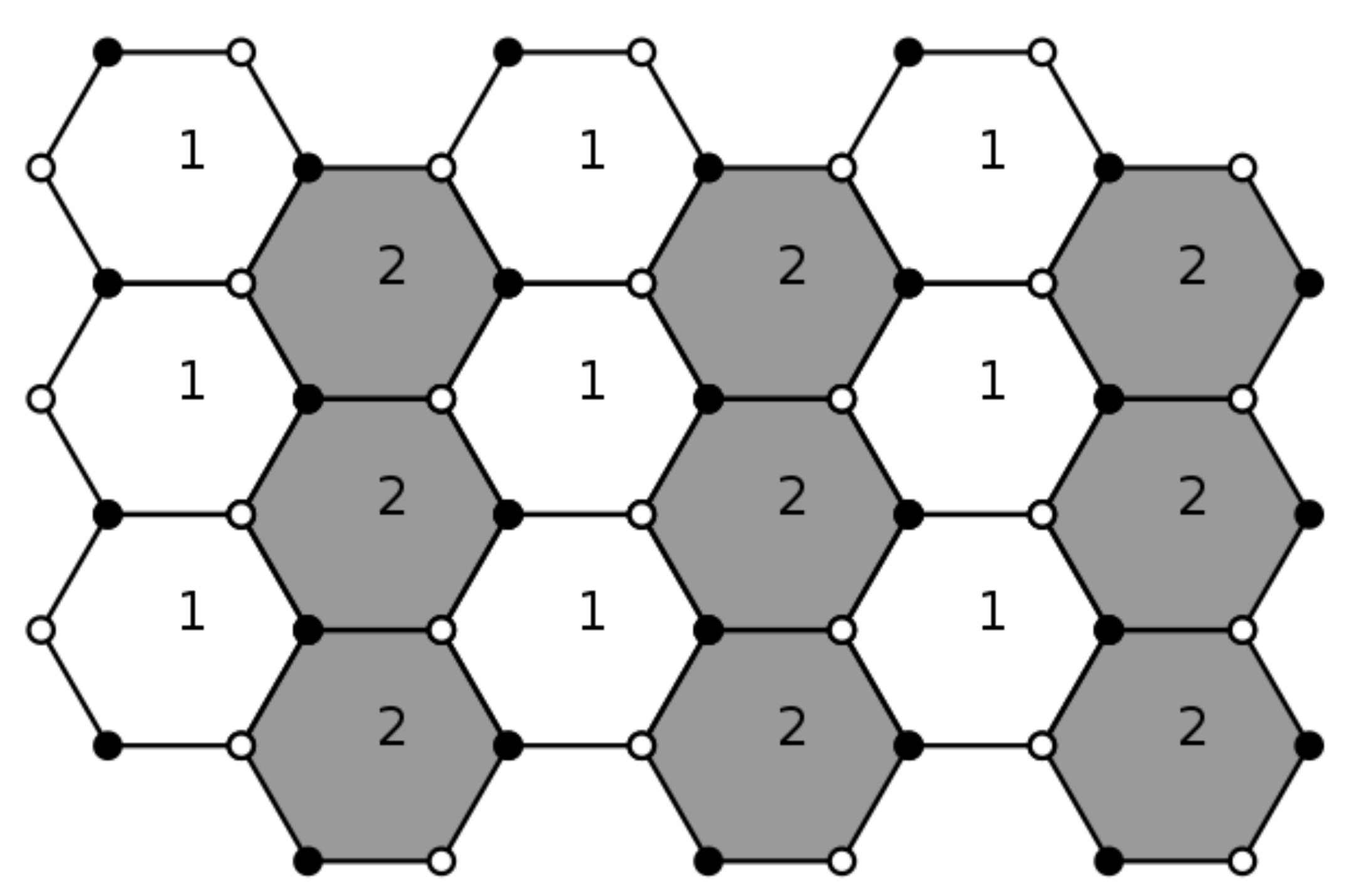}}
		\caption{The dimer of the $\mathbb{C}^2/\mathbb{Z}_2 \times \mathbb{C}$ singularity with its ${\cal N}=2$ fractional branes.}
		\label{fig:boat12}
	\end{figure}
	
	Note that, {\it e.g.} in figure \ref{fig:boat10} the sequence 1-2-3 is not a strip, since going along it white nodes are not always kept on the left. Equivalently, the sub-dimer 1-2-3 is anomalous, since there is not the same number of ingoing and outgoing arrows for faces 1 and 3. The same applies to the sequence 4-5-6. In fact, the $dP3$ singularity does not admit ${\cal N}=2$ branes at all. 
		
	\item DSB branes
	
	Any other kind of anomaly free rank assignment usually leads to a dynamically generated superpotential and hence breaks supersymmetry dynamically \cite{Affleck:1983mk} (usually into runaway directions \cite{Berenstein:2005xa,Franco:2005zu,Bertolini:2005di}). These branes are called DSB fractional branes.
\end{itemize}

Generically, combinations of fractional branes may provide other types of fractional branes. For example, the combination of a ${\cal N}=2$  fractional brane with a deformation fractional brane may correspond to a DSB one. Similarly, the combination of two deformation fractional branes can be a DSB brane, when the corresponding complex structure deformations are incompatible with each other. 

\subsection{Orientifold rules} 
\label{OrientRules}

Let us now see how to construct an orientifold field theory, starting from a dimer diagram. We closely follow \cite{Franco:2007ii}.

To construct an orientifold, one should mod out the dimer by a $\mathbb{Z}_2$ involution, provided a $\mathbb{Z}_2$ symmetry exists in the dimer. There are two ways to obtain an orientifold projection. There can be point reflections or line reflections. The rules one should follow to get the corresponding gauge theory are similar and are summarized below. 

\begin{enumerate}
	\item Every face reflected onto itself becomes an $SO(N)$ or a $USp(N)$ group, depending on the O-plane charge, + or $-$ respectively. Such faces are those on top of orientifold points or lines.  All other faces get identified with their reflection and remain associated to $SU(N)$ groups.
	
	\item Every edge on top of an orientifold point or line becomes a chiral superfield in symmetric or antisymmetric representations (or their conjugate), depending again on the O-plane charge. All other edges get identified with their images, and remain bifundamental fields. Finally, for point reflections each white node is reflected onto a black node and viceversa such that the orientifold projection produces an orientation reversal in general.  
	
	\item O-plane charges cannot be arbitrary. In case of point reflection, the number of plus signs depends on the number of vertices of the parent theory. If the number of white vertices in the unit cell is even, allowed orientifold projections need an even number of + signs, otherwise an odd one. In the case of line reflection we have either two fixed loci (``horizontal" lines) or one (``diagonal" lines), and there are no constraints on the signs. 
	
	\item The superpotential is given by projecting all fields with the orientifold and then keeping only half of the terms. 
	\end{enumerate}

As a concrete example, we work out again the $PdP4$ model, whose dimer, including fixed points under the orientifold projection, is reported in figure \ref{fig:PdP42}. 
\begin{figure}[h!]
	\centerline{\includegraphics[width=0.27\linewidth]{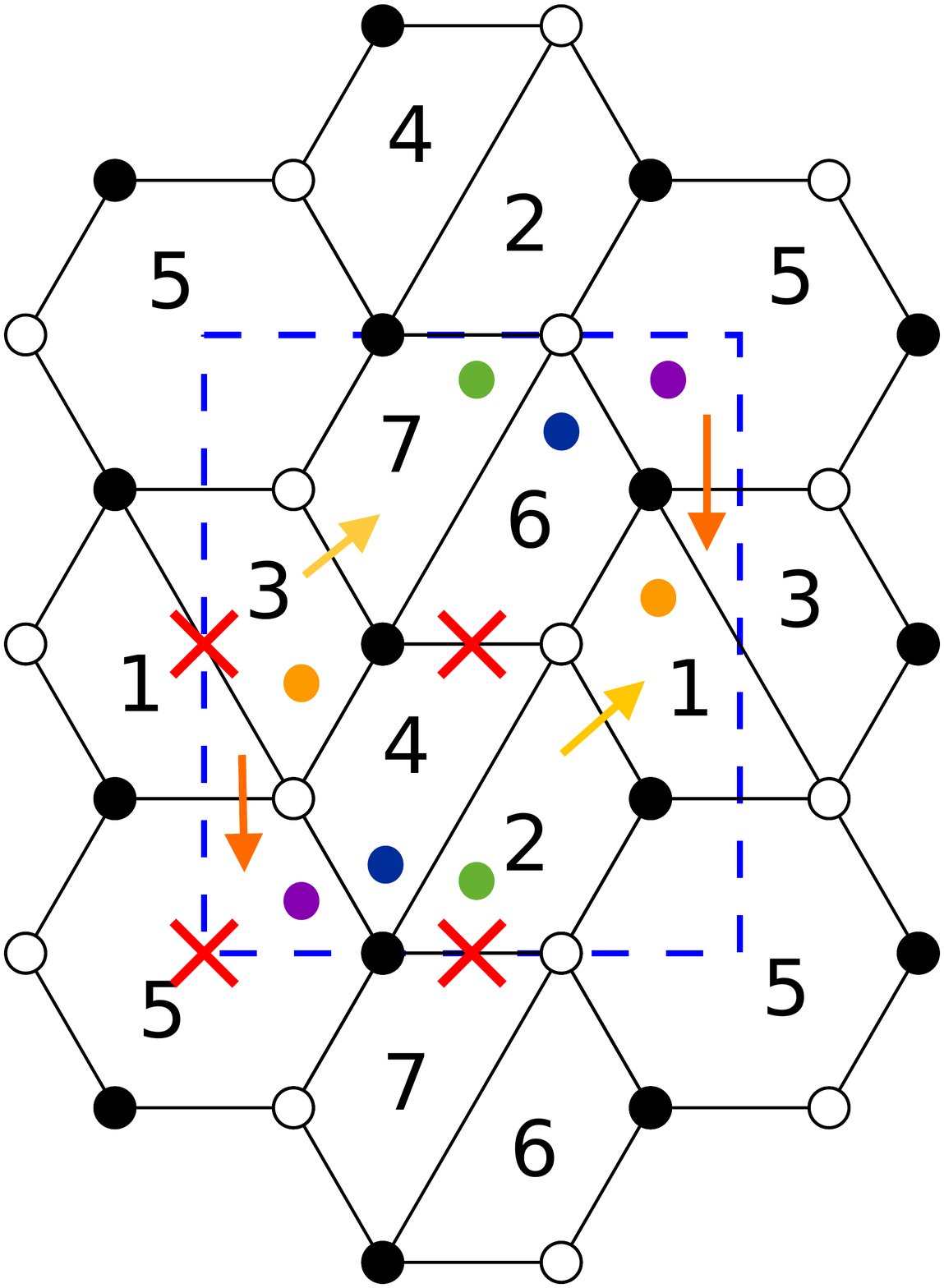}}
	\caption{The dimer of the PdP4 orientifold point singularity. Spots with the same color are mapped into each other (and so the corresponding faces).}
	\label{fig:PdP42}
\end{figure}

In a dimer, there are four points which are fixed by the $\mathbb{Z}_2$ action, if any. Such fixed points are indicated with crosses $\times$ in the figure. To facilitate the operation, faces which are identified under the reflection operation are indicated with spots of the same color in the figure. We see that face 1 is mapped to face 3 and hence they become a unique gauge factor in the orientifold theory, and similarly faces 2-7 and 4-6. Face 5, instead, is reflected into itself.

The orientifold theory will then have a $ SU(N_1)\times SU(N_2)\times SU(N_4) \times SO/USp(N_5)$ gauge group, where the last gauge group is orthogonal or symplectic depending on the charge of the O-plane,\footnote{In our conventions $USp(N) = Sp(N/2)$, with $N$ even. In this way $N$ is always the number of D-branes in the parent theory.} ranks can be different but should be compatible with anomaly cancellation conditions, and we have chosen faces 1, 2, 4 and 5 to survive the orientifold projection.

As for matter fields, in the parent theory we have the following bifundamentals
\begin{eqnarray}
&X_{15},\hspace{0,2 cm} X_{54},\hspace{0,2 cm} X_{43},\hspace{0,2 cm} X_{31}, &\nonumber
\\
&X_{27},\hspace{0,2 cm} X_{76},\hspace{0,2 cm} X_{65},\hspace{0,2 cm} X_{52}, &\nonumber
\\
&X_{16},\hspace{0,2 cm}X_{64},\hspace{0,2 cm}X_{42},\hspace{0,2 cm}X_{21}, &\nonumber
\\
&X_{37},\hspace{0,2 cm} X_{75},\hspace{0,2 cm} X_{53}. &\nonumber
\end{eqnarray}

Let us consider, for example, $X_{15}$ and $X_{53}$. Since faces 1 and 3 are identified, these two fields correspond to one single field in the orientifold theory, transforming as $(\wb {\tiny  \yng(1)}_{\, 1} , \Box_5)$. Similar identifications holds for the following fields
\begin{eqnarray}
& X_{52}-X_{75},\hspace{0,2 cm}X_{54}-X_{65} &\nonumber
\\
& X_{76}-X_{42},\hspace{0,2 cm} X_{21}-X_{37},\hspace{0,2 cm}X_{43}-X_{16} &\nonumber
\end{eqnarray}
From these identifications we then get the following bifundamentals
\begin{eqnarray}
&&(\wb {\tiny  \yng(1)}_{\, 1} , {\tiny  \yng(1)}_{\, 5}), \hspace{1 cm} ({\tiny  \yng(1)}_{\, 5} , {\tiny  \yng(1)}_{\, 2}), \hspace{1 cm} ({\tiny  \yng(1)}_{\, 5} , {\tiny  \yng(1)}_{\, 4})\nonumber
\\
&&(\wb {\tiny  \yng(1)}_{\, 4} , {\tiny  \yng(1)}_{\, 2}),   \hspace{1 cm} (\wb {\tiny  \yng(1)}_{\, 2} , {\tiny  \yng(1)}_{\, 1}),  \hspace{1 cm} (\wb {\tiny  \yng(1)}_{\, 4} ,\wb {\tiny  \yng(1)}_{\, 1})~. \nonumber
\end{eqnarray}
We have neglected the bar, if any, for group 5, since that will be either a $SO$ or $USp$ group. For the other groups, we have followed the convention that a field $X_{ij}$ transforms in the  $(\wb {\tiny  \yng(1)}_{\, i} , {\tiny  \yng(1)}_{\, j} )$ and identified the  representation accordingly. 

A different class of identifications holds for the following fields
\begin{eqnarray}
&X_{27}~,\hspace{0,2 cm} X_{31}~,\hspace{0,2 cm} X_{64}~.&\nonumber
\end{eqnarray}
These fields are self-identified under the orientifold projection and become symmetric or antisymmetric of the surviving gauge group, depending on the O-plane charge. Specifically, we get
\begin{eqnarray}
{\tiny  \wb {\yng(1,1)}_{\, 2}} ~~\text{or}~~ {\tiny  \wb {\yng(2)}_{\,2}}~, \hspace{1 cm} {\tiny \yng(1,1)_{\,1}} ~~\text{or}~~ {\tiny \yng(2)_1}~, \hspace{1 cm} {\tiny \yng(1,1)_{\,4}} ~~\text{or}~~ {\tiny \yng(2)_{\,4}}~. \nonumber
\end{eqnarray}
The conjugate arises, again, depending on whether the field in the parent theory transforms in the fundamental or anti-fundamental representation. 

Orientifold field theories may admit fractional branes. Looking at the dimer of the mother theory and at the orientifold action, one can read off the dimer the fractional branes surviving the orientifold projection. However, the nature of fractional branes can change upon orientifolding \cite{Argurio:2017upa}. The basic reason is that orientifold planes may carry fractional brane charge and can then affect the brane dynamics.


\end{document}